\documentclass[review, brazil, english]{elsarticle}
\biboptions{numbers,sort&compress,semicolon}

\makeatletter
\def\ps@pprintTitle{%
  \let\@oddhead\@empty
  \let\@evenhead\@empty
  \let\@oddfoot\@empty
  \let\@evenfoot\@oddfoot
}
\makeatother

\UseRawInputEncoding
\usepackage[titletoc]{appendix}
\usepackage{supertabular}
\usepackage{lineno,hyperref} %
\usepackage{siunitx}
\usepackage{soul}
\usepackage{listings}
\usepackage{setspace,lipsum}
\usepackage{comment}
\usepackage{multirow}
\usepackage{xcolor}
\usepackage{lscape}
\usepackage{bm}

\usepackage[T1]{fontenc}
\hypersetup{pdfauthor=author}
\usepackage[utf8]{inputenc}
\usepackage{tikz}
\usetikzlibrary{shapes.multipart}
\usepackage{xcolor}
\usepackage{framed}
\usepackage{pdflscape}
\usepackage{appendix}
\usepackage{booktabs}

\colorlet{punct}{red!60!black}
\definecolor{background}{HTML}{EEEEEE}
\definecolor{delim}{RGB}{20,105,176}
\colorlet{numb}{magenta!60!black}

\modulolinenumbers[5]

\usepackage[skip=0pt, margin = 0pt, font=normal]{subcaption}
\usepackage{bm,amsmath,amssymb}
\usepackage{tikz}
    \usetikzlibrary{arrows, positioning, shapes}

\lstdefinelanguage{json}{
	basicstyle=\small \ttfamily,
	numberstyle=\scriptsize,
	stepnumber=1,
	numbersep=8pt,
	showstringspaces=false,
	breaklines=true,
	frame=lines,
	backgroundcolor=\color{background},
	literate=
	*{0}{{{\color{numb}0}}}{1}
	{1}{{{\color{numb}1}}}{1}
	{2}{{{\color{numb}2}}}{1}
	{3}{{{\color{numb}3}}}{1}
	{4}{{{\color{numb}4}}}{1}
	{5}{{{\color{numb}5}}}{1}
	{6}{{{\color{numb}6}}}{1}
	{7}{{{\color{numb}7}}}{1}
	{8}{{{\color{numb}8}}}{1}
	{9}{{{\color{numb}9}}}{1}
	{:}{{{\color{punct}{:}}}}{1}
	{,}{{{\color{punct}{,}}}}{1}
	{\{}{{{\color{delim}{\{}}}}{1}
	{\}}{{{\color{delim}{\}}}}}{1}
	{[}{{{\color{delim}{[}}}}{1}
	{]}{{{\color{delim}{]}}}}{1},
}

\begin{document}

\begin{frontmatter}

\title{CFD-DEM study of mixing in a monodispersed solid-liquid fluidized bed}

\author[ppgeq,polytechnique]{Victor Oliveira Ferreira}
\author[polytechnique]{Bruno Blais}
\author[ppgeq,deq]{Gabriela Cantarelli Lopes\corref{mycorrespondingauthor}}
\cortext[mycorrespondingauthor]{Corresponding author.}
\ead{gclopes@ufscar.br}

\address[ppgeq]{Chemical Engineering Graduate Program, Federal University of S\~ao Carlos, Rod. Washington Luiz, Km 235 - SP 310, S\~ao Carlos, SP 13565-905, Brazil}

\address[deq]{Chemical Engineering Department, Federal University of S\~ao Carlos, Rod. Washington Luiz, Km 235 - SP 310, S\~ao Carlos, SP 13565-905, Brazil}

\address[polytechnique]{Chemical engineering High-performance Automatization Optimization and Simulation (CHAOS), Department of Chemical
Engineering, Polytechnique Montr\'eal, PO Box 6079, Stn Centre-Ville, Montr\'eal,
QC, Canada, H3C 3A7}

\begin{abstract}

In this work, we assess the internal dynamics of particles in liquid-solid fluidized beds using an unresolved CFD-DEM model. We use the Nearest Neighbors Method (NNM) and the mixing index based on the principal component analysis proposed by \citet{Doucet_2008} to quantify the bed mixing for several flow regimes and particle properties. Discussions on the advantages and drawbacks of each method are provided. Using results for several inlet flow rates, we show that the dimensionless mixing time reaches a plateau and does not increase significantly with the inlet velocity. The principal component analysis shows that the studied fluidized bed has no preferential mixing component. Additionally, we show that, except for the sliding friction coefficient, the collision properties have almost negligible influence on the mixing behavior.

\end{abstract}


\end{frontmatter}



\section{Introduction}

Liquid-solid fluidized beds (LSFB) are applicable to various processes in which enhanced heat and/or mass transfer between particles and liquids are desired \cite{Epstein_2003}. In these equipment, the drag force counterbalances the apparent mass of the particles, agitating particles in a dynamically stable and uniform fluidized state \cite{Epstein_2003_BOOK}. Though the interphase momentum exchange plays the most important role in the fluidization state, the collisions between particles are an important part of the fluidization dynamics and should be taken into account. Anaerobic fluidized bed reactors (AFBRs), for example, rely on uniform fluid-particle and particle-particle interactions through the equipment's volume to control the removal rate of the treated matter from the particle's surface, consequently, mixing plays an important role in the process's efficiency \cite{Zamani_2015}.

Measuring the degree of mixing of particles \textit{in-situ} is not trivial, especially in liquid-particle systems. It is challenging to track the movement of individual particles, and, consequently, detect abnormal distributions of particles or inefficient mixing through the equipment's volume. 

One alternative to studying the mixing behavior in liquid-solid systems avoiding experimental limitations is to use simulations. The unresolved Computational Fluid Dynamics-Discrete Elements Method (CFD-DEM) \cite{Zhou_2010, Berard_2020}, validated to several fluid-particle systems \cite{DiRenzo_2011, Pepiot_2012, Blais_2016, Geitani_2023, Ferreira_2023}, stands out for being able to simulate fluid-particle and particle-particle interactions in systems with hundreds of thousands of particles at a reasonable computational cost \cite{Berard_2020, Norouzi_2016_BOOK}. 

The unresolved CFD-DEM is particularly powerful in the study of particles' mixing compared to other approaches, such as the Two-Fluid Model (TFM), since the DEM allows for individual particle tracking \cite{Norouzi_2016_BOOK}. Knowing the positions of particles over time allows the use of several methods for mixing measurement, including variance reduction ratio \cite{Danckwerts_1953}, nearest neighbors method (NNM), neighbor distance method, Average height method \cite{Fan_1970}, mixing entropy \cite{Arntz_2008}, mixing segregation index \cite{Eitzlmayr_2015}, Lacey mixing index \cite{Lacey_1954}, Doucet mixing index \cite{Doucet_2008}, and others \cite{Bhalode_2020}.

In these methods, mixing is expressed as an index varying between 0 (for unmixed systems) and 1 (for fully mixed systems). Usually, the system is considered fully mixed when the mixing index is $> 0.90$, and the time it takes for the value to be achieved ($t_{90}$) is used to express mixing time, allowing for the comparison between mixing methods, regimes, and equipment designs. Recently, we have validated the unresolved CFD-DEM method \cite{Ferreira_2023} against liquid-solid fluidized bed experiments \cite{Ferreira_2023a} at a wide range of flow regimes. However, to the best of our knowledge, unresolved CFD-DEM was not used to study the mixing of particles within the bed. In fact, few studies on mixing in liquid-solid fluidized beds have been reported, and most of those efforts focus on bidisperse systems \cite{Escudie_2006, Peng_2016, Khan_2020, Xie_2021, Xie_2022}.

Given the importance of solid agitation and mixing for the conversion in liquid-solid reactors and given the lack of reports on the matter, in this work, we assess the mixing behavior of particles in a liquid-solid fluidized bed using techniques based on the discrete information of particles obtained from unresolved CFD-DEM simulations. To do so, we measure the degree of mixing of three different monodispersed beds at a wide range of regimes ($\mathrm{Re}$ from 534 to
19021) using NNM \cite{Fan_1970} and the mixing index by \citet{Doucet_2008}. We discuss the effect of the regime on the mixing time and assess the principal component of mixing. Additionally, the role of particles' collisional properties (Young's modulus, coefficient of restitution, sliding and rolling friction) in the mixing behavior was also investigated. Finally, we show the influence of regimes and particle properties on the solid mixing rate.

\section{Governing equations}

We briefly describe the model formulation used in this work. A full description of the model can be found in previous papers \cite{Geitani_2023a, Geitani_2023, Geitani_2023b, Ferreira_2023, Geitani_2023c}. For a more fundamental description of the Volume-Average Navier-Stokes equations, we refer the reader to \citet{Zhou_2010}, \citet{Berard_2020}, \citet{Gidaspow_1994_BOOK}, and \citet{Geitani_2023a}; and for the DEM formulation, we refer to \citet{Zhu_2007}, \citet{Blais_2019}, and \citet{Golshan_2022}.

\subsection{Solid phase formulation}

For each particle, we apply Newton's second law of motion. The translational and angular components of the momentum are, respectively:
\begin{align}
\label{eqn3::DEM_linear}
m_i \frac{d\bm{v}_i}{dt} &= \sum_{j=1,j\neq i}(\bm{f}_{c, ij}) + \sum_{j,j\neq i}(\bm{f}_{nc, ij}) + \bm{f}_{pf, i} + \bm{f}_{g, i} \\
\label{eqn3::DEM_tangential}
I_i \frac{d\bm{\omega}_i}{dt} &= \sum_{j, j\neq i}(\bm{M}_{t, ij} + \bm{M}_{r, ij})
\end{align}
where $i$ and $j$ are particles, $\bm{v}$ is the particle's velocity, $\bm{f}_c$ is the set of contact forces, $\bm{f}_{g}$ is the gravitational force, $\bm{f}_{pf}$ the summation of fluid-particle interacting forces, and $\bm{M}_t$ and $\bm{M}_r$ are the tangential and rolling friction torques.

Particles' collisions are simulated using the soft-sphere model proposed by \citet{Cundall_1979}:
\begin{equation}
\label{eqn3::DEM_contact}
\bm{f}_{c,ij} = \bm{f}_{cn,ij} + \bm{f}_{ct,ij} = - k_{n, ij} \bm{\delta}_{n, ij} - \gamma_{t,ij} \dot{\bm{\delta}}_{n, ij} - k_{t, ij} \bm{\delta}_{t, ij} - \gamma_{t,ij} \dot{\bm{\delta}}_{t, ij}
\end{equation}
where the subscripts $n$ and $t$ mean normal and tangential, respectively. $k_{n, ij}$ and $k_{t, ij}$ are the stiffness, while $\gamma_{n,ij}$ and $\gamma_{n,ij}$ are the damping coefficients of the colliding pair, calculated according to the equations in Table \ref{tab3::DEM_equations}.

\begin{table}
\centering
\caption{DEM equations}
\begin{tabular}{lr}
\hline
\textbf{Property}                    & \textbf{Equation}                                                                                                         \\ \hline
Radius of particle i                 & $R_i$                                                                                                                     \\
Distance from the contact point & $\bm{r}_{i,j}$                                                                                                        \\
Equivalent mass                      & $\frac{1}{m^{*}_{ij}} = \frac{1}{m_{i}} + \frac{1}{m_{j}}$                                                                \\
Equivalent radius                    & $\frac{1}{R^{*}_{ij}} = \frac{1}{R_{i}} + \frac{1}{R_{j}}$                                                                \\
Equivalent Young's modulus           & $\frac{1}{Y^{*}_{ij}} = \frac{1-\nu^{2}_i}{Y_i} + \frac{1-\nu^{2}_j}{Y_j}$                                                \\
Equivalent shear modulus             & $\frac{1}{G^{*}_{ij}} = \frac{2(2+\nu^{2}_i)(1-\nu^{2}_i)}{G_i} + \frac{2(2+\nu^{2}_j)(1-\nu^{2}_j)}{G_j}$                \\
Normal stiffness                     & $k_{n, ij} = \frac{4}{3} Y^{*}_{ij} \sqrt{R^{*}_{ij} \bm{\delta_{n,ij}}}$                                                     \\
Tangential stiffness                 & $k_{t, ij} = 8 G^{*}_{ij} \sqrt{R^{*}_{ij} \bm{\delta_{n,ij}}}$                                                  \\
Normal damping                       & $\gamma_{n,ij} = -2 \sqrt{\frac{5}{6}} \frac{ln(e)}{\sqrt{ln^2(e)+\pi^2)}} \sqrt{\frac{2}{3}k_{n,ij}m^{*}_{ij}}$          \\
Tangential damping                   & $\gamma_{t,ij} = -2 \sqrt{\frac{5}{6}} \frac{ln(e)}{\sqrt{ln^2(e)+\pi^2)}} \sqrt{k_{t,ij}m^{*}_{ij}}$ \\ \hline
\end{tabular}
\label{tab3::DEM_equations}
\end{table}

The tangential and rolling friction torques are calculated, respectively, as:
\begin{align}
\label{eqn3::DEM_torque_tan}
\bm{M}_{t,ij} &= \bm{r}_{i,j} \times (\bm{f}_{ct,ij})\\
\label{eqn3::DEM_torque_rol}
\bm{M}_{r,ij} &= -\mu_{r,ij} \left | \bm{f}_{ct,ij} \right | \frac{\bm{\omega}_{ij}}{\left | \bm{\omega}_{ij} \right |} R^{*}_{ij}
\end{align}
where the coefficient of rolling friction ($\mu_{r,ij}$) and the equivalent radius ($R^{*}_{ij}$) are calculated by the equations in Table \ref{tab3::DEM_equations}.

The particle-fluid interaction forces are calculated as:
\begin{equation}
\label{eqn3::DEM_fpf}
\bm{f}_{pf, i} = \bm{f}_{d,i} + \bm{f}_{\nabla p,i} + \bm{f}_{\nabla \cdot \bm{\tau},i} + \bm{f}_{Ar,i} + \bm{f}_{\mathrm{Saff},i}
\end{equation}
where $\bm{f}_{d,i}$ is the drag force, $\bm{f}_{\nabla p,i}$ is the pressure gradient force, $\bm{f}_{\nabla \cdot \bm{\tau},i}$ is the particle-fluid shear force, and $\bm{f}_{\mathrm{Saff},i}$ is the Saffman lift force. Each of these terms is explained in subsection \ref{sec3::interphase_momentum_transfer}. Other forces, namely Basset, Magnus, and virtual mass, were neglected \cite{DiRenzo_2007, Ferreira_2023}.

\subsection{Liquid phase modeling}

The liquid phase in the unresolved CFD-DEM approach is modeled using the Volume-Averaged Navier-Stokes equations (VANS). The model type A \cite{Gidaspow_1994_BOOK} (also known as model set II \cite{Zhou_2010}) of the VANS equations is used in the present work \cite{Geitani_2023a, Ferreira_2023}.
\begin{equation}
\label{eqn3::VANS_continuity}
\frac{\partial \varepsilon_f }{\partial t} + \nabla\cdot (\varepsilon_f \bm{u}) = 0
\end{equation}
\begin{equation}
\label{eqn3::VANS_model}
\rho_{f} \left [  \frac{\partial \varepsilon_f \bm{u}}{\partial t} + \nabla \cdot (\varepsilon_f \bm{u} \otimes \bm{u})  \right ] = -\varepsilon_f \nabla p + \varepsilon_f \nabla \cdot \bm{\tau} - \bm{F}_{pf}
\end{equation}

In Eq. \eqref{eqn3::VANS_model}, $\bm{\tau}$ is the deviatoric stress tensor
\begin{equation}
\label{eqn3::VANS_Viscous}
\bm{\tau} = \mu \left [ \left ( \nabla \cdot \bm{u} \right ) + \left ( \nabla \cdot \bm{u} \right )^T  - \frac{2}{3}  \left ( \nabla \cdot \bm{u} \bm{I} \right) \right ]
\end{equation}
where $\mu$ is the dynamic viscosity and $\bm{I}$ is the identity tensor, and $\bm{F}_{pf}$ is the fluid-particle momentum exchange (source) term
\begin{equation}
\label{eqn3::VANS_Fpf}
\bm{F}_{pf} =\frac{1}{\Delta V_{\Omega_C}}\sum_{i}^{N_{p,C}}\left ( \bm{f}_{d,i} \right ) = \frac{1}{\Delta V_{\Omega_C}}\sum_{i}^{N_{p,c}}\left ( \bm{f}_{pf,i} - \bm{f}_{\nabla p} - \bm{f}_{\nabla \cdot \bm{\tau}} - \bm{f}_{Ar} \right )
\end{equation}
where the index $N_{p,C}$ stands for the number of particles inside the cell $\Omega_C$ over which the averaging is applied. The force terms in Eq. \eqref{eqn3::VANS_Fpf} are the same as in Eq. \eqref{eqn3::DEM_fpf}.

\subsection{Interphase momentum exchange} \label{sec3::interphase_momentum_transfer}

In this subsection, the terms in Eqs. \eqref{eqn3::VANS_Fpf} and \eqref{eqn3::DEM_fpf} are individually described. Details about each of the terms can be found in \citet{Crowe_2011_BOOK}.

\subsubsection{Undisturbed flow and buoyancy forces}

In Lethe, $\nabla p$ does not account for the hydrostatic pressure, hence, buoyancy forces ($\bm{f}_{Ar,i}$) need to be added separately in Eq. \eqref{eqn3::DEM_fpf}. The expressions that represent the pressure gradient and buoyancy forces are, respectively:
\begin{align}
	\label{eqn3::DEM_f_pressure}
	\bm{f}_{\nabla p,i} &= V_{p,i} \nabla p \\
	\label{eqn3::DEM_f_buoyancy}
	\bm{f}_{Ar,i} &= V_{p,i} \rho_{f} \bm{g}
\end{align}
where $V_{p,i}$ is the volume of the particle $i$, $\rho_f$ is the density of the fluid, $p$ stands for dynamic pressure, $\bm{g}$ is the gravity acceleration vector. Note that Eq. \eqref{eqn3::DEM_f_pressure} accounts for the interchange force due to the undisturbed pressure only.

The force resulting from the fluid shear stress is written as:
\begin{equation}
	\label{eqn3::DEM_f_shear}
	\bm{f}_{\nabla \cdot \bm{\tau},i} = V_{p,i} \bm{\nabla \cdot \bm{\tau}}
\end{equation}

\subsubsection{Drag force}

The drag force is defined as:
\begin{equation}
	\label{eqn3::CFD-DEM_f_drag1}
	\bm{F}_{d} = \sum_{i}^{N_{p,C}} \bm{f}_{d,i} = \sum_{i}^{N_{p,C}} \beta_i \left ( \bm{u} - \bm{v}_i \right )
\end{equation}
For model A, $\beta_i$ is the interphase momentum transfer coefficient, given by correlations in the literature \cite{Gidaspow_1994_BOOK, DiFelice_1994, Beetstra_2007, Mazzei_2007, Rong_2013}. The correlation by \citet{Rong_2013} was applied to all simulations in this work for presenting the best correspondence with the experimental data in the validation campaign \cite{Ferreira_2023a, Ferreira_2023}. In the correlation, $\beta_i$ is calculated as:
\begin{equation}
	\label{eqn3::CFD-DEM_Beta1}
	\beta_i = \frac{1}{2} C_{D0} \frac{\pi d_p^2}{4} \rho_f \left| \bm{u} - \bm{v}_i \right| G(\varepsilon_f, \mathrm{Re}_{p,i})
\end{equation}
where $C_{D0}$ is the drag coefficient for a single particle, given by the correlation by \citet{Dallavalle_1948_BOOK}
\begin{equation}
	\label{eqn3::CFD-DEM_Dallavalle}
	C_{D0} = \left( 0.63 + \frac{4.8}{\sqrt{\mathrm{Re}_p}} \right)^2
\end{equation}
and $G(\varepsilon_f, \mathrm{Re}_{p,i})$ is
\begin{equation}
	\label{eqn3::CFD-DEM_Rong}
	G(\varepsilon_{f}, \mathrm{Re}_{p,i}) = \frac{C_{D,i}}{C_{D0}} = \varepsilon_f ^ {2-\left\{ 2.65 \left( \varepsilon_f + 1 \right) - \left( 5.3 - 3.5 \varepsilon_f \right) \varepsilon_f^2 exp \left[ - \frac{\left( 1.5 - log_{10} \mathrm{Re}_{p,i} \right)^2}{2} \right] \right\}}
\end{equation}

\subsubsection{Saffman lift force}

The Saffman lift force is calculated using the Saffman-Mei lift force model, which is a combination of the equations proposed by \citet{Saffman_1965, Saffman_1968} and \citet{Mei_1992}
\begin{equation}
	\label{eqn3::saffman_lift_force}
	\bm{f}_{\mathrm{Saff},i} = 1.161 C_{\mathrm{Saff},i} d_p^2(\mu \rho_f)^{1/2} \left | \boldsymbol{\omega}_{c,i} \right |^{-1/2}\left [ (\bm{u} - \bm{v}_{i}) \times \boldsymbol{\omega}_{c,i} \right ]
\end{equation}
where $\boldsymbol{\omega}_{c,i}$ is the fluid vorticity ($\nabla \times \bm{u}$), $C_{\mathrm{Saff},i}$ is the Saffman lift coefficient
\begin{equation}
	\label{eqn3::lift_coefficient}
	\begin{aligned}
	C_{\mathrm{Saff},i} = \left\{\begin{matrix}
		\left ( 1 - 0.3314 \alpha^{1/2} \right ) \textup{exp}\left ( \frac{-\mathrm{Re}_{p,i}}{10} \right ) + 0.3314 \alpha^{1/2} \textrm{, for } \mathrm{Re}_{p,i} \leq 40
		\\
		0.0524(\alpha_l \mathrm{Re}_{p,i})^{1/2} \textrm{, for } \mathrm{Re}_{p,i} > 40
	\end{matrix}\right.
	\end{aligned}
\end{equation}
and $\alpha$ is given by:
\begin{equation}
	\label{eqn3::alpha_lift}
	\alpha = \frac{d_p}{2 \left | \bm{u} - \bm{v}_i \right |} \left | \boldsymbol{\omega}_{c,i} \right |
\end{equation}

We refer the reader to \citet{Crowe_2011_BOOK} for further details on the Saffman lift force model. Previous works show that the Saffman lift force prevents fluid preferential path through the walls, which are critical to the particle dynamics when simulating liquid fluidized beds \cite{Ferreira_2023}.


\section{Methods}

All simulations were carried out using Lethe \cite{Blais_2020, Golshan_2022, Geitani_2023, Geitani_2023a, Geitani_2023b}, an open-source CFD/DEM/CFD-DEM tool based on deal.II Finite Elements Method (FEM) library \cite{Arndt_2021, Arndt_2022}. The methodology applied to this work was previously validated against pilot-scale experimental data by us \cite{Ferreira_2023a}, as described in \citet{Ferreira_2023}. Lethe's unresolved CFD-DEM module was also validated for other gas-solid and liquid-solid systems, including fluidized beds, spouted beds, and rotary kilns \cite{Geitani_2023, Geitani_2023c}.

\subsection{Simulation setup}

The simulation setup applied to this work follows what is described in the validation study conducted by our group \cite{Ferreira_2023}. We used the pilot-scale circulating liquid-solid fluidized bed described in \citet{Ferreira_2023a} and \citet{Ferreira_2023} as a reference to the simulations. The experimental unit is composed of a 1 \si{\meter} height, 10 \si{\centi \meter} diameter cylindrical vessel, where particles are fluidized in water.

Unless mentioned, simulations follow the parameters described in Table \ref{tab::simulation_params}. Particles' collisional properties, namely Young's modulus ($Y$), coefficient of restitution ($e$), coefficient of rolling friction ($\mu_r$), and coefficient of sliding friction ($\mu_f$) were the only parameters related to particles modified throughout the study (Section \ref{sec::part_props}).

\begin{table}[!htpb]
	\centering
	\caption{Simulation parameters.}
    \label{tab::simulation_params}
\begin{tabular}{lr}
\hline
\textbf{Parameter}                                     & \textbf{Value}                                                                  \\ \hline
Time integration method                                & BDF1                                                                            \\
CFD time-step and coupling interval ($\Delta t_{CFD}$) & \SI{0.001}{\second}                           \\
DEM time-step ($\Delta t_{DEM}$)                       & \SI{0.00001}{\second}                         \\
Diameter of the cylinder ($D_C$)                       & \SI{10}{\centi\meter}          \\
Height of the cylinder ($H_C$)                         & \SI{1.10}{\meter}                             \\
Height of the inlet portion                            & \SI{10}{\centi\meter}                             \\
Mesh ($n_r \ \times \ n_\theta \ \times \ n_z$)        & $6 \ \times \ 16 \ \times \ 132$                                                \\
Liquid density ($\rho_f$) &
  \SI{996.78}{\kilo\gram\per\cubic\meter} \\
Liquid dynamic viscosity ($\mu$)                       & \SI{8.352d-4}{\pascal \second} \\
Young's modulus ($Y$)                                  & \SI{10}{\mega \pascal}         \\
Coefficient of restitution ($e$)                       & 0.9                                                                             \\
Poisson ratio ($\nu$)                                  & 0.3                                                                             \\
Coefficient of rolling friction ($\mu_r$)              & 0.2                                                                             \\
Coefficient of sliding friction ($\mu_f$)              & 0.1                                                                             \\
VANS model type                                        & A                                                                               \\
Void fraction calculation scheme                       & PCM                                                                             \\
Void fraction smoothing length                         & 2.0 $\times \ d_p$                                                              \\
Boundary conditions at the walls                       & Free slip                                                                       \\
Drag model                                             & \citet{Rong_2013}                                  \\
Lift force model                                       & Saffman-Mei\footnote{\cite{Saffman_1965, Saffman_1968, Mei_1992, Crowe_2011_BOOK}.}     \\
Gravity ($g$) &
  \SI{9.81}{\meter\per\second\squared} \\ \hline
\end{tabular}
\end{table}

\subsubsection{Mesh and initial packing of particles}

\begin{figure}[!htpb]
	\centering
	\includegraphics[width=1\textwidth]{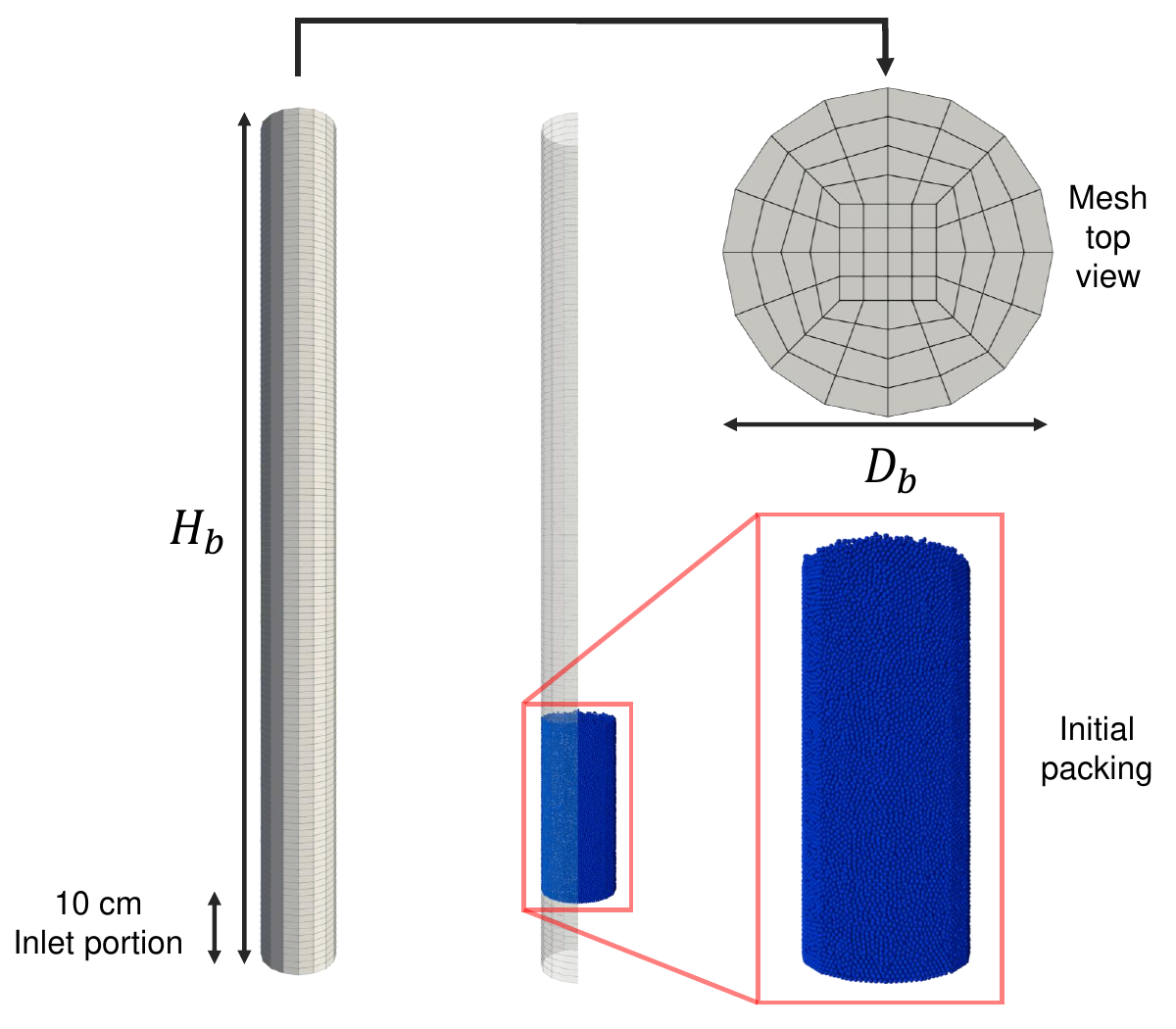}
	\caption{Schematic representation of the mesh and initial packing of particles.} 
	\label{fig3::schematic_simulation}
\end{figure}

We used deal.II's built-in tool to generate the mesh used in all simulations. The mesh is composed of a 1.10 \si{\meter} height, 10 \si{\centi \meter} diameter cylinder, as shown in Figure \ref{fig3::schematic_simulation}. The number of elements in cylindrical coordinates ($n_r \ \times \ n_\theta \ \times \ n_z$) is $6 \ \times \ 16 \ \times \ 132$. An analysis of the mesh can be found in \citet{Ferreira_2023}.

During simulations, particles are held 10 \si{\centi \meter} above the bottom of the mesh by a floating wall, leaving a free-of-particles inlet portion to the fluid (referred to as inlet portion in Figure \ref{fig3::schematic_simulation}), preventing sharp void fraction changes in the proximity of the boundaries, which can lead to numerical instabilities \cite{Geitani_2023a}. In all simulations, the fluid injection only starts after particles are fully settled above the floating wall. This is achieved by running a DEM simulation, carried out on Lethe-DEM \cite{Golshan_2022}, prior to the CFD-DEM simulation.

\subsubsection{Particles}

We used 3 different groups of particles, named A, B, and C. Particles A and C are referred to as "alginate" and "alumina" in the validation study \cite{Ferreira_2023}, as shown in Figure \ref{fig::validation_results}. For particles A and C, the parameters are the same as the experimental study. Since particle C is more than three times denser than particle A, we created a fictitious intermediary particle B with the same size as particle A and a density close to the acrylonitrile butadiene styrene (ABS) particle used in \citet{Ferreira_2023a}. The particular parameters associated with each particle are presented in Table \ref{tab::particle_params}.

\begin{landscape}
\vspace*{\fill}
\begin{table}[!htpb]
	\centering
	\caption{Particle associated parameters.}
    \label{tab::particle_params}
\begin{tabular}{lccc}
\hline
\textbf{Parameter} &
  \textbf{Particle A} &
  \textbf{Particle B} &
  \textbf{Particle C} \\ \hline
Total real-time of simulations &
  \SI{35}{\second} &
  \SI{20}{\second} &
  \SI{20}{\second} \\
Mixing index reference time ($t_0$) &
  \SI{10}{\second} &
  \SI{5}{\second} &
  \SI{5}{\second} \\
Output time-step ($\Delta t_{DEM}$) &
  \SI{0.50}{\second} &
  \SI{0.20}{\second} &
  \SI{0.20}{\second} \\
Number of particles ($N_p$) &
  107960 &
  107960 &
  72400 \\
Particle terminal velocity ($U_0$) &
  \SI{2.9}{\centi \meter \per \second} &
  \SI{20.8}{\centi \meter \per \second} &
  \SI{48.2}{\centi \meter \per \second} \\
Lowest and highest inlet velocities ($U$) &
  0.44 - 1.09 \SI{}{\centi \meter \per \second} &
  4.24 - 11.00 \SI{}{\centi \meter \per \second} &
  8.50 - 15.70 \SI{}{\centi \meter \per \second} \\
Number of inlet velocities &
  7 &
  2 &
  8 \\
Diameter of the particles ($d_p$) &
  \SI{2.66}{\milli\meter} &
  \SI{2.66}{\milli\meter} &
  \SI{3.09}{\milli\meter} \\
Density of the particles ($\rho_p$) &
  \SI{1029}{\kilo\gram\per\cubic\meter} &
  \SI{1822}{\kilo\gram\per\cubic\meter} &
  \SI{3586}{\kilo\gram\per\cubic\meter} \\ \hline
\end{tabular}
\end{table}
\vspace*{\fill}
\end{landscape}

\subsubsection{Validation}

\begin{figure}[!htpb]
	\centering
	\begin{subfigure}{.50\textwidth}
		\centering
		\includegraphics[width=1\linewidth]{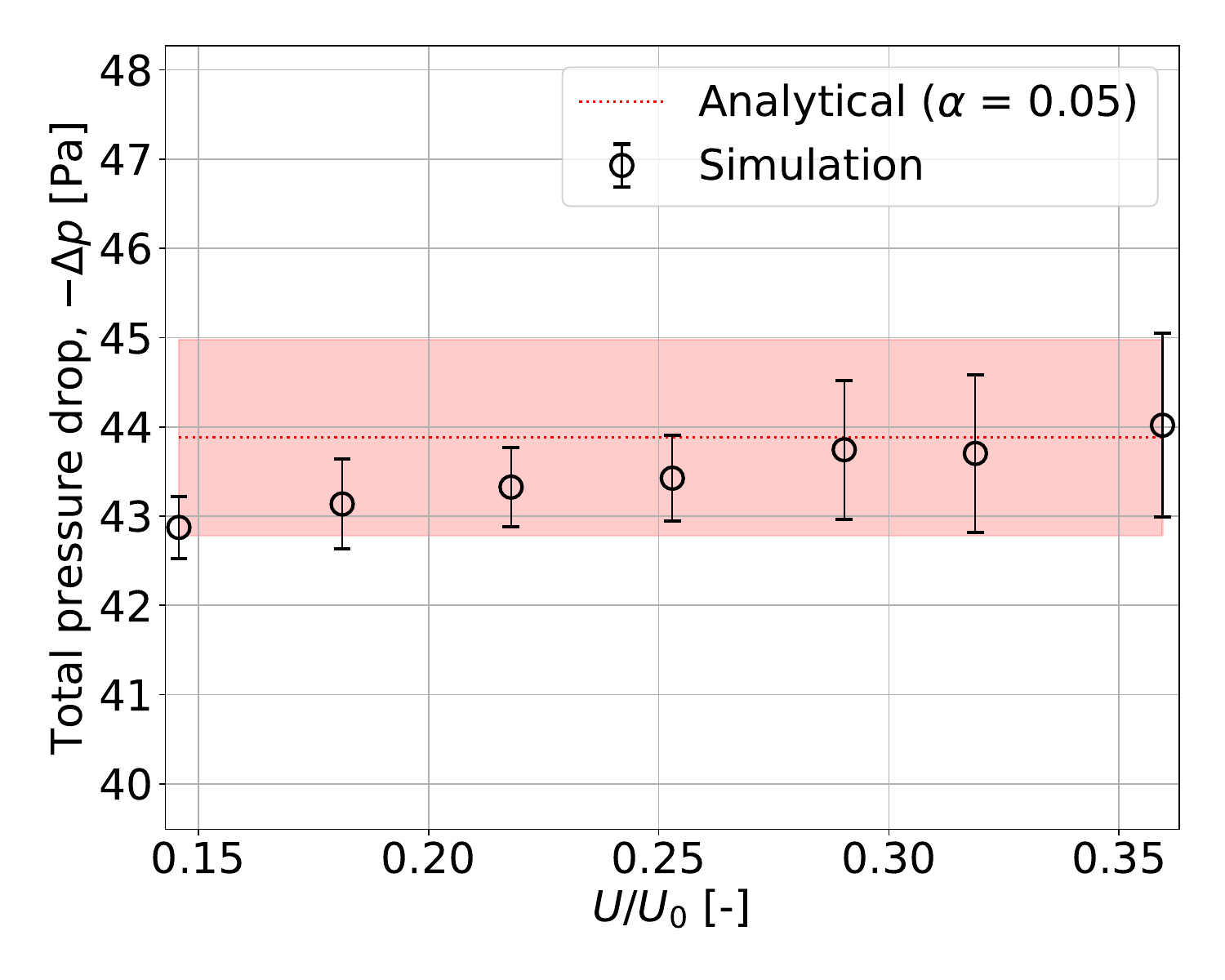}
		\label{fig::valid_pressure_alginate}
	\end{subfigure}%
	\begin{subfigure}{.50\textwidth}
		\centering
		\includegraphics[width=1\linewidth]{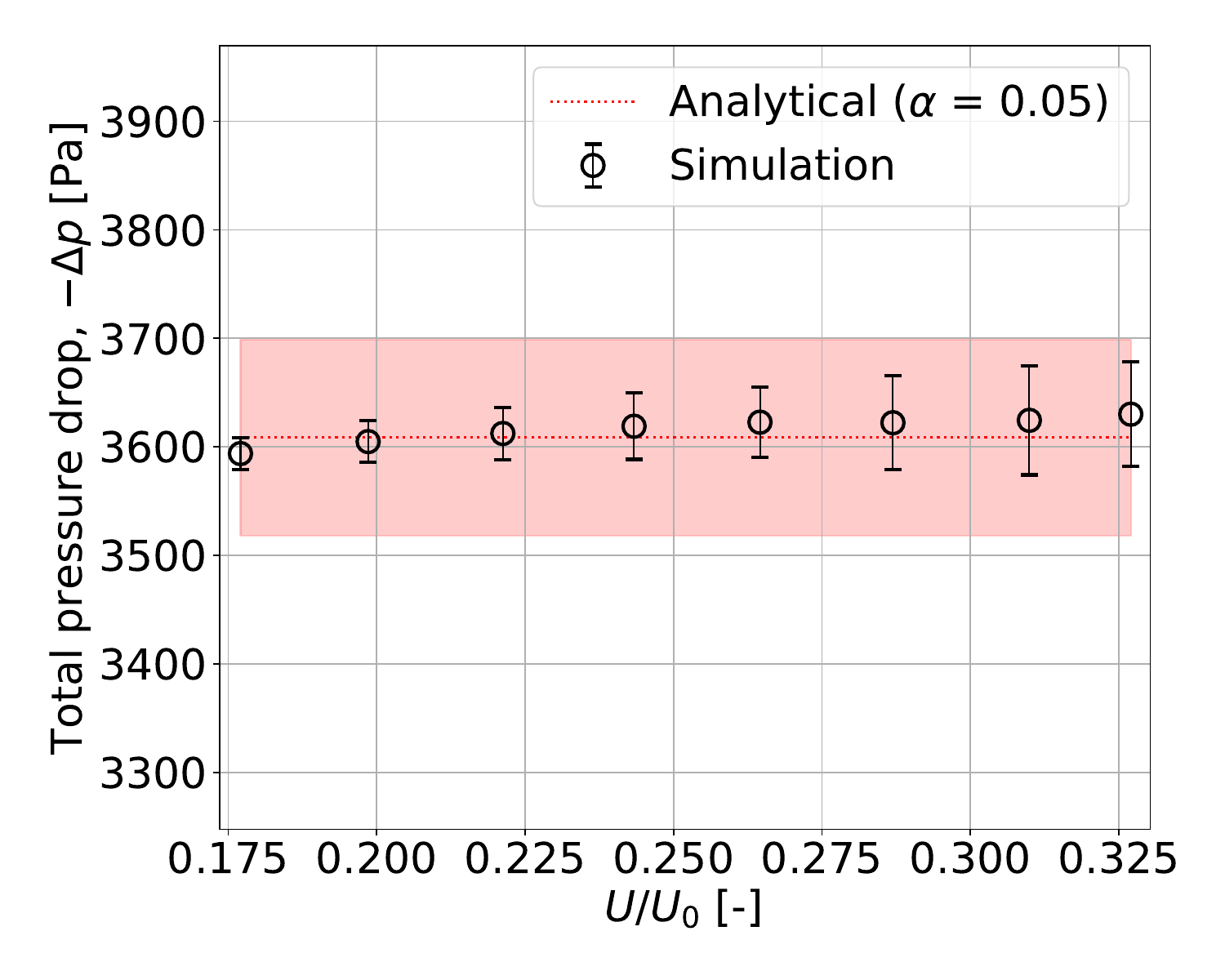}
		\label{fig::valid_pressure_alumina}
	\end{subfigure}
    \begin{subfigure}{.50\textwidth}
		\centering
		\includegraphics[width=1\linewidth]{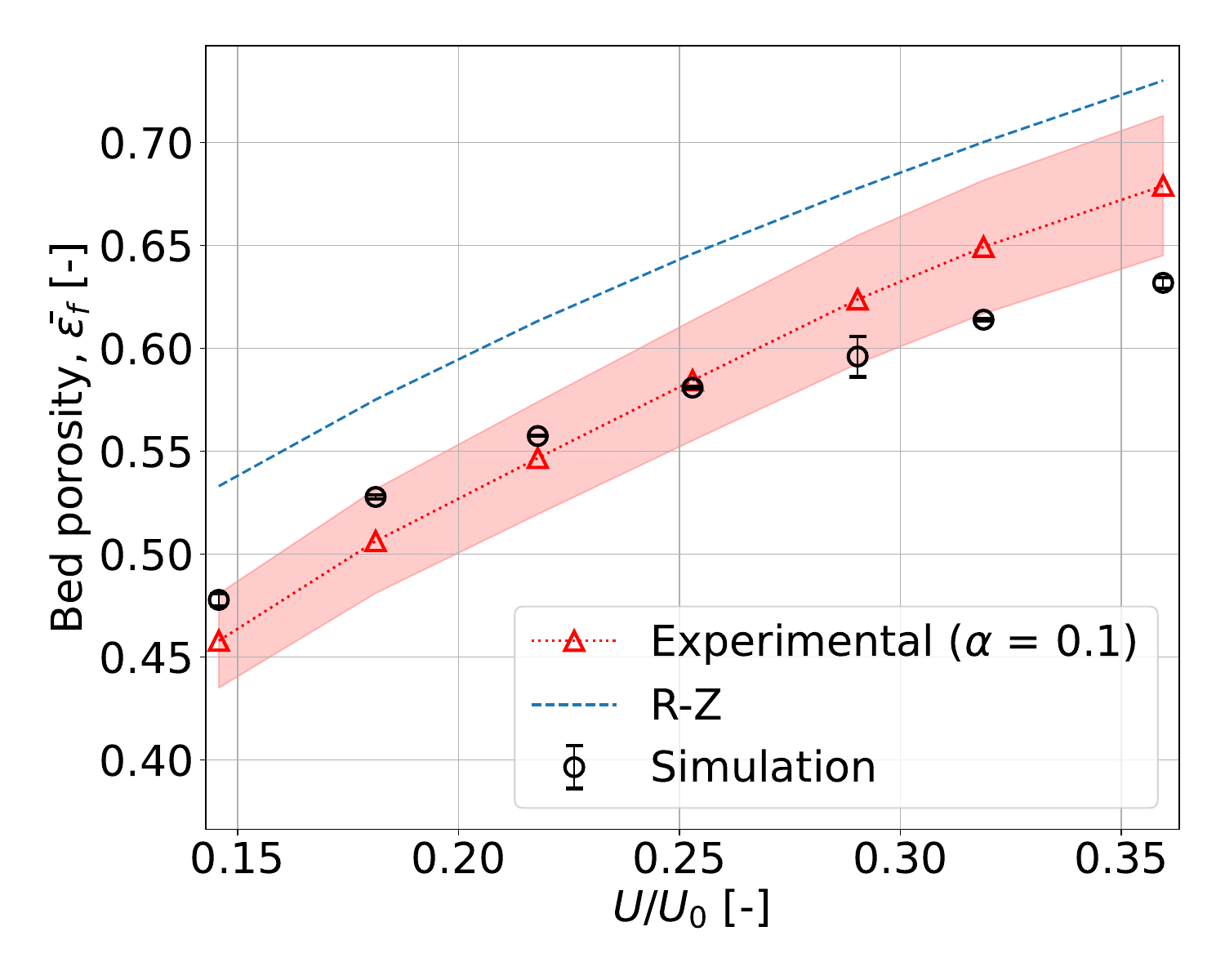}
		\caption{Alginate (Particle A)}
		\label{fig::valid_eps_alginate}
	\end{subfigure}%
	\begin{subfigure}{.50\textwidth}
		\centering
		\includegraphics[width=1\linewidth]{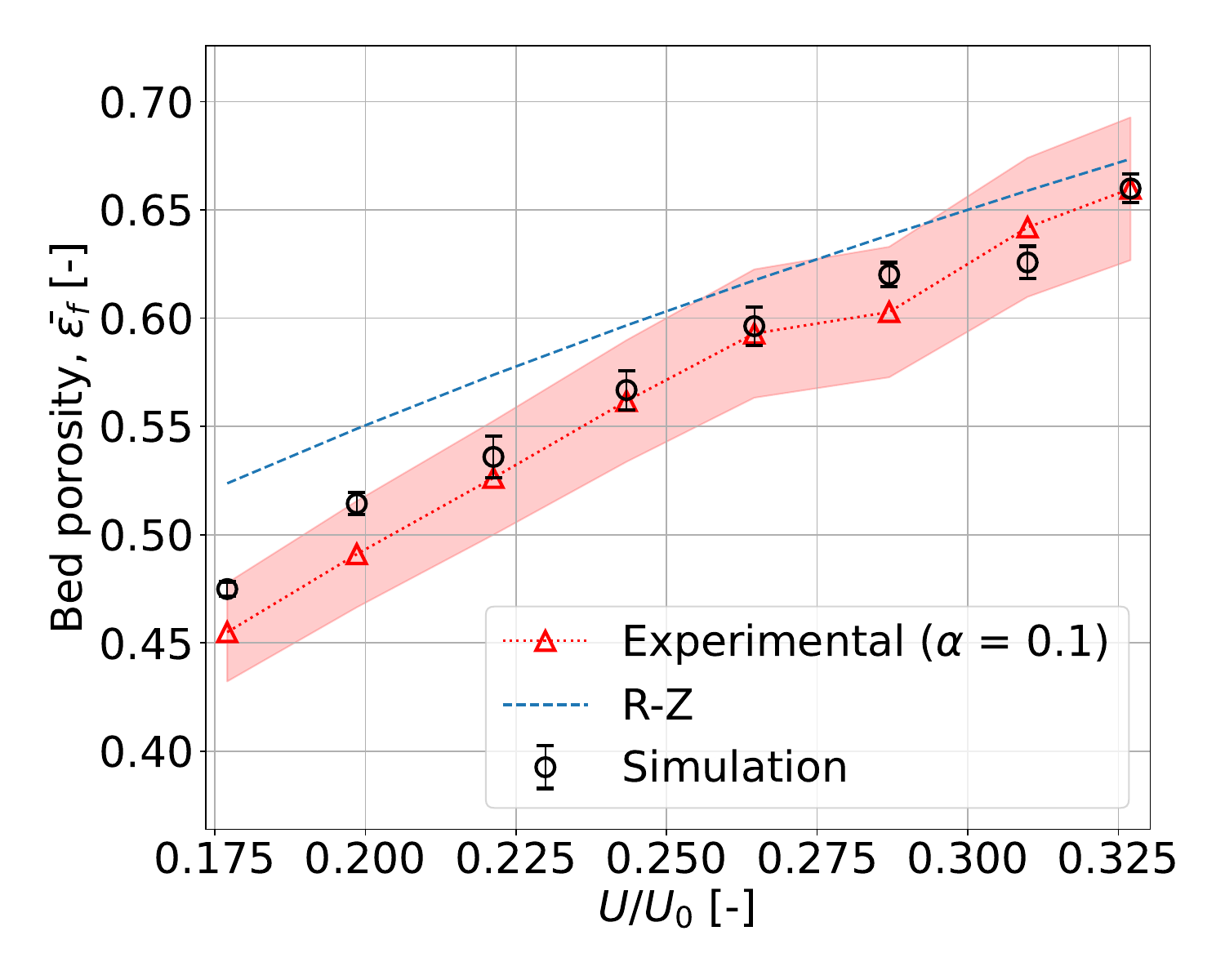}
		\caption{Alumina (Particle C)}
		\label{fig::valid_eps_alumina}
	\end{subfigure}
	\caption{Average of the total pressure difference ($-\Delta p$) and bed expansion ($\bar{\varepsilon}_f$) of the validation campaign for the (a) alginate (particle A in this work) and (b) alumina particles (particle C in this work). Adapted from \citet{Ferreira_2023}.}
	\label{fig::validation_results}
\end{figure}

The main results of the validation campaign are presented in Figure \ref{fig::validation_results}. All results of the validation step were obtained using the parameters in Table \ref{tab::simulation_params}. In the Figure, the total pressure drop is compared to the analytical pressure drop calculated by \cite{Epstein_2003_BOOK}:
\begin{equation}
	\label{eqn3::total_pressure}
	-\Delta p = \frac{M_p \left ( \rho_p - \rho_f \right )g}{\rho_p A_b}
\end{equation}
where $M_p$ is the total mass of particles in the bed, $\rho_p$ is particle's density, $\rho_f$ is fluid's density, $g$ is gravity acceleration constant, and $A_b$ is the bed's cross section area. Bed expansion results (expressed as bed average porosity, $\bar{\varepsilon}_f$) are compared to experiments and estimations using the equation by \citet{Richardson_1954} (R-Z). In the figures, $\alpha$ corresponds to a double-tail confidence interval. The simulation results correspond to time averages in the pseudo-steady-state regime, and the error bars are equivalent to one temporal standard deviation.

The results in Figure \ref{fig::validation_results} were obtained for particles A and C (Table \ref{tab::particle_params}) using the parameters in Table \ref{tab::simulation_params} and the mesh in Figure \ref{fig3::schematic_simulation}. The simulations were capable of accurately reproducing the fluidized bed behavior. For further details on the importance of the Saffman lift force, the choice of the drag correlation, and the mesh, we refer the reader to \citet{Ferreira_2023}.

\subsection{Mixing index}

Several methods can be applied to track particles' mixing using DEM information. In this work, we applied two: the nearest neighbors method (NNM) \cite{Fan_1970, Godlieb_2007} and Doucet's mixing index \cite{Doucet_2008}. Both methods were applied to Lethe-DEM results using a post-processing module written in Python using the PyVista library \cite{Sullivan_2019} (available on \href{https://github.com/lethe-cfd/lethe}{Lethe's official GitHub webpage}, explained in details on \href{https://lethe-cfd.github.io/lethe/index.html}{Lethe's official documentation}). Different from Lacey's mixing index \cite{Lacey_1954, Godlieb_2007}, the methods are grid-independent, meaning that only the particle position is required. NNM and Doucet's methods follow different principles, thus, providing different information about the particles mixing.

\subsubsection{Nearest neighborsbors method}

\begin{figure}[!htpb]
	\centering
	\frame{\includegraphics[width=0.6\textwidth]{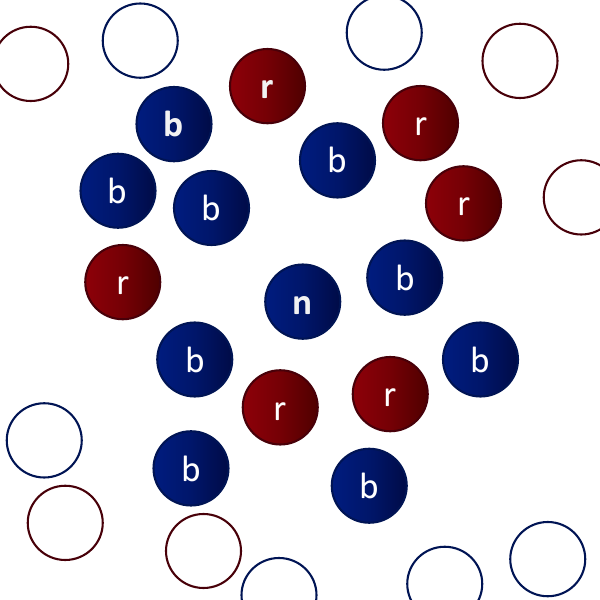}}
	\caption{Schematic bi-dimensional representation of the nearest neighbors method for mixing index calculation \cite{Godlieb_2007}. In the Figure, blue ("b") and red ("r") particles are members of two different groups. The $n^{th}$ particle, represented by the letter "n" on it, is a blue particle to which the NNM mixing index is calculated. Hollow particles are not part of the 15 nearest neighbors of the particle and, hence, are not used in the mixing index calculation.} 
	\label{fig::nnm_schematic}
\end{figure}

The method consists of splitting the particles into two groups according to their position at a given moment and tracking the number of neighbors per particle that are part of the other group. A schematic representation of the method is presented in Figure \ref{fig::nnm_schematic}.

The NNM mixing index is calculated by \cite{Godlieb_2007} as:
\begin{equation}
	\label{eqn::nnm}
	\bar{M}^{\text{\text{NNM}}}_t = 1/N_{p} \sum_{n = 0}^{N_{p}}{M_n^{\text{\text{NNM}}}} = 1/N_{p} \sum_{n = 0}^{N_{p}}{\frac{2 N_{n,\text{diff}}}{N_{\text{neighbors}}}}
\end{equation}
where $\bar{M}^{\text{\text{NNM}}}_t$ is the average mixing index of the system at time $t$, $M_n^{\text{\text{NNM}}}$ is the mixing index of the $n^{th}$ particle, $N_{p}$ is the total number of particles in the system, $N_{\text{neighbors}}$ is the number of nearest neighbors used in the method (user-defined parameter), and $N_{n,\text{diff}}$ is the number of neighbor particles that are of the opposite group of the $n^{th}$ particle. In the present work, we chose $N_{\text{neighbors}} = 15$. The KD-tree nearest neighbors method implemented in the Scikit-learn Python module \cite{Pedregosa_2011} was used to find particles' nearest neighbors.

In Figure \ref{fig::nnm_schematic}, blue and red particles are part of different groups. Filled particles are part of the 15 nearest neighbor particles of particle $n$, while hollow particles are not. Since $N_{n,\text{diff}}$ is 6 (number of filled red particles), according to Eq. \eqref{eqn::nnm}, $M_n^{\text{\text{NNM}}} = 0.8$. Note that $0 \leq M_n^{\text{\text{NNM}}} \leq 2$, so that $0 \leq \bar{M}^{\text{\text{NNM}}} \leq 1$.

In NNM, the particles must be split in half so that the coordinate components can be analyzed individually. Since we are working in a cylindrical geometry, particles are split according to their position in cylindrical coordinates (namely $r$, $\theta$, and $z$).

The moment of splitting is chosen so that the initial bed expansion, which presents different bed dynamics from the fully expanded pseudo-stationary bed, is not considered. This reference time ($t_0$) is 10 \si{\second} for all particles, which is considered the time needed for Particle A to reach the pseudo-steady state \cite{Ferreira_2023}. An example of the bed splitting per coordinate is presented in Figure \ref{fig::nnm_split}.

\begin{landscape}
\vspace*{\fill}
\begin{figure}[!htpb]
	\centering
        \subcaptionbox{Radial coordinate ($r$).\label{fig::nnm_split_radius}}
        {\includegraphics[width=.3\linewidth]{{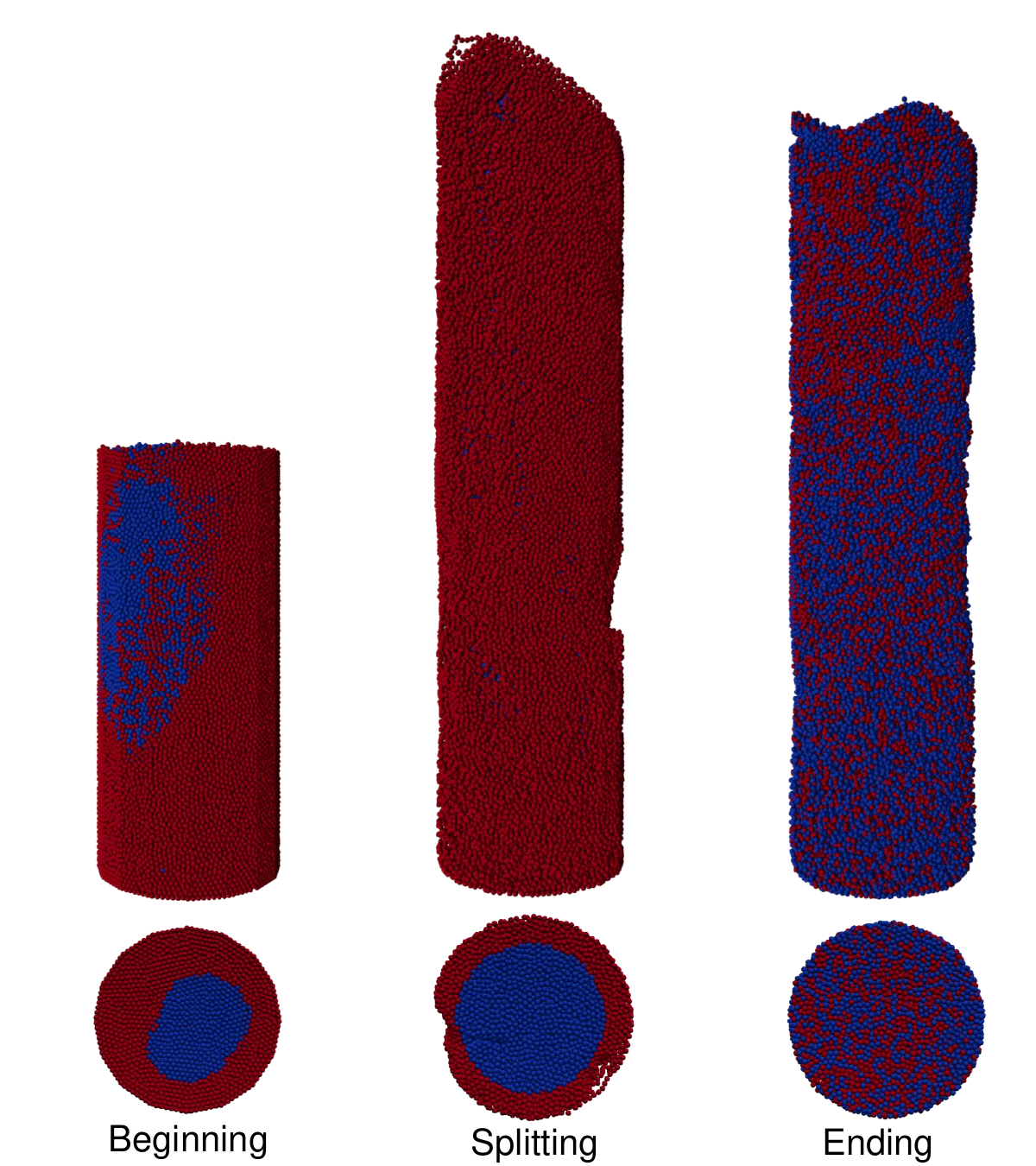}}}
        \hfill
        \subcaptionbox{Angular coordinate ($\theta$).\label{fig::nnm_split_theta}}
        {\includegraphics[width=.3\linewidth]{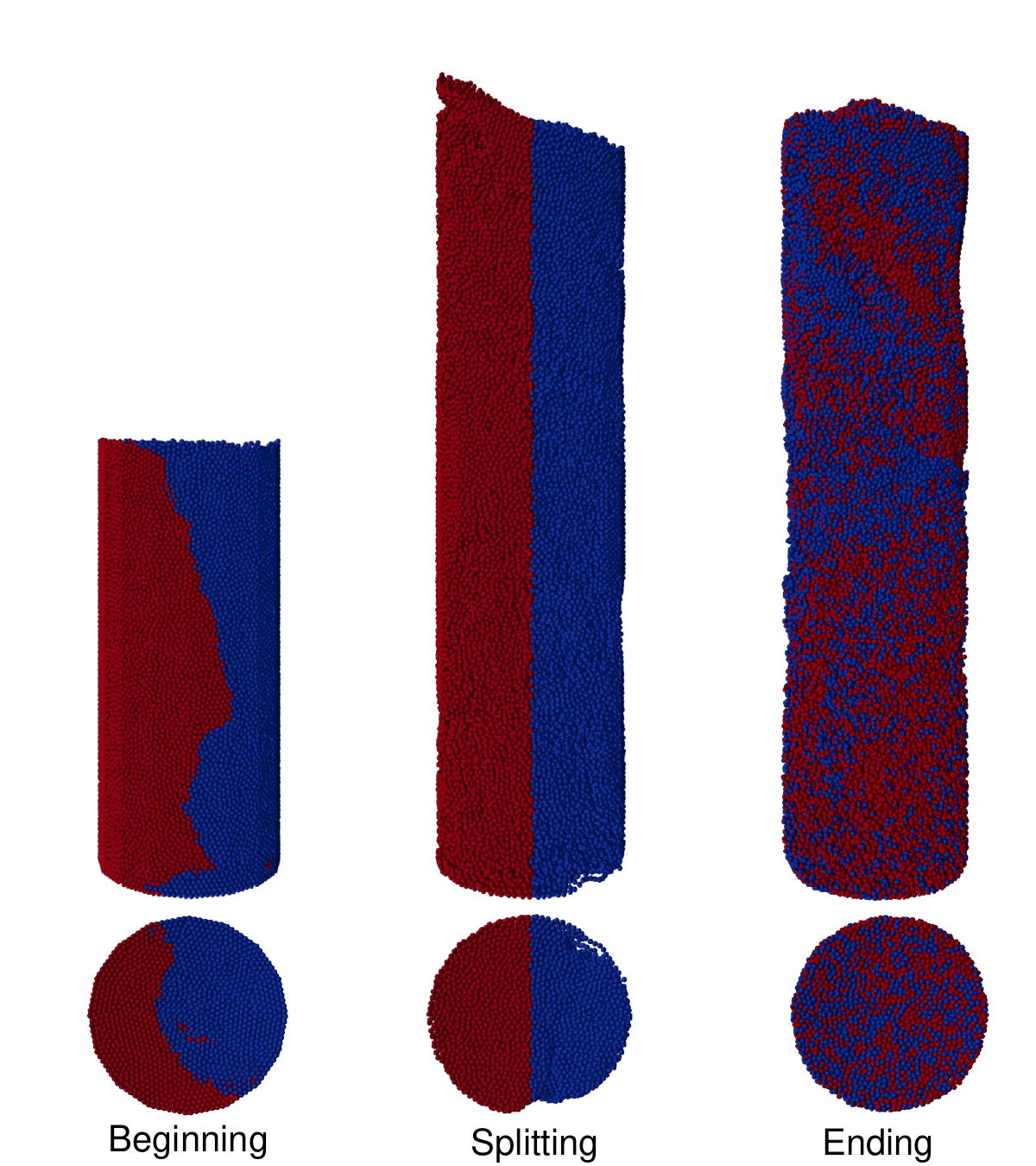}}
        \hfill
        \subcaptionbox{Axial coordinate ($z$).\label{fig::nnm_split_height}}
        {\includegraphics[width=.3\linewidth]{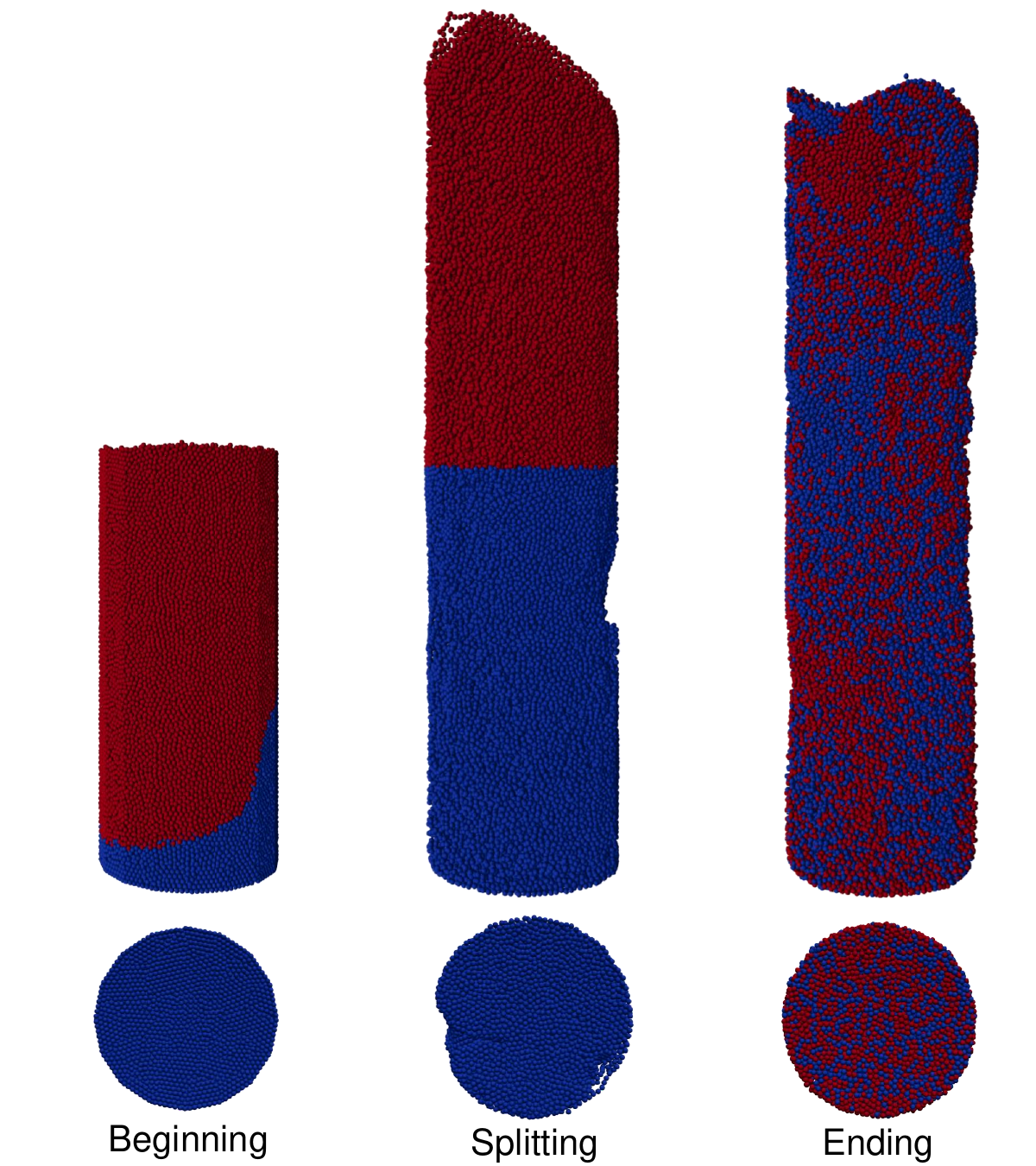}}

	\caption{Example of side and bottom views of the beginning, splitting, and ending time-steps of simulations per coordinate. Particles are split according to their position so that NNM can be applied. Splitting directions follow the cylindrical coordinates system, i.e.,  radius ($r$), angle ($\theta$), and height ($z$). Particles are split at the splitting time step such that half of the particles will be part of each group.}
	\label{fig::nnm_split}
\end{figure}
\vspace*{\fill}
\end{landscape}

As shown in Figure \ref{fig::nnm_split}, NNM results do not start from zero because, at $t_0$, the particles close to the splitting boundary will have neighbor particles on the opposite side. As a consequence, the higher the surface area of the splitting layer, the higher the initial mixing index will be. In our particular case, $z < \theta < r$. For this reason, we normalize $\bar{M}^{\text{NNM}}_t$ by calculating:
\begin{equation}
	\label{eqn::nnm_norm}
	\bar{M}^{\text{NNM}}_t = \frac{\bar{M}^{\text{NNM}}_t - \bar{M}^{\text{NNM}}_{t_{ref}}}{1 - \bar{M}^{\text{NNM}}_{t_{ref}}}
\end{equation}

\subsubsection{Doucet's mixing index}

Different from NNM, the method by \citet{Doucet_2008} does not require domain splitting. Instead, the method applies a Principal Component Analysis (PCA) \cite{Shlens_2014} on the correlation between the particles' position ($\bm{x} \in \mathbb{R}^d$, where $d$ stands for the number of dimensions) at a given time-step ($\bm{x}_t$) and its position at a reference time-step ($\bm{x}_{t_{ref}}$). The method consists of the following steps:

\begin{enumerate}
    \item Find the correlation matrix $\bm{C}_t$ given $C_{t,ij} = \rho(\bm{x}_{t,i}, \bm{x}_{t_{ref},j}), \forall i, \forall j$;
    
    \item Compute $\bm{B}_t = \bm{C}_t \bm{C}_t^T$;

    \item Diagonalize $\bm{B}_t$ and find its maximum eigenvalue $\lambda_t = (\bm{\lambda}_t)_{max}$;

    \item Find eigenvector associated to $\lambda_t$, $\bm{w}_t$;

    \item Do the same to all t.
\end{enumerate}

The more mixed particles are, the lower the correlation between the current and its position at the reference time-step and, consequently, the lower $\lambda_t$ is going to be. The highest component of $\bm{w}_t$ gives the direction to which the mixing is poorest.

As shown by \citet{Blais_2017}, it is important to respect the natural coordinate system of the problem. Similar to what was done to NNM, the cylindrical coordinate system was used to represent the particles' positions. Also as in NNM, we used 10 \si{\second} as the reference time ($t_0$) for the mixing index calculation.

To establish a comparison between NNM and Doucet's mixing index, the latter was normalized so that the index increases with mixing (contrary to \citet{Doucet_2008}, and \citet{Blais_2017}).

\begin{equation}
	\label{eqn::doucet}
	\bar{M}_t^{\text{Doucet}} = 1 - \frac{\lambda_t}{\lambda_{t_{ref}}}
\end{equation}

\subsection{Interaction characteristics study} \label{sec::part_props}

In this work, we also investigated the role of particles' interaction properties on the mixing. To do so, we simulated the liquid-solid fluidized bed dynamics varying particles' Young's modulus ($Y$), coefficient of restitution ($e$), coefficient of rolling friction ($\mu_r$), coefficient of sliding friction ($\mu_f$). The properties were varied individually while the others are kept constant (e.g. when $\mu_f$ varies from the one presented in Table \ref{tab::simulation_params}, $e$, $Y$, and $\mu_r$ are equal to those in Table \ref{tab::simulation_params}). The values for each of the properties are shown in Table \ref{tab::particle_interaction}.

\begin{table}[!htpb]
	\centering
	\caption{Particle properties applied to the particle interaction study.}
    \label{tab::particle_interaction}
\begin{tabular}{lccc}
\hline
\textbf{Property} &
  \textbf{Values} \\ \hline
Young's modulus ($Y$) &
  \{10$^5$, 10$^7$, 10$^9$\} \si{\pascal} \\
Coefficient of restitution ($e$) &
  \{0.1, 0.3, 0.6, 0.9\} \\
Coefficient of sliding friction ($\mu_f$) &
  \{0.1, 0.2, 0.6, 0.9\} \\
Coefficient of rolling friction ($\mu_r$) &
  \{0.1, 0.2, 0.6, 0.9\} \\ \hline
\end{tabular}

\end{table}

\section{Results and Discussion}

To establish comparisons among regimes, particles, and methods, the mixing time results are presented as the dimensionless number of flows through:
\begin{equation}
	\label{eqn::n_flows}
	N_{\text{flows}} = (t - t_{ref})\frac{U}{H_C}
\end{equation}
where $t_{ref}$ is the mixing index reference time (Table \ref{tab::particle_params}), $U$ is the fluid inlet velocity, and $H_C$ is the height of the cylinder. To avoid the influence of the initial bed expansion, all results were obtained in the fully developed pseudo-steady state. Additionally, the results of the study of the particle interaction properties can be found in the tables provided in the \href{art3::supplementary}{Supplementary Material}, while in this section only the main features are presented and discussed. The data includes the number of flows through to reach 95, 90, 80, and 70\% mixing ($N_{\text{flows}}^{95}$, $N_{\text{flows}}^{90}$, $N_{\text{flows}}^{80}$, and $N_{\text{flows}}^{70}$, respectively), the maximum mixing ($\bar{M}_{\text{max}}$), and the maximum mixing divided by the number of flows through to reach it for each mixing index ($\bar{M}_{\text{max}}/N_{\text{flows}}^{\text{max}}$).

\subsection{Mixing indices comparison}

\begin{figure}[!htpb]
	\centering
     \begin{subfigure}{0.9\textwidth}
		\centering
		\includegraphics[width=1\linewidth]{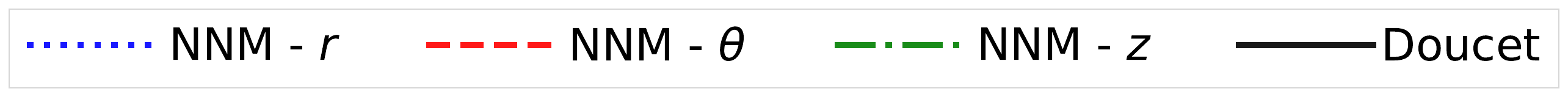}
		\label{fig::mix_nflowslegend}
	\end{subfigure}
	\begin{subfigure}{.50\textwidth}
		\centering
		\includegraphics[width=1\linewidth]{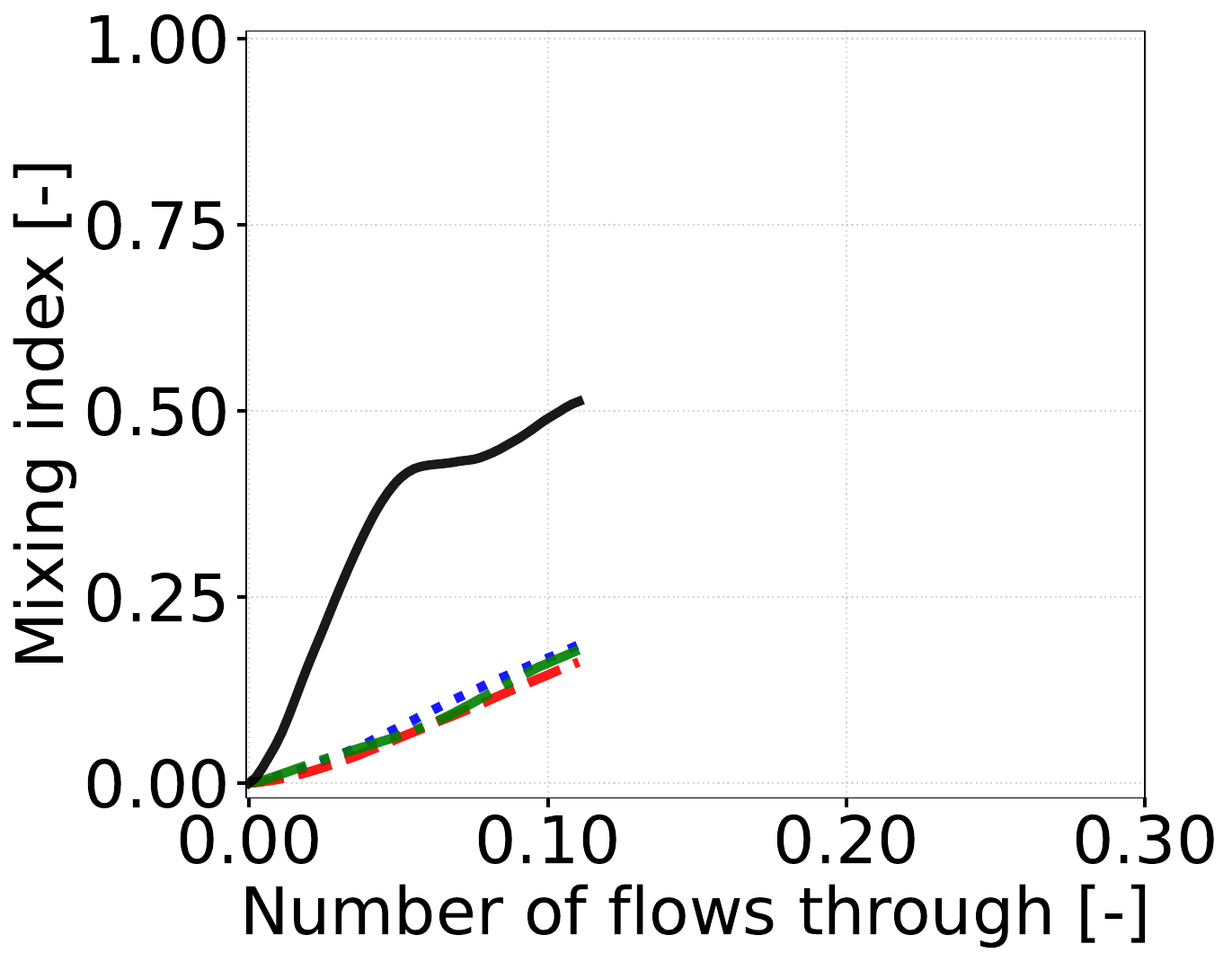}
        \caption{Particle A - Low}
		\label{fig::mix_nflows_alginate_low}
	\end{subfigure}%
    \bigskip
	\begin{subfigure}{.50\textwidth}
		\centering
		\includegraphics[width=1\linewidth]{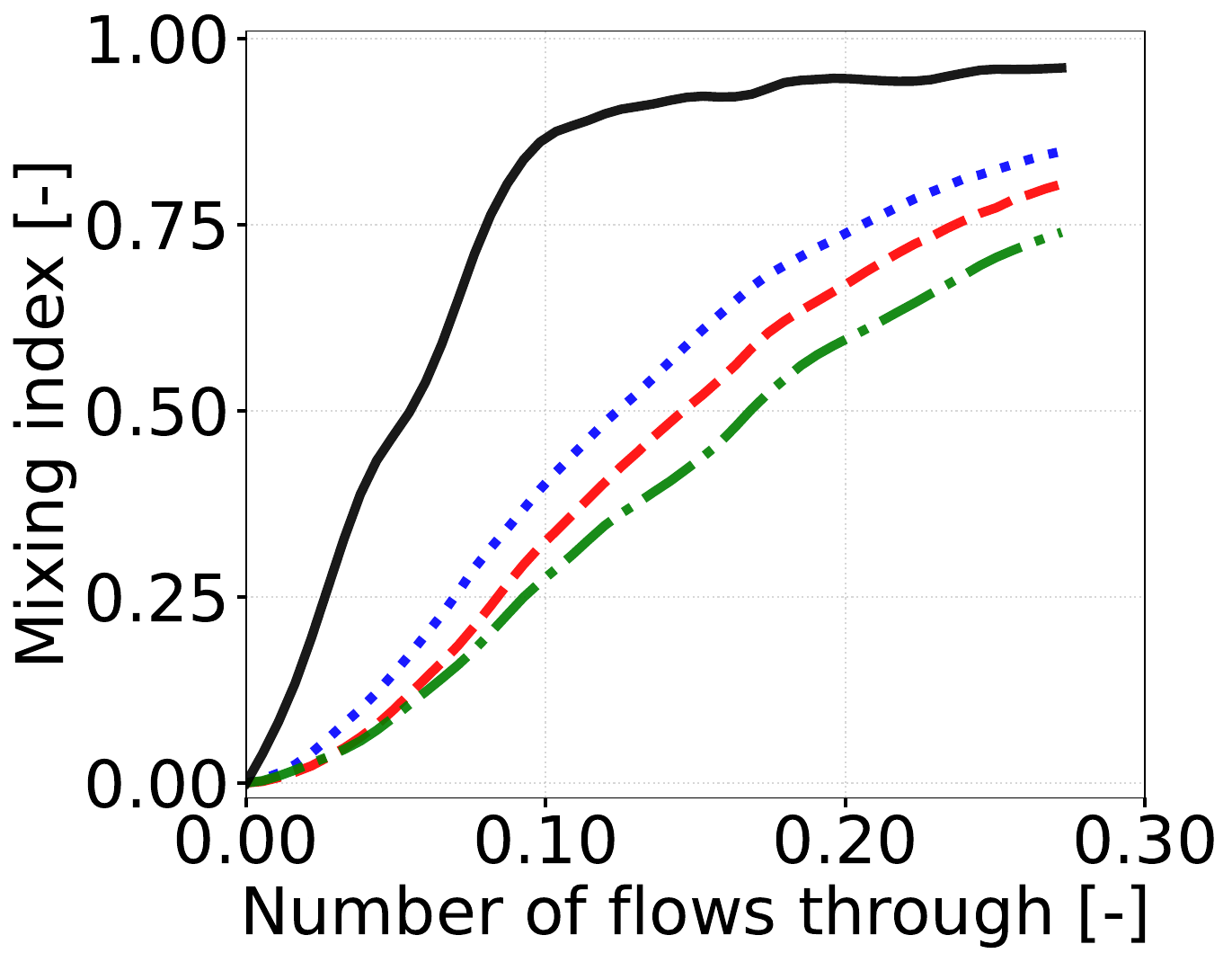}
        \caption{Particle A - High}
		\label{fig::mix_nflows_alginate_high}
	\end{subfigure}
    \begin{subfigure}{.50\textwidth}
		\centering
		\includegraphics[width=1\linewidth]{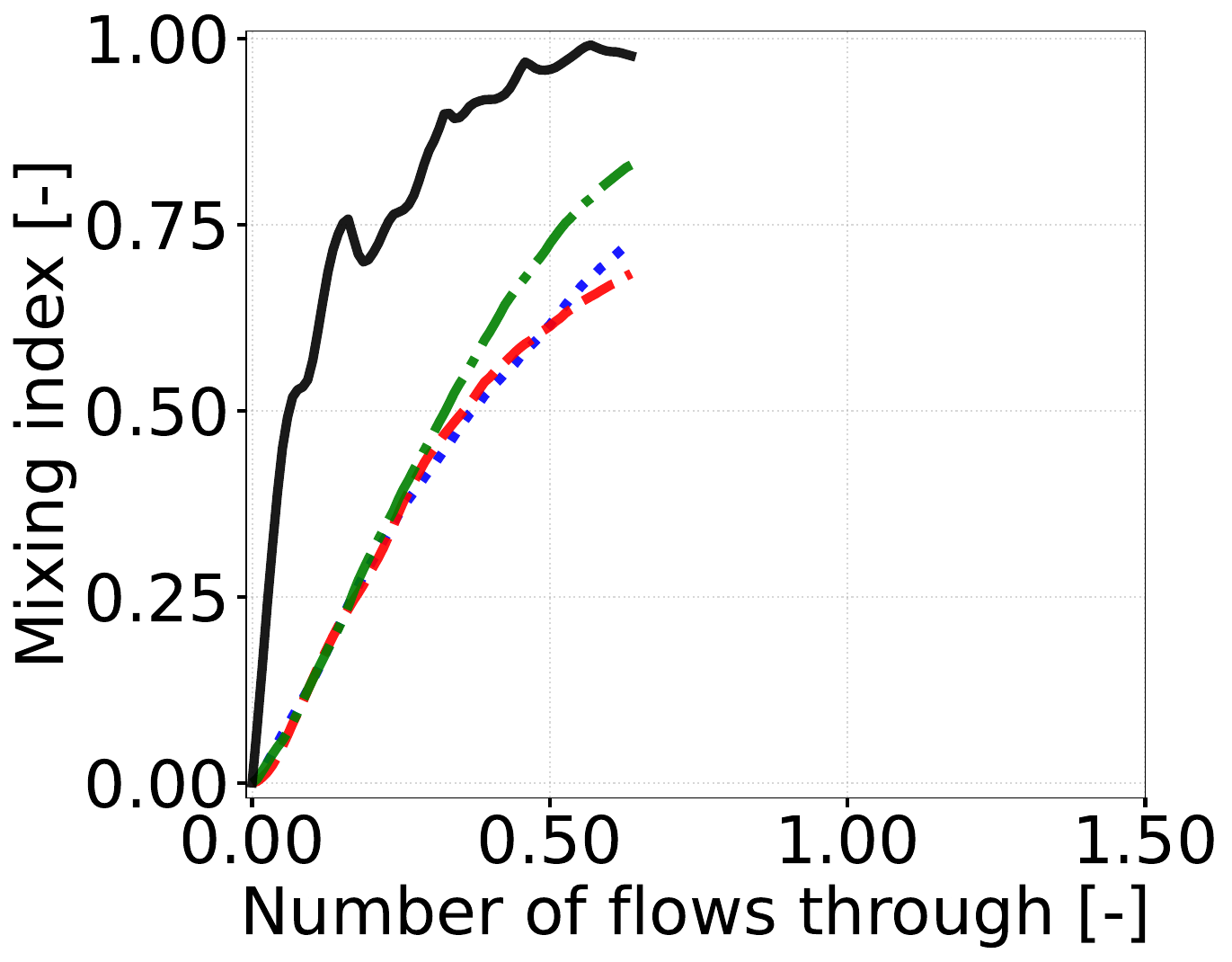}
        \caption{Particle B - Low}
		\label{fig::mix_nflows_abs_low}
	\end{subfigure}%
    \bigskip
	\begin{subfigure}{.50\textwidth}
		\centering
		\includegraphics[width=1\linewidth]{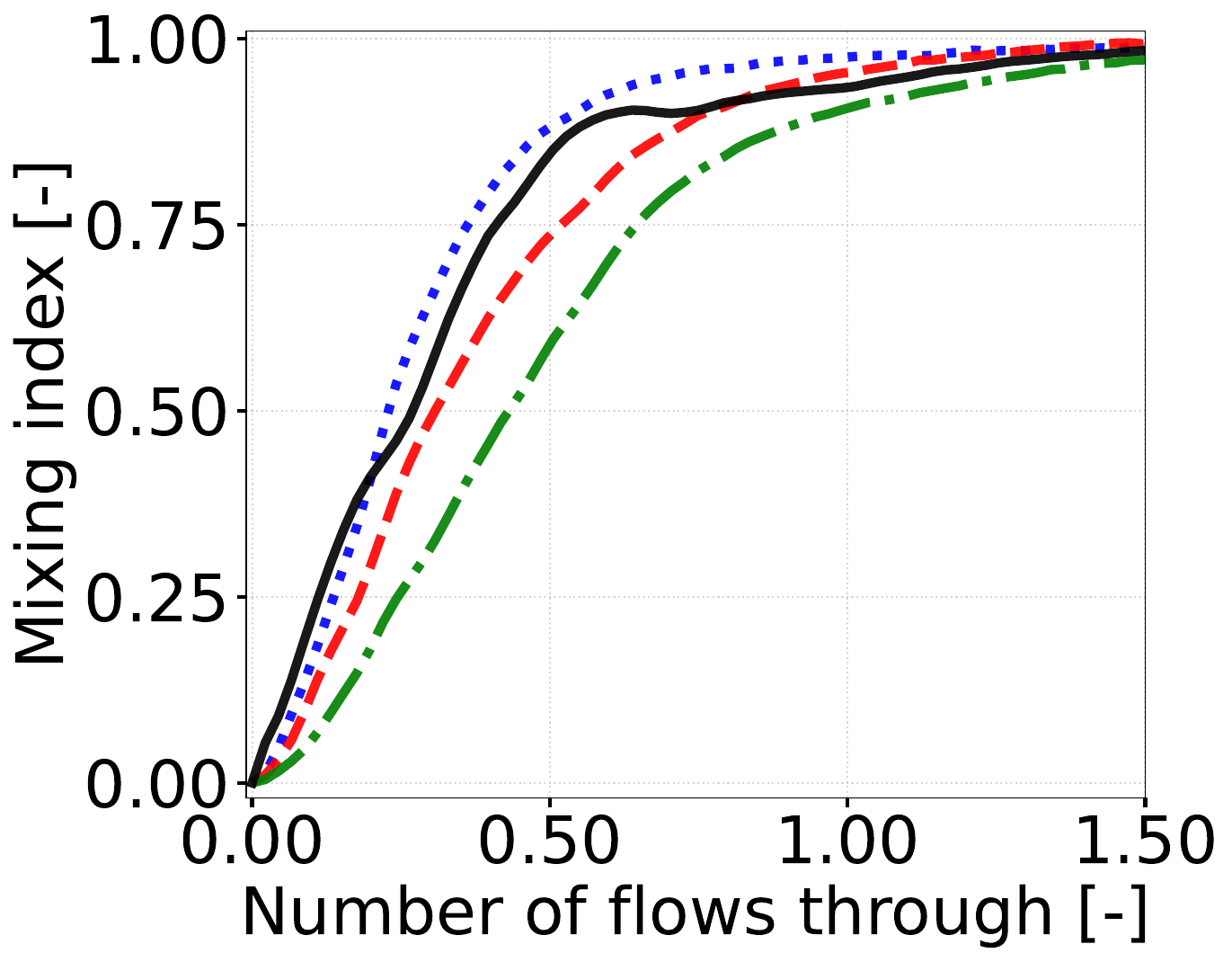}
        \caption{Particle B - High}
		\label{fig::mix_nflows_abs_high}
	\end{subfigure}
    \begin{subfigure}{.50\textwidth}
		\centering
		\includegraphics[width=1\linewidth]{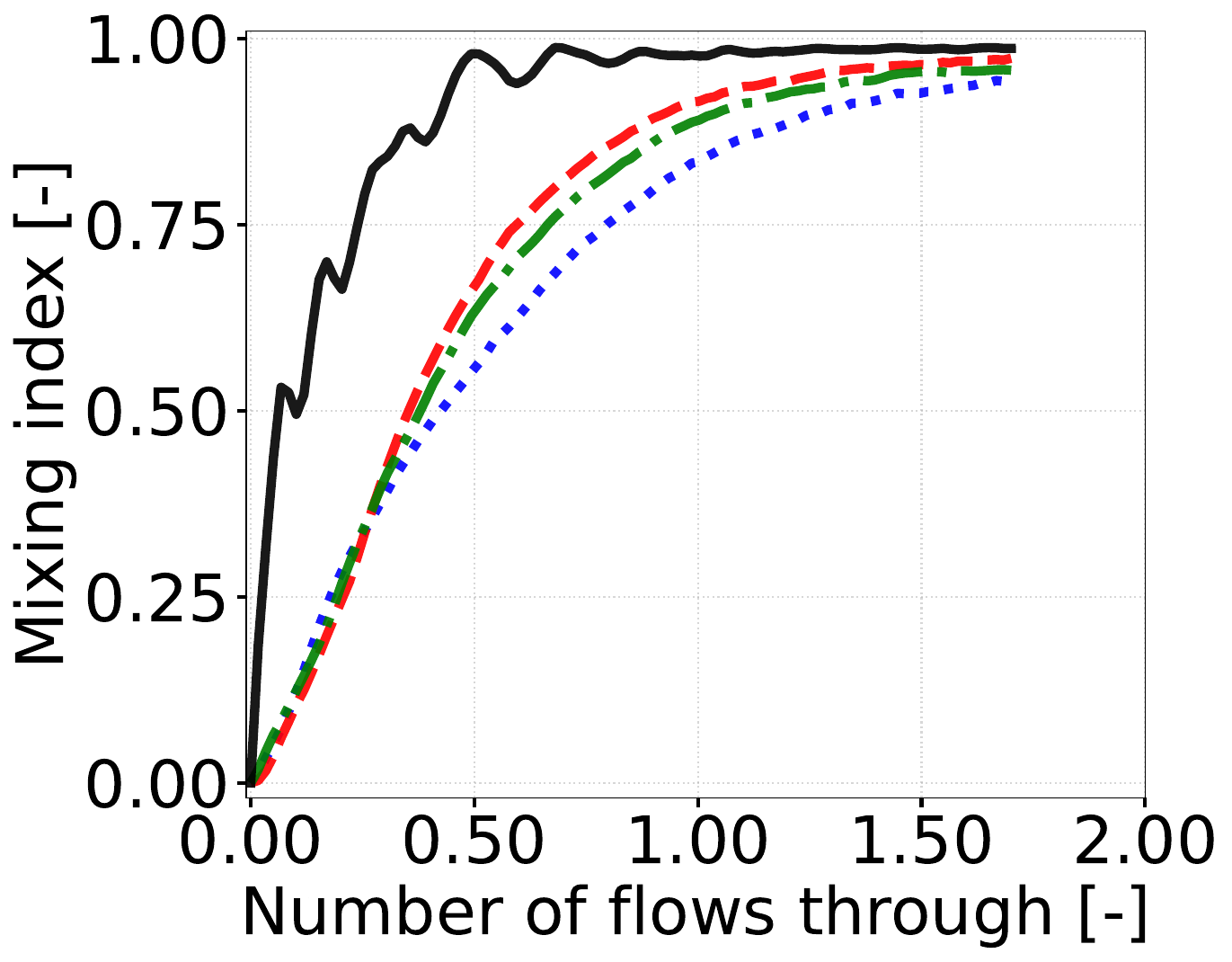}
        \caption{Particle C - Low}
		\label{fig::mix_nflows_alumina_low}
	\end{subfigure}%
	\begin{subfigure}{.50\textwidth}
		\centering
		\includegraphics[width=1\linewidth]{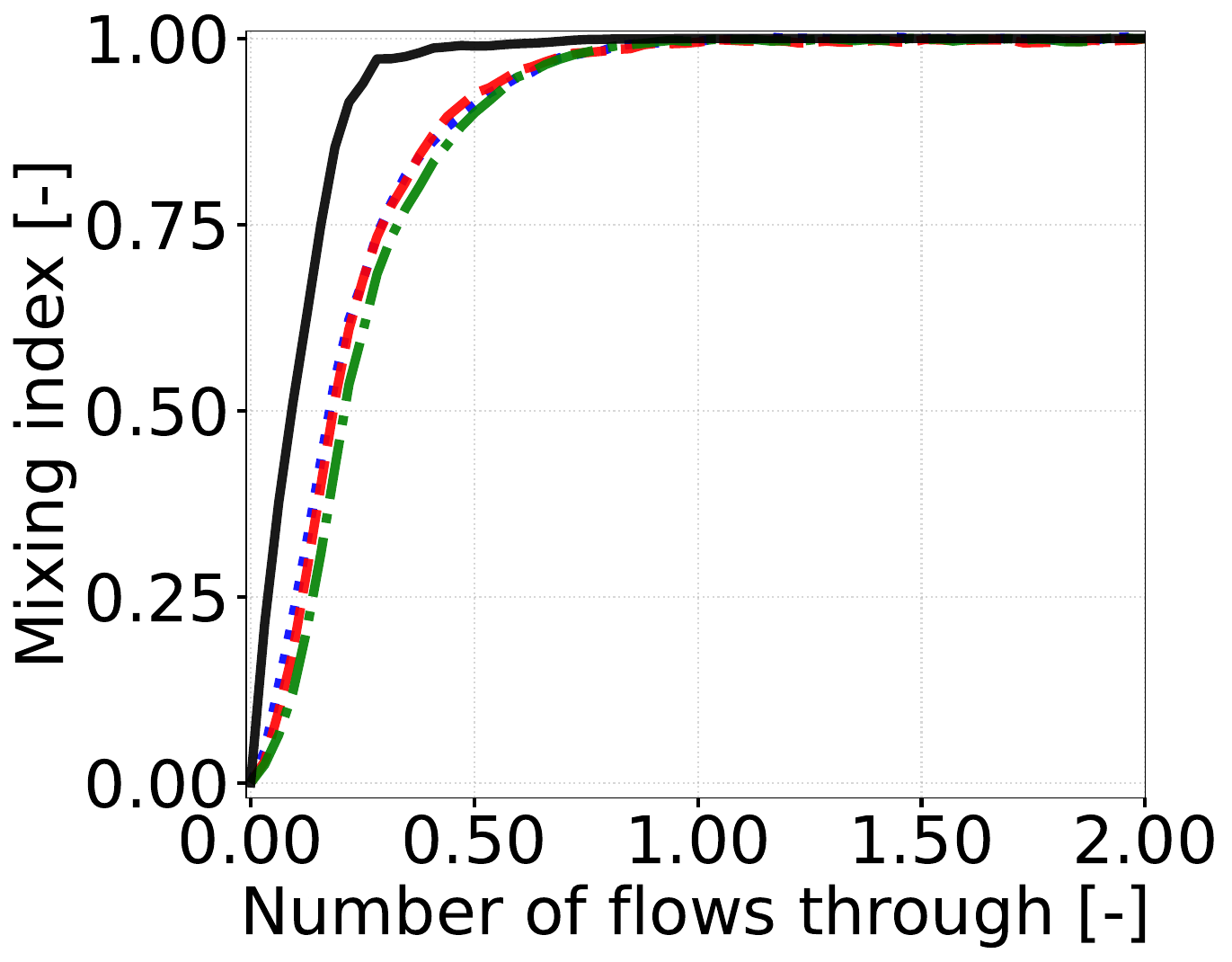}
        \caption{Particle C - High}
		\label{fig::mix_nflows_alumina_high}
	\end{subfigure}
	\caption{Mixing index as a function of $N_{\text{flows}}$ for the three particles at lowest (low) and highest (high) fluid inlet flow rates.}
	\label{fig::mix_nflows_high_low}
\end{figure}

In Figure \ref{fig::mix_nflows_high_low} we show how the mixing index evolves with respect to the number of flows through using the different methods, for the three particles, and under the highest (high) and the lowest (low) fluid inlet flow rates.

NNM and Doucet's mixing indices reflect very different aspects of particle mixing. In NNM, we split the domain and count the neighbors originally on the opposite side, which provides a straightforward view of mixing in each direction. The method is very versatile, given that the domain split can be done in any coordinate system. Additionally, the method allows for the evaluation of individual mixing indices to each particle, having more localized responses to each of the components. Such results can point to regions in the domain where the mixing is poorer, and not only returning in a global response. It is important to state that the method is not suitable to any geometry, as splitting requires symmetry.

The main disadvantages of the method are the lack of information about the global mixing without accounting for individual direction components, and the need for a manually defined $N_{neighbors}$. Furthermore, the maximum mixing index of 1 can only be reached if the domain is split such that half of the particles lie on each side, consequently, the need for a precise split is a limitation.

Oppositely, Doucet's method does not require splitting since particles are compared to themselves at a reference time-step only. In addition, the poorest mixing component at a time $t$ can be determined by the eigenvector $\bm{w}_t$, which directly compares the mixing components and highlights time-dependent effects. It is a convenient and efficient approach to estimate mixing using Lagrangian information.

\subsection{Inlet flow rate}

\begin{figure}[!htpb]
	\centering
    \begin{subfigure}{1\textwidth}
		\centering
		\includegraphics[width=1\linewidth]{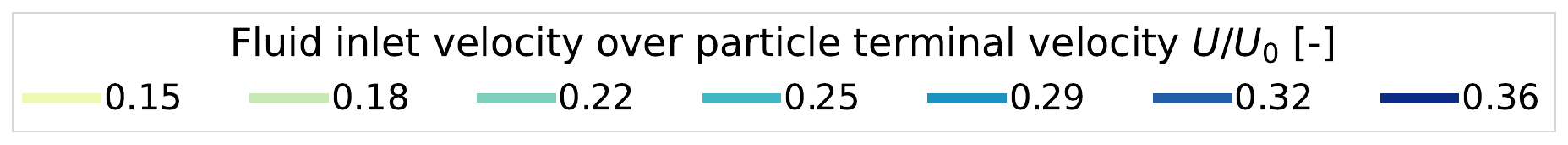}
		\label{fig::mix_time_alginate_legend}
    \end{subfigure}
	\begin{subfigure}{.50\textwidth}
		\centering
		\includegraphics[width=1\linewidth]{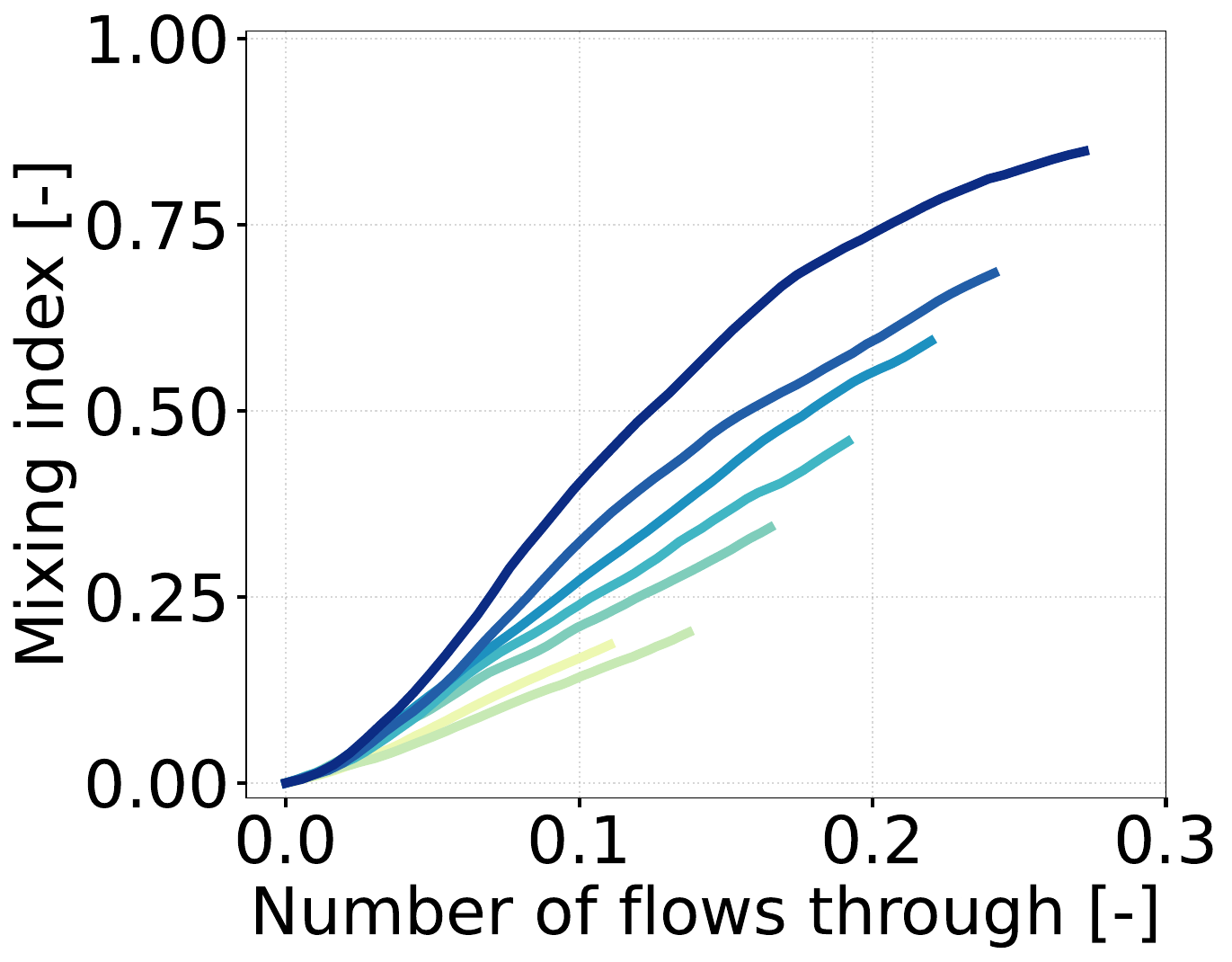}
        \caption{NNM - $r$}
		\label{fig::mix_time_algiante_radius}
	\end{subfigure}%
    \bigskip
	\begin{subfigure}{.50\textwidth}
		\centering
		\includegraphics[width=1\linewidth]{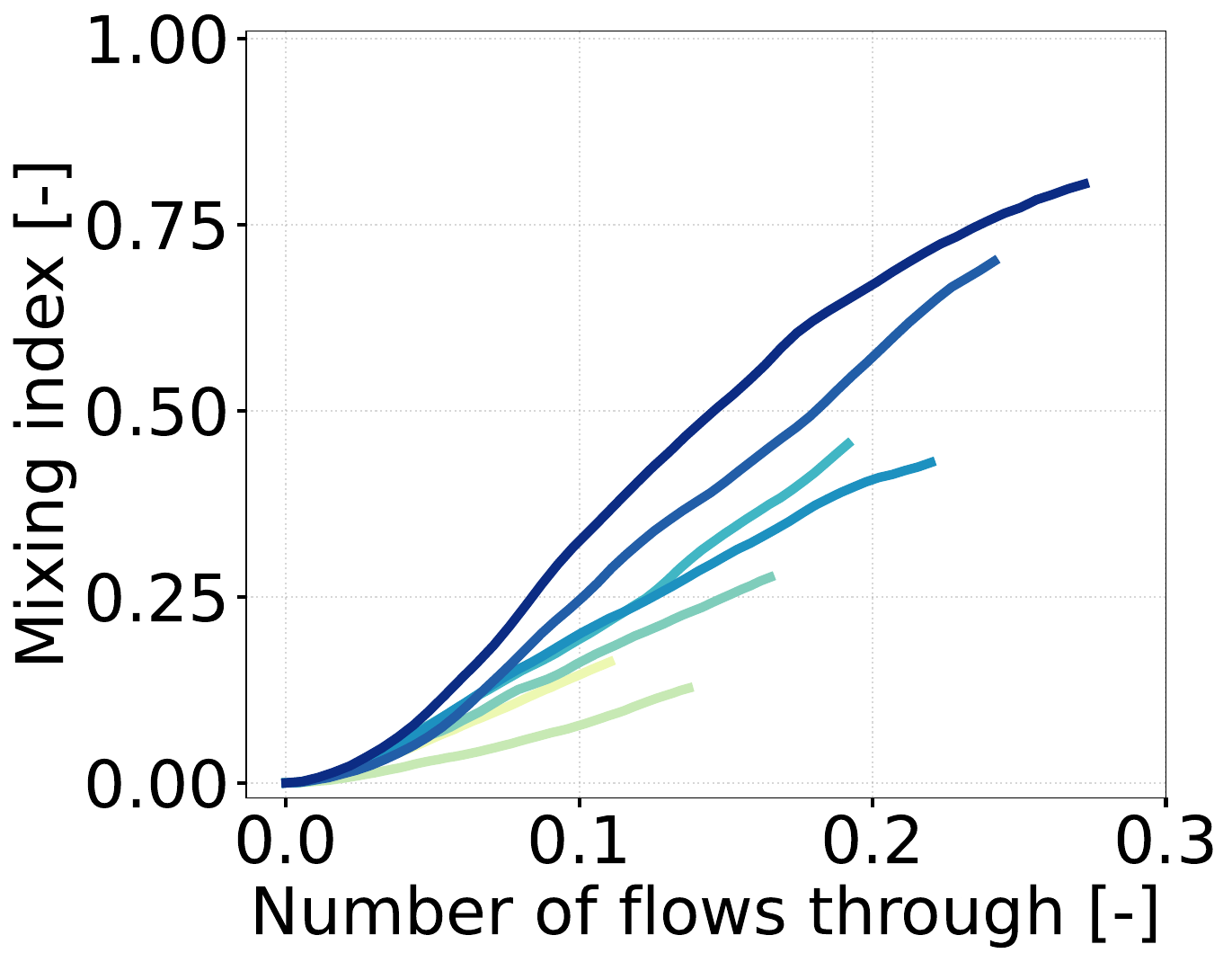}
        \caption{NNM - $\theta$}
		\label{fig::mix_time_algiante_theta}
	\end{subfigure}
    \begin{subfigure}{.50\textwidth}
		\centering
		\includegraphics[width=1\linewidth]{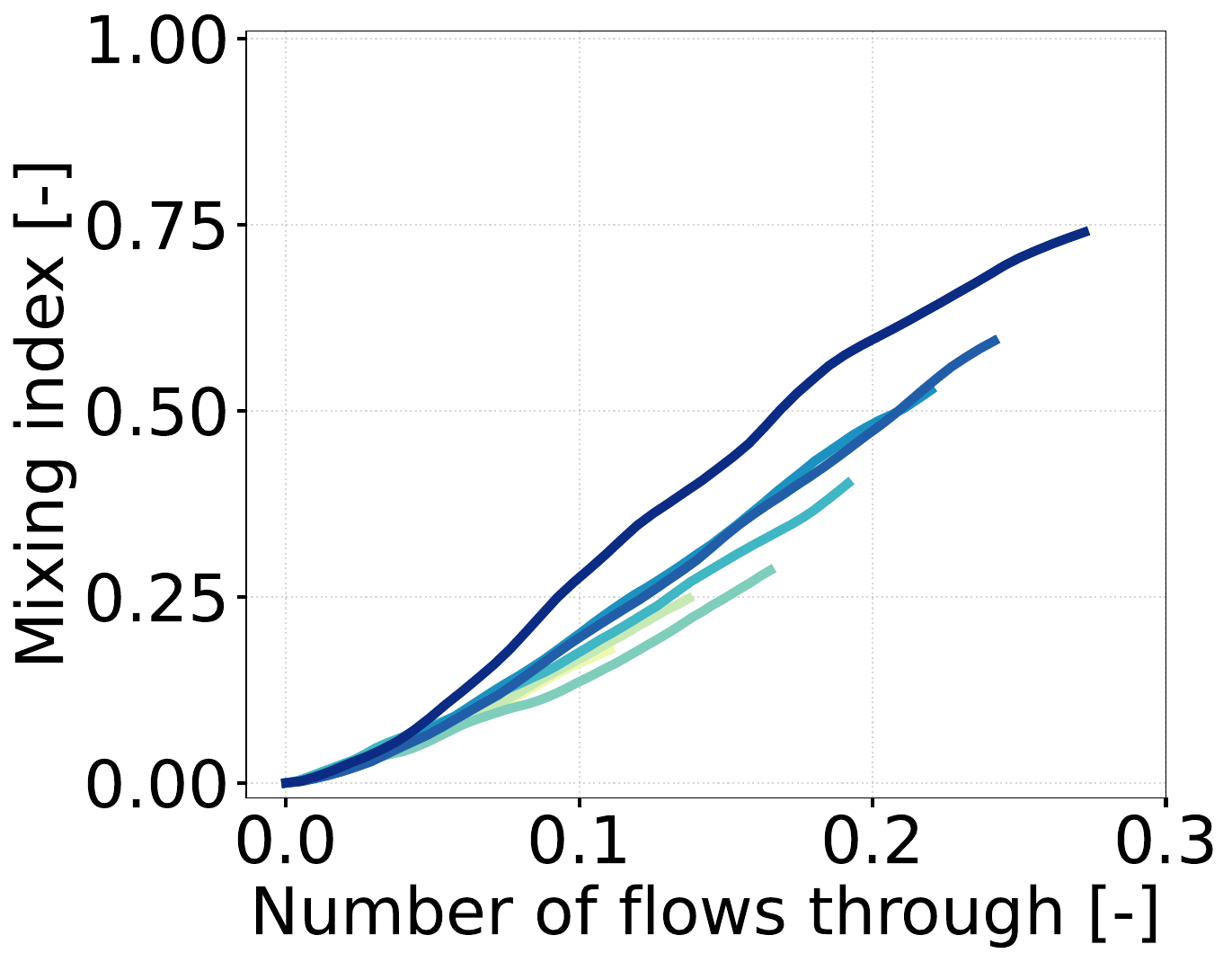}
        \caption{NNM - $z$}
		\label{fig::mix_time_algiante_height}
	\end{subfigure}%
	\begin{subfigure}{.50\textwidth}
		\centering
		\includegraphics[width=1\linewidth]{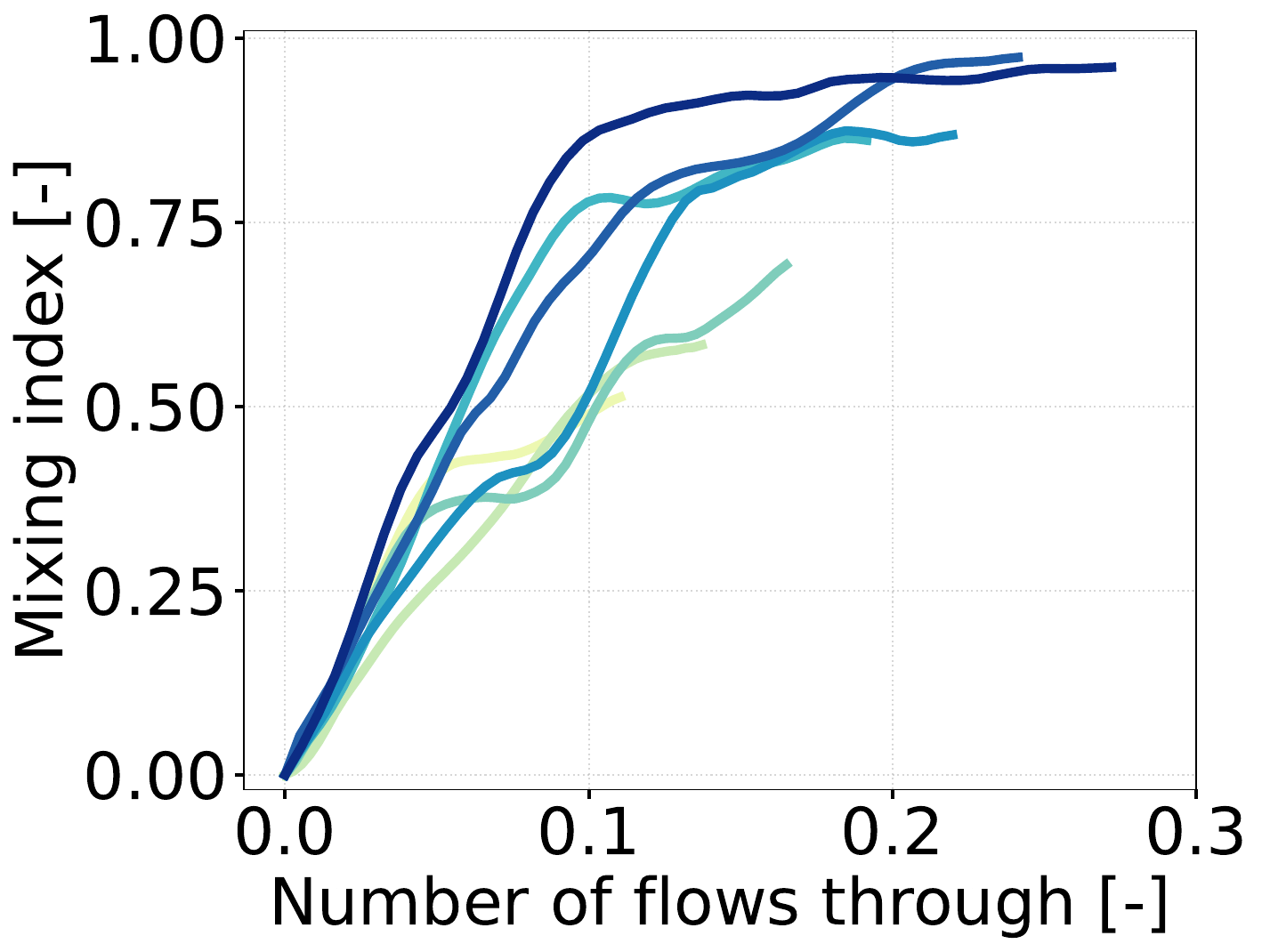}
        \caption{Doucet}
		\label{fig::mix_time_algiante_doucet}
	\end{subfigure}    
	\caption{Mixing index as a function of time for particle A at several inlet flow rates.}
	\label{fig::mix_time_algiante}
\end{figure}

\begin{figure}[!htpb]
	\centering
    \begin{subfigure}{1\textwidth}
		\centering
		\includegraphics[width=1\linewidth]{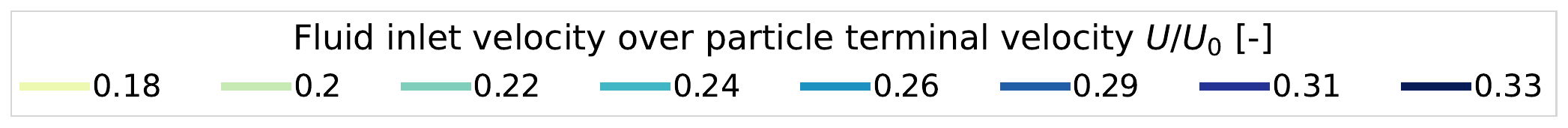}
		\label{fig::mix_time_alumina_legend}
    \end{subfigure}
    \bigskip
    \begin{subfigure}{.50\textwidth}
		\centering
		\includegraphics[width=1\linewidth]{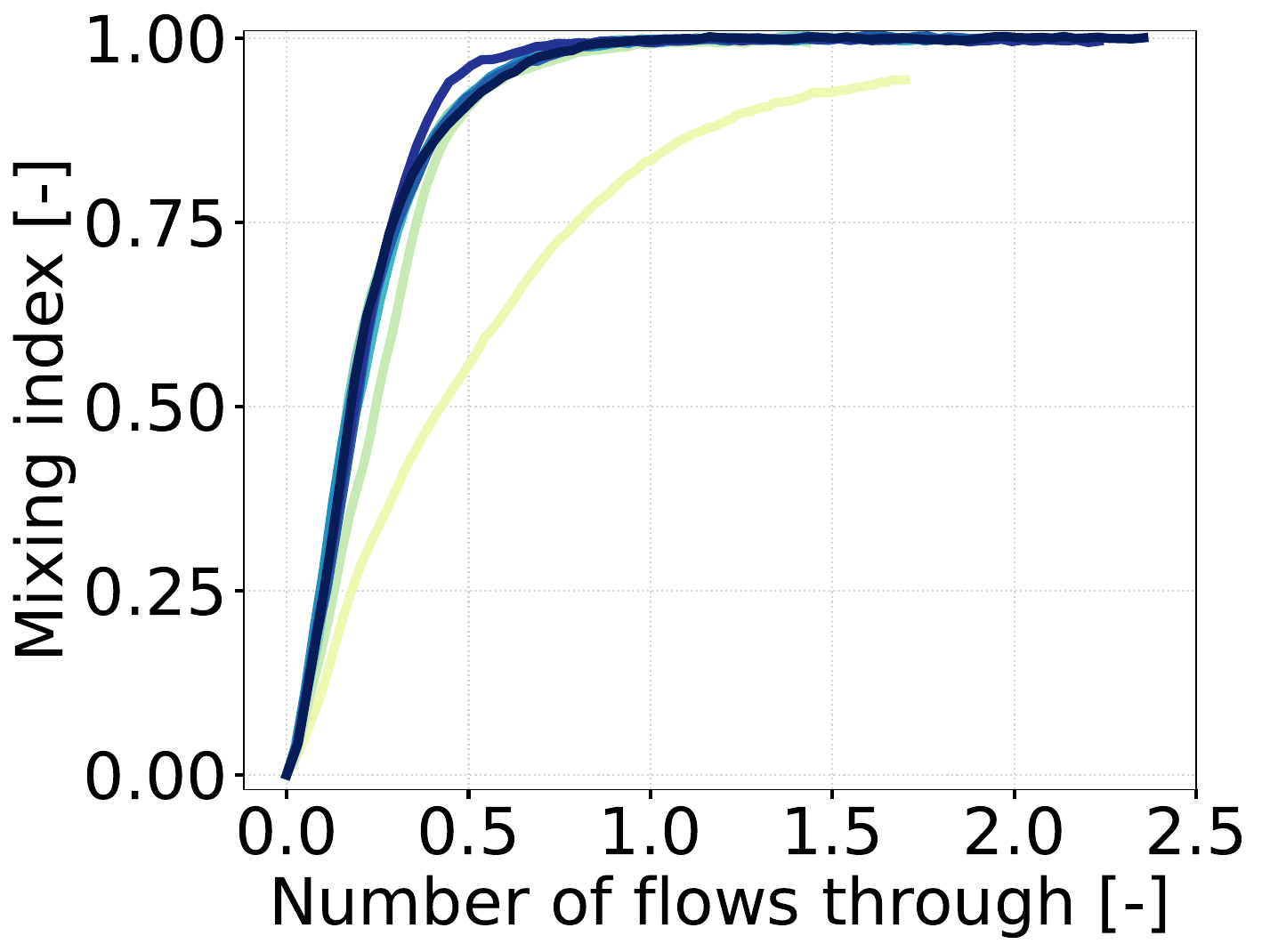}
        \caption{NNM - $r$}
		\label{fig::mix_time_alumina_radius}
	\end{subfigure}%
	\begin{subfigure}{.50\textwidth}
		\centering
		\includegraphics[width=1\linewidth]{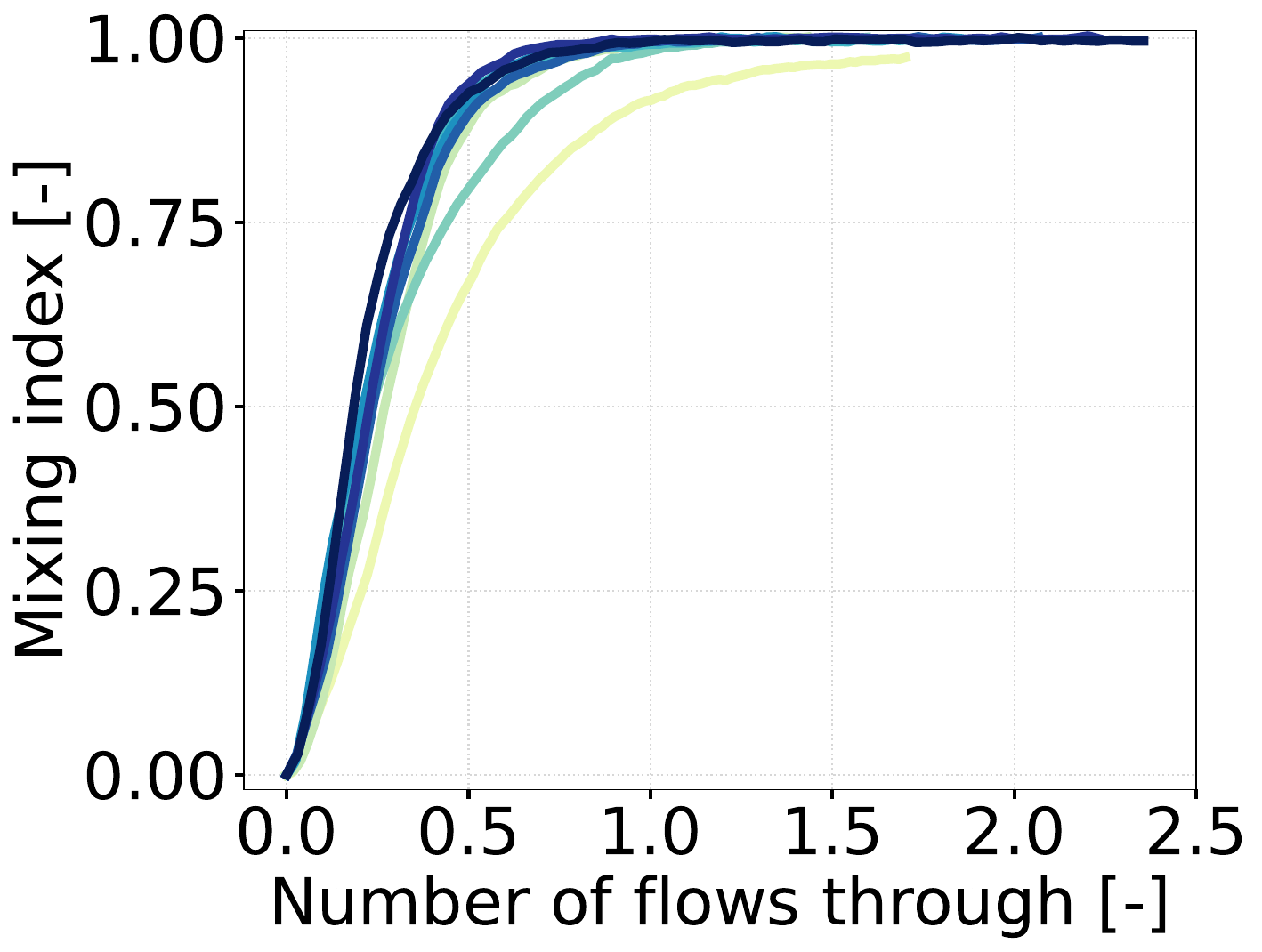}
        \caption{NNM - $\theta$}
		\label{fig::mix_time_alumina_theta}
	\end{subfigure}
    \begin{subfigure}{.50\textwidth}
		\centering
		\includegraphics[width=1\linewidth]{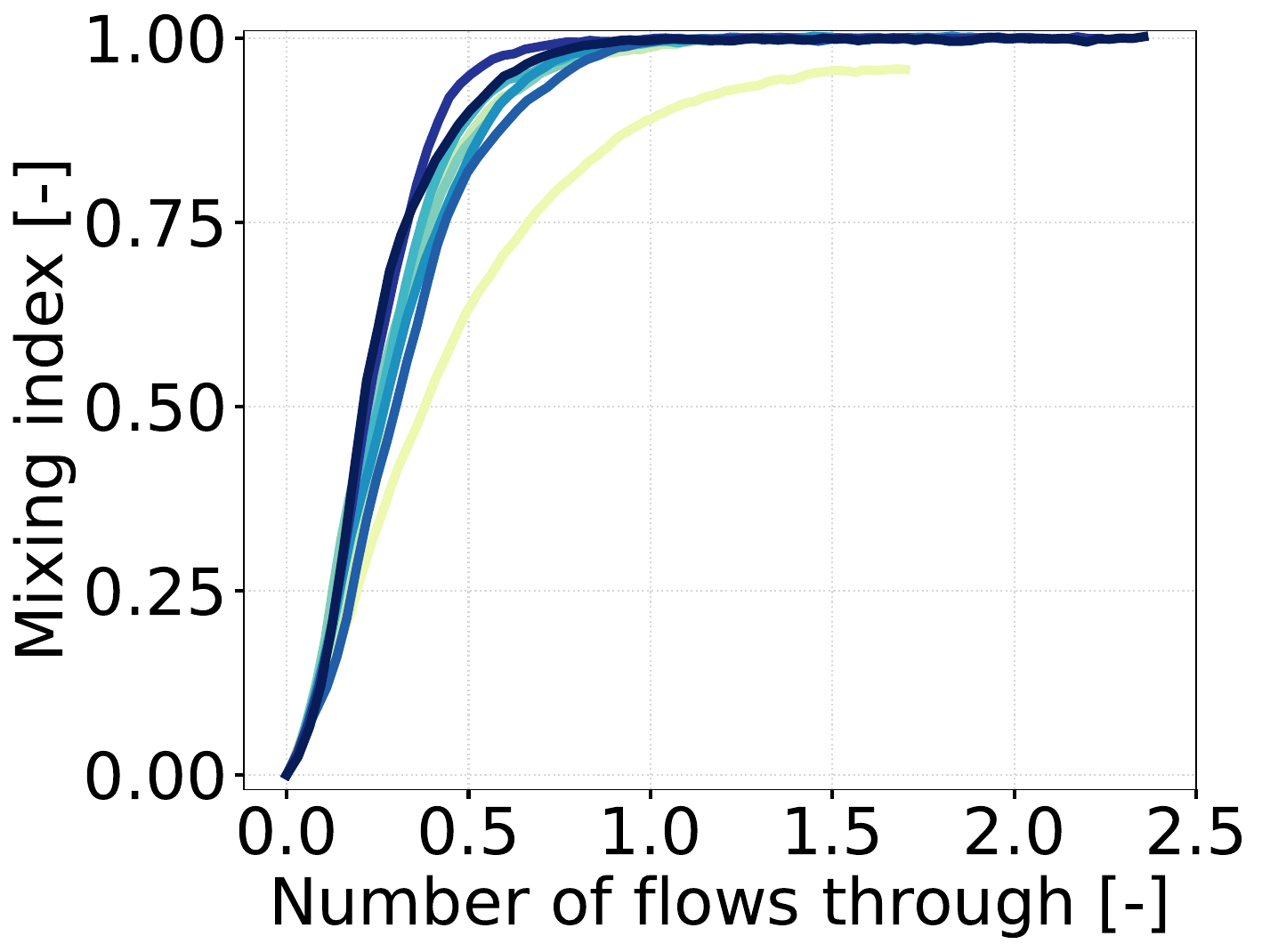}
        \caption{NNM - $z$}
		\label{fig::mix_time_alumina_height}
	\end{subfigure}%
	\begin{subfigure}{.50\textwidth}
		\centering
		\includegraphics[width=1\linewidth]{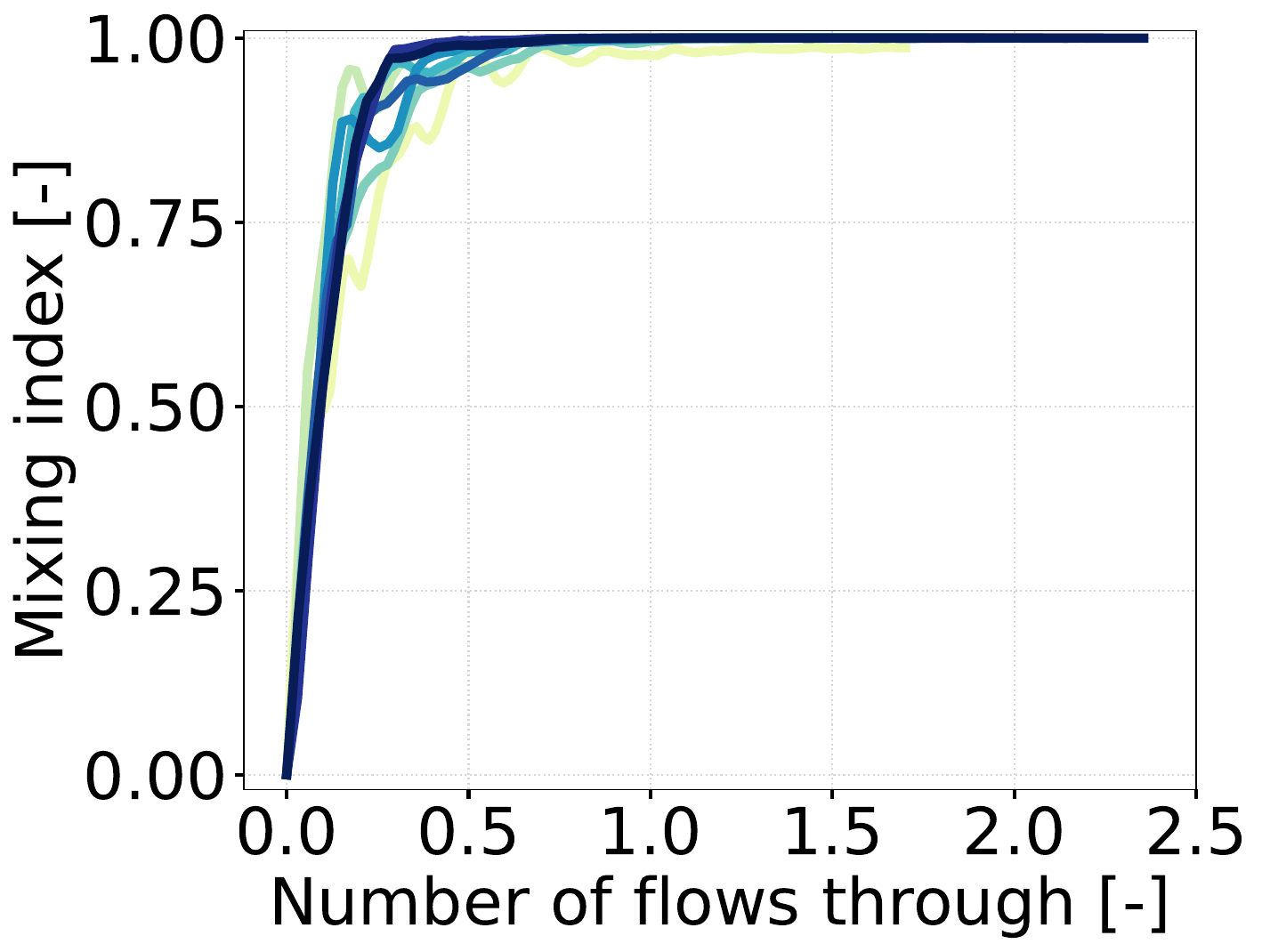}
        \caption{Doucet}
		\label{fig::mix_time_alumina_doucet}
	\end{subfigure}
	\caption{Mixing index as a function of time for particle C at several inlet flow rates.}
	\label{fig::mix_time_alumina}
\end{figure}

In Figures \ref{fig::mix_time_algiante} and \ref{fig::mix_time_alumina}, we show the effect of the inlet flow rate on the mixing of particles A and C, respectively.

It is intuitive to think that the inlet flow rate has a major impact on mixing in a fluidized bed. The higher the inlet flow rate is, the higher the particle-fluid forces are going to be, which increases particle momentum. In addition, increasing the average void fraction $\bar{\varepsilon}_f$ will increase the particles' mean free path. In other words, in theory, the mixing rate is improved because particles move faster with less resistance due to interactions. This behavior is clearly demonstrated in the results in Figures \ref{fig::mix_time_algiante} and \ref{fig::mix_time_alumina}. In general, cases with $U/U_0$ closer to $1.00$ presented lower mixing times, regardless of the component.

In the case of particle A, increasing the inlet flow rate had a positive impact on mixing. For the same number of flows through, higher flow rate simulations achieved higher mixing indices than lower ones. This was observed on all indices. For instance, the mixing index was twice higher at maximum inlet flow rate compared to the minimum when the number of flows through was close to 0.1. In other words, at very low Reynolds, increasing the flow rate had a positive impact on the mixing rate.

\begin{figure}[!htpb]
	\centering
    \begin{subfigure}{.50\textwidth}
		\centering
		\includegraphics[width=1\linewidth]{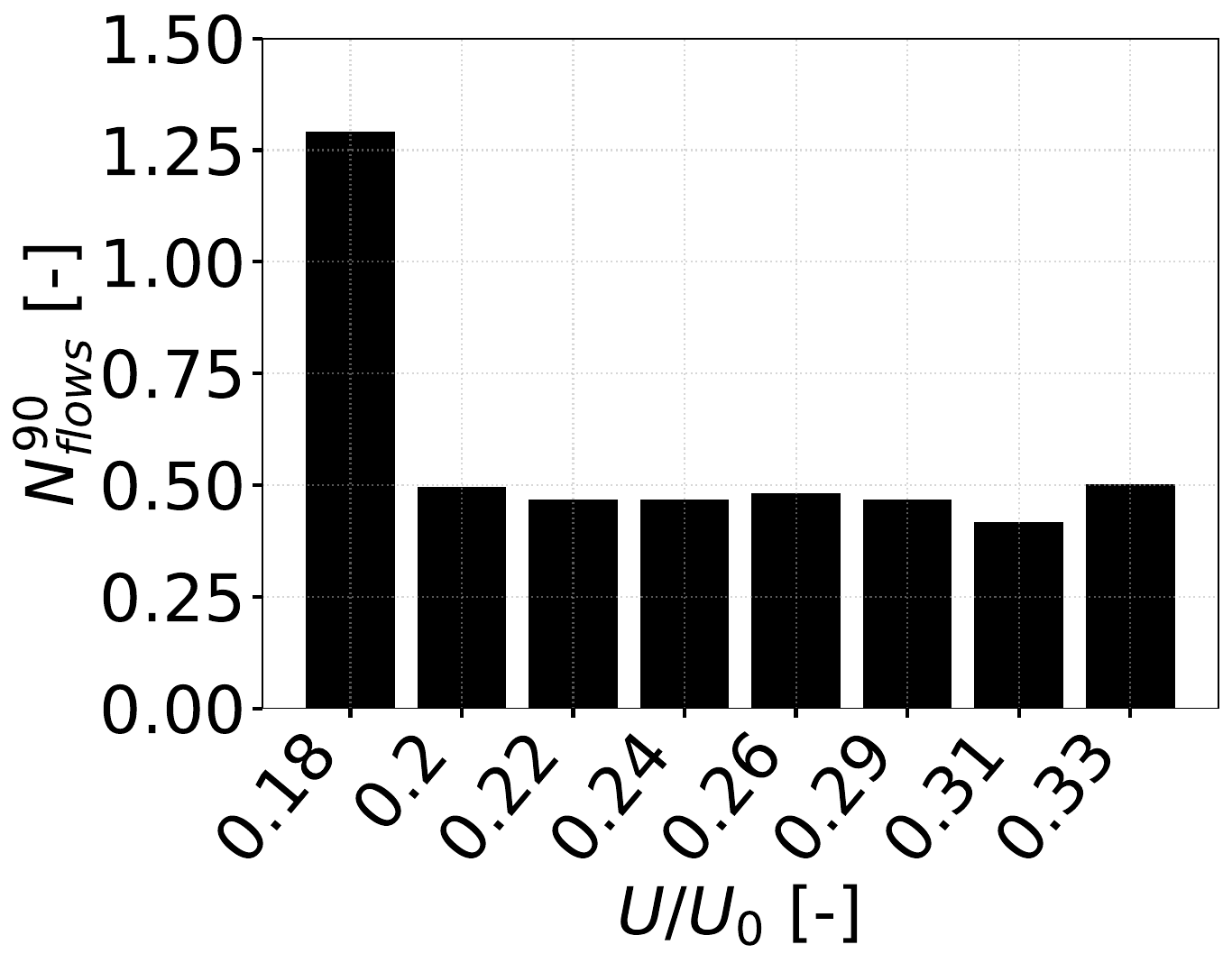}
        \caption{NNM - $r$}
		\label{fig::n_90_alumina_radius}
	\end{subfigure}%
	\begin{subfigure}{.50\textwidth}
		\centering
		\includegraphics[width=1\linewidth]{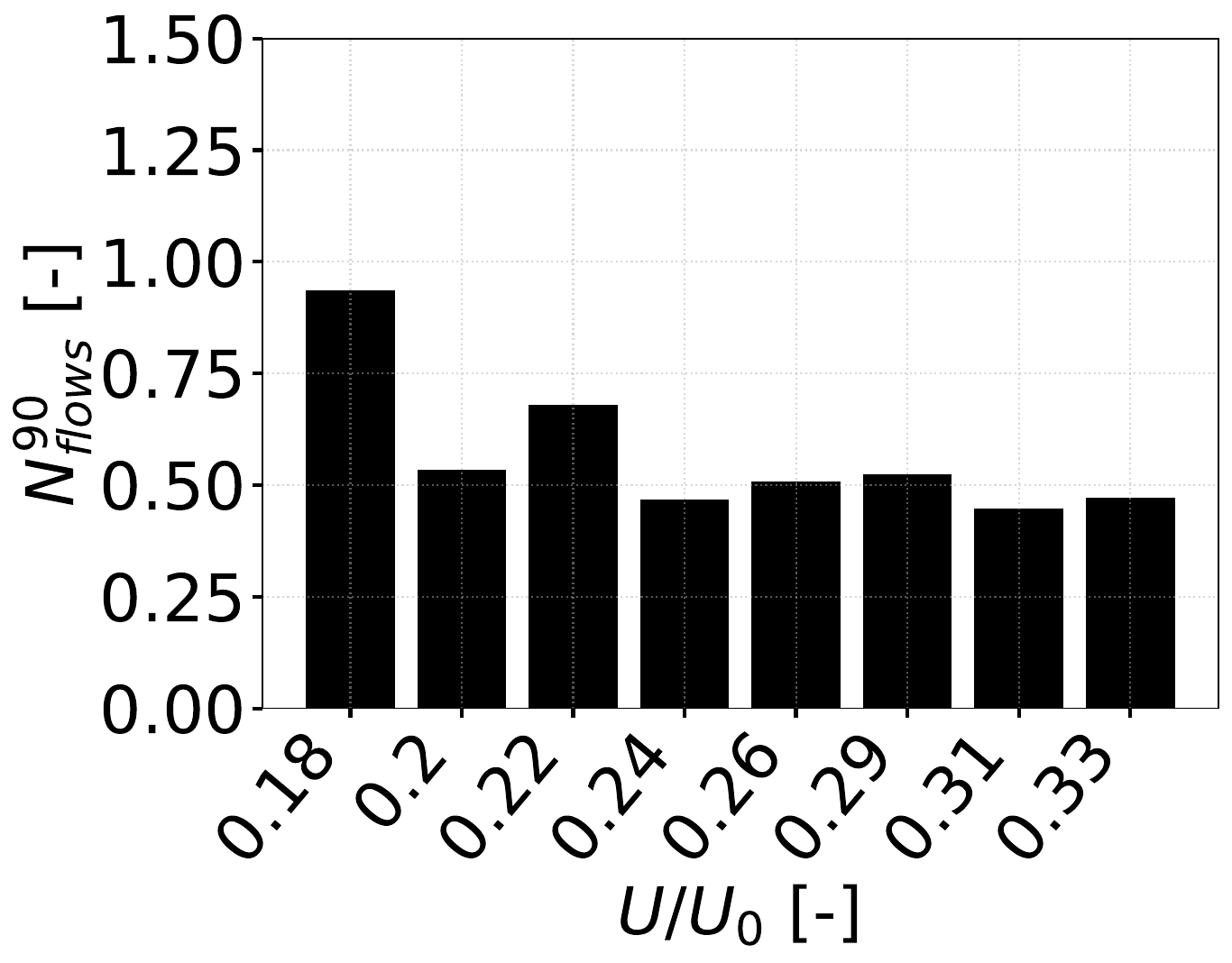}
        \caption{NNM - $\theta$}
		\label{fig::n_90_alumina_theta}
	\end{subfigure}
    \begin{subfigure}{.50\textwidth}
		\centering
		\includegraphics[width=1\linewidth]{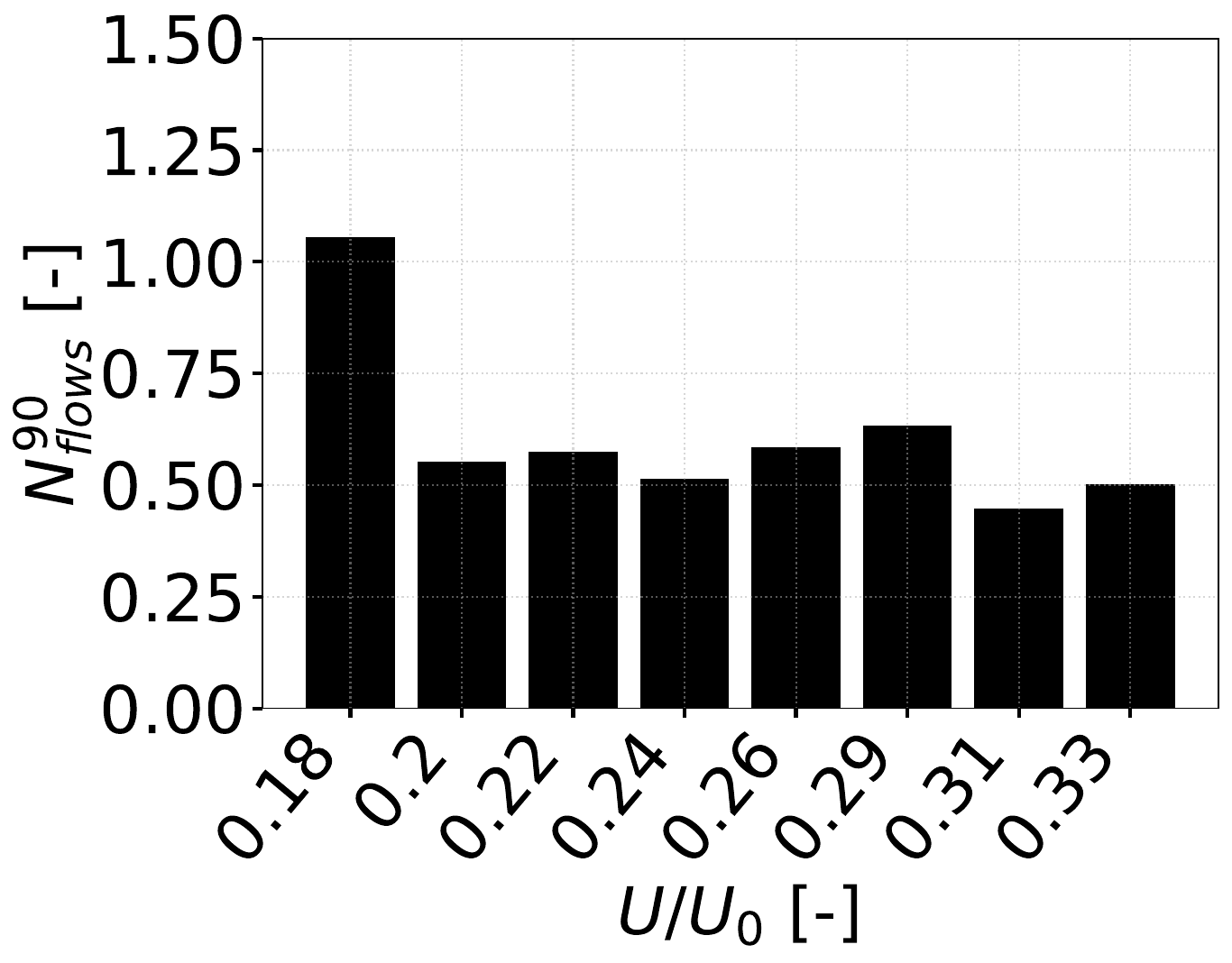}
        \caption{NNM - $z$}
		\label{fig::n_90_alumina_height}
	\end{subfigure}%
	\begin{subfigure}{.50\textwidth}
		\centering
		\includegraphics[width=1\linewidth]{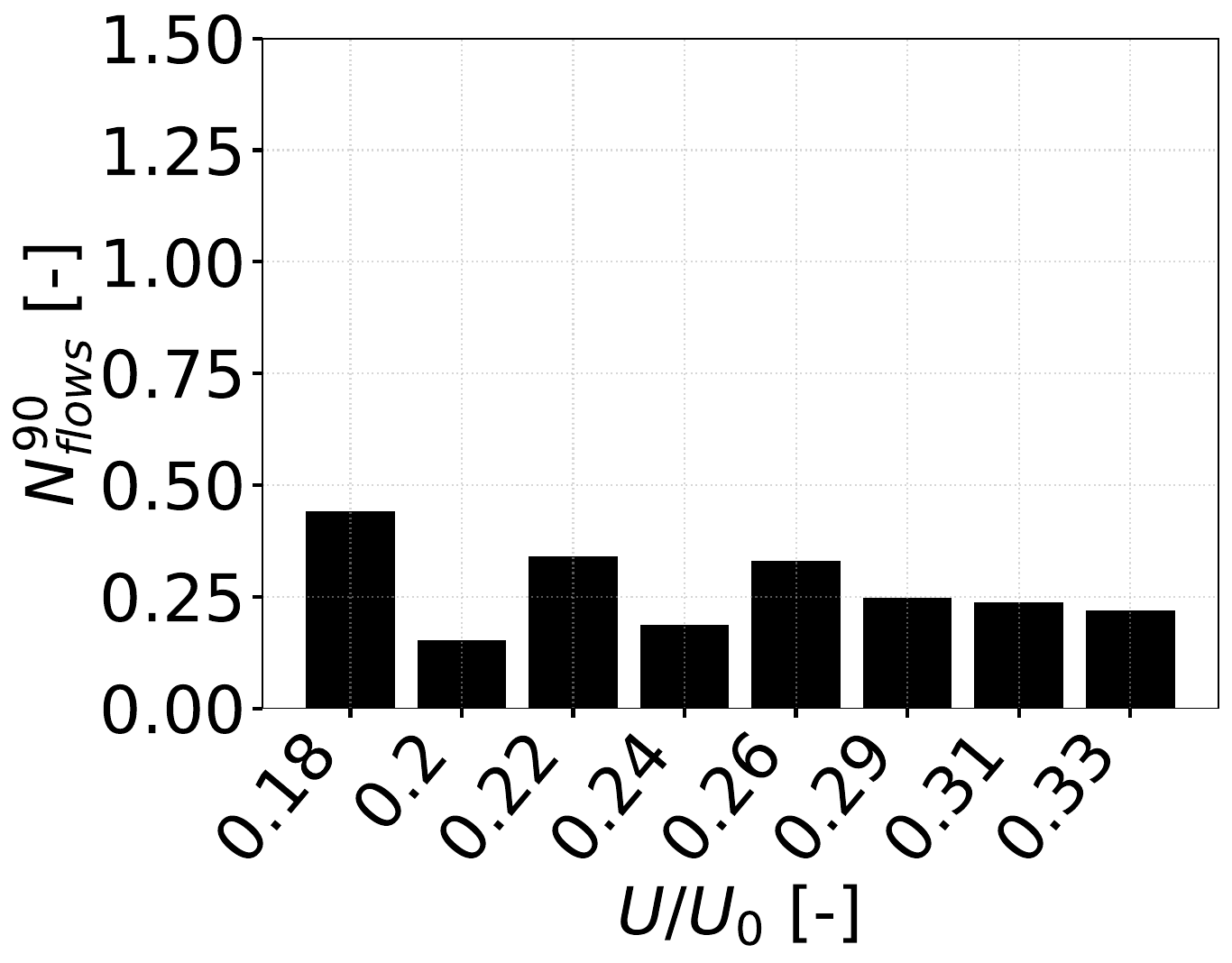}
        \caption{Doucet}
		\label{fig::n_90_alumina_doucet}
	\end{subfigure}
	\caption{$N_{\text{flows}}^{90}$ particle C at several inlet flow rates.}
	\label{fig::mix_90_alumina}
\end{figure}

Conversely, observing the evolution of mixing with $N_{\text{flows}}$, this increase seems to reach a \textit{plateau}, as demonstrated in Figure \ref{fig::mix_90_alumina}. In the Figure, the number of flows through for the system to reach 90\% mixing is presented as a function of the inlet flow rate.

It is not clear whether there is a gain in the mixing rate with the increase of $U/U_0$. Apparently, apart from velocities very close to the minimum fluidization, the effect of the bed expansion on the evolution of mixing with the number of flows through is mild. All simulations presented very similar mixing performances, except for the very concentrate bed. Overall, in the case of particle C, there was no gain in the mixing rate when the fluid inlet velocity was increased.

The difference between regimes can be due to their difference in Stokes numbers. Particle A has a very low density. As such, Particle A responds almost immediately to differences in the fluid flow, while Particle C has a way higher inertia. This difference has an important impact on the particles' agitation in the fluidized bed. In the case of particle A, small increments in the fluid velocity reflected in a higher agitation state, while for particle C this was not the case.

\subsection{Principal mixing component}

Apart from the initial bed expansion, the movement of particles in a pseudo-steady state LSFB does not necessarily follow the flow direction. The resultant force acting over a particle (collisions and interphase momentum exchange) changes direction very frequently. Consequently, the principal mixing component given by the $\bm{w}_t$ is not the same throughout a simulation, even at a well-established pseudo-steady state.

\begin{figure}[!htpb]
	\centering
 \includegraphics[width=0.7\linewidth]{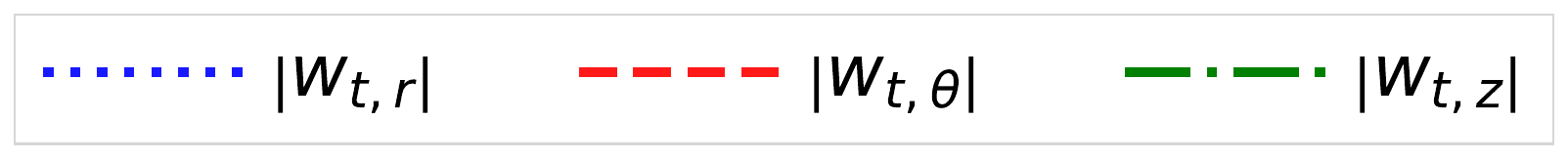}
 \hfill
	\includegraphics[width=.9\textwidth]{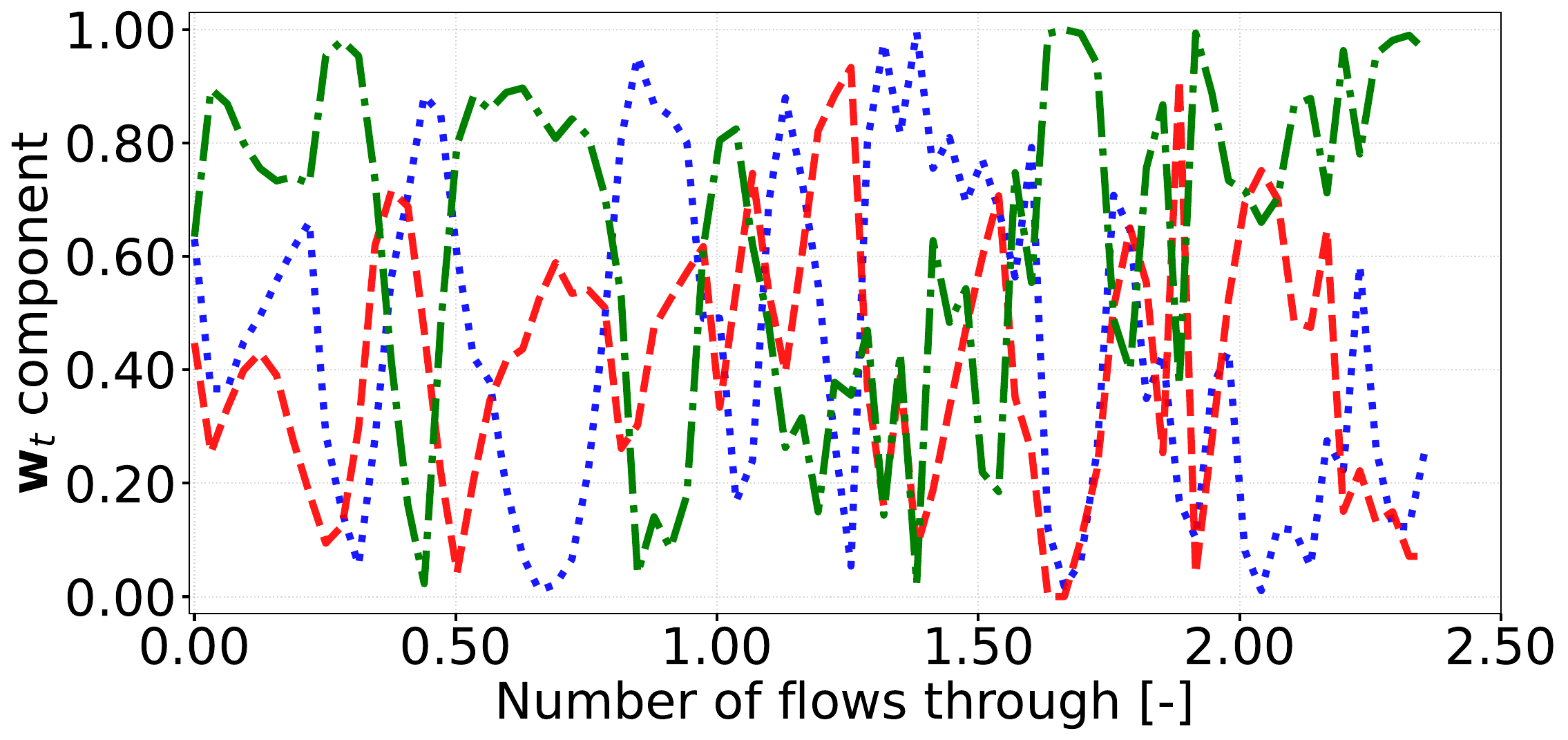}
	\caption{Main eigenvector extracted from the principal component analysis as a function of time.}
	\label{fig::rate_flows_alumina_doucet}
\end{figure}

In Figure \ref{fig::rate_flows_alumina_doucet}, we show Doucet's eigenvector analysis for particle C results for the highest inlet flow rate. From Figure \ref{fig::rate_flows_alumina_doucet}, it can be seen that the components are not predominantly slower at all times. This behavior was observed in all simulations and particles. As such, the mixing was isotropic for all particles and regimes. A similar behavior can be observed by performing the principal component analysis on a Brownian motion system. An illustration of this PCA in a Brownian motion system is provided in the \ref{brownian}.

\subsection{Interaction properties}

The discussions in this section are based on the data in the Supplementary Material. For the present analysis, we use the maximum mixing index achieved at the end of the simulation, defined as $\bar{M}_{\text{max}}$, to have an assessment of the mixing level regardless of the regime. Overall, changing the particle interaction properties did not imply a significant change in the mixing behavior. We could not observe significant implications when the coefficient of restitution or the coefficient of rolling friction were varied. The highest standard deviation among the mixing dimensionless time for the high inlet flow rate was approximately 9.2\% of the average (Doucet's mixing index, $N_{\text{flows}}^{95}$), while $\bar{M}_{\text{max}}$ presented a maximum standard deviation of 3.2\% (NNM, component $z$). For the low inlet flow rate, $N_{\text{flows}}^{70}$ was achieved by neither of the simulations (except for outliers due to oscillations in Doucet's mixing index).

The maximum standard deviation among $\bar{M}_{\text{max}}$ was 19\% (NNM, component $\theta$). Despite the high standard deviation, the maximum NNM mixing index in this direction was too small to draw any conclusion (0.16). As for particles B and C, the interaction properties were even less influential, especially for the high inlet flow rate. It is an expected result given that fluidized beds are driven by the interphase momentum change, while collisions play a secondary role \cite{Grace_2020_BOOK}. Yet, one feature can be pointed out.

\begin{figure}[!htpb]
    \centering

    \includegraphics[width=0.7\linewidth]{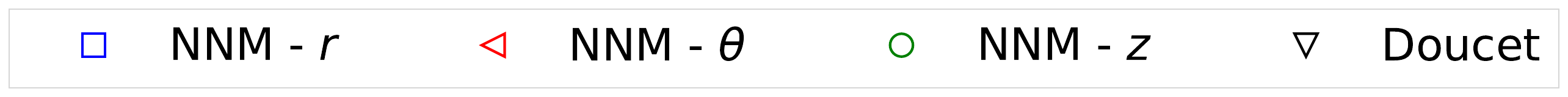}

    \hfill
    \subcaptionbox{Particle A - Low - NNM.\label{fig::props_alginate_low_nnm}}{\includegraphics[width=.49\textwidth]{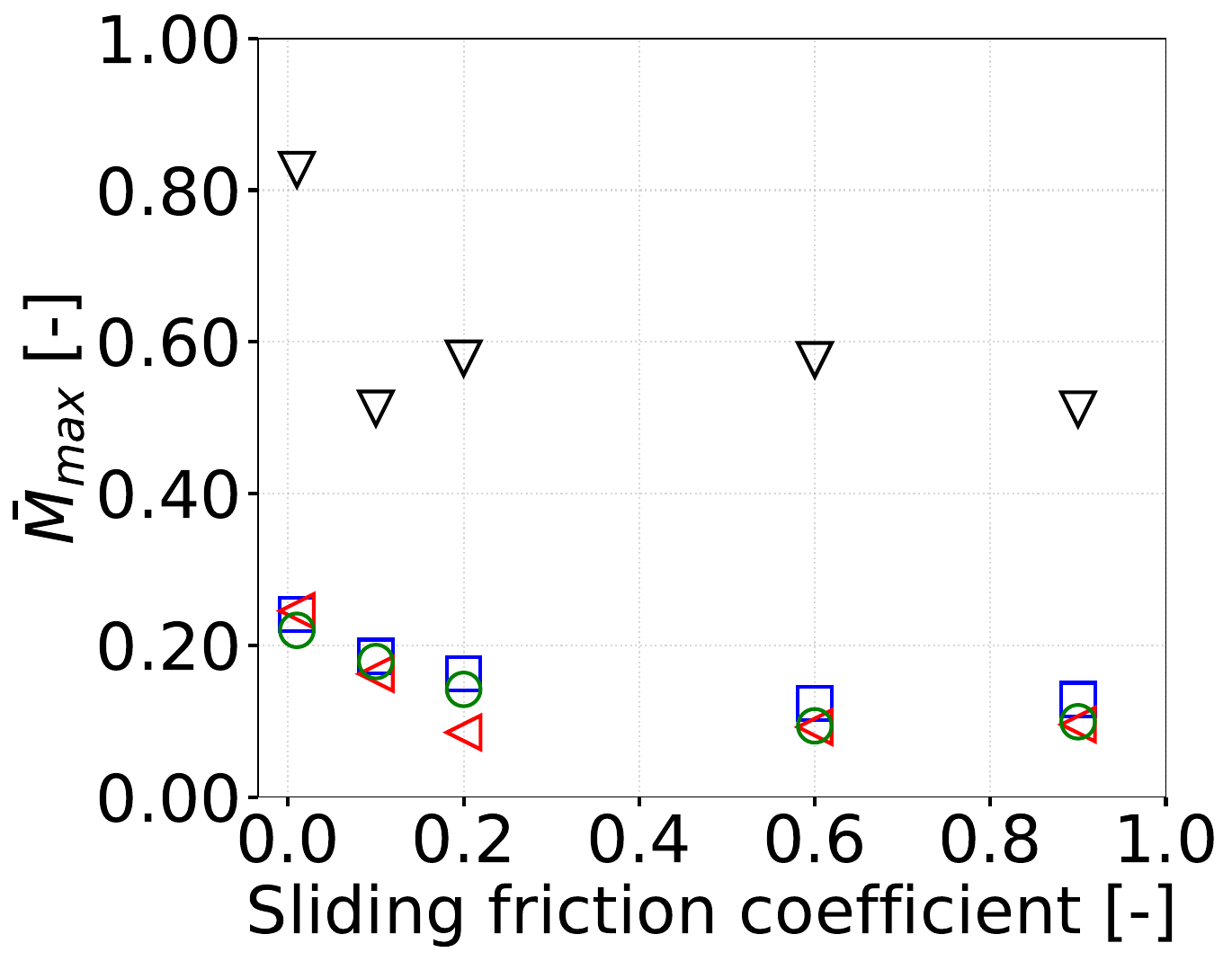}}
    \hfill
    \subcaptionbox{Particle A - High - NNM.\label{fig::props_alginate_nnm}}{\includegraphics[width=.49\textwidth]{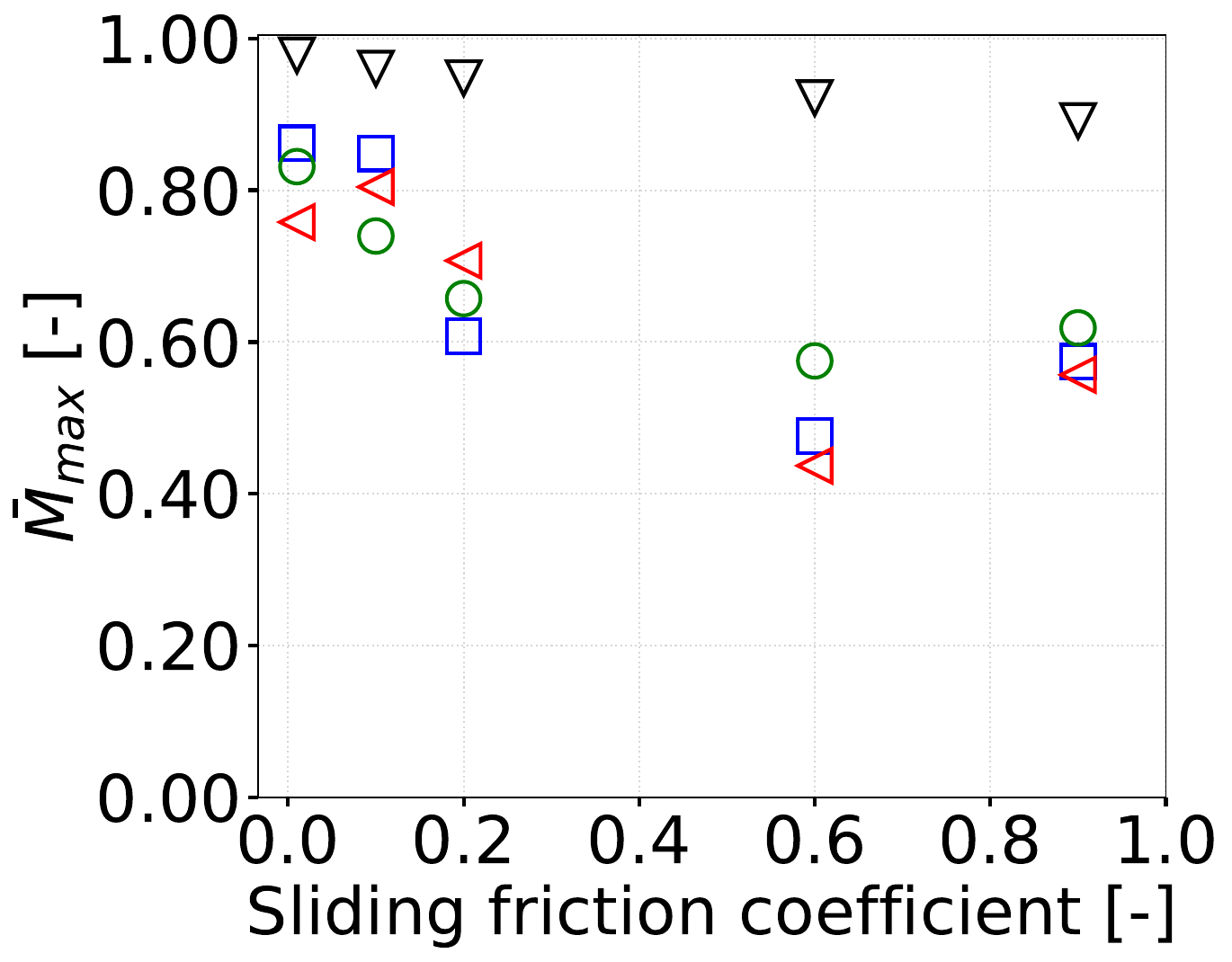}}
    \hfill

        \vspace*{+5mm}

    \hfill
    \subcaptionbox{Particle B - Low - NNM.\label{fig::props_abs_nnm}}{\includegraphics[width=.49\textwidth]{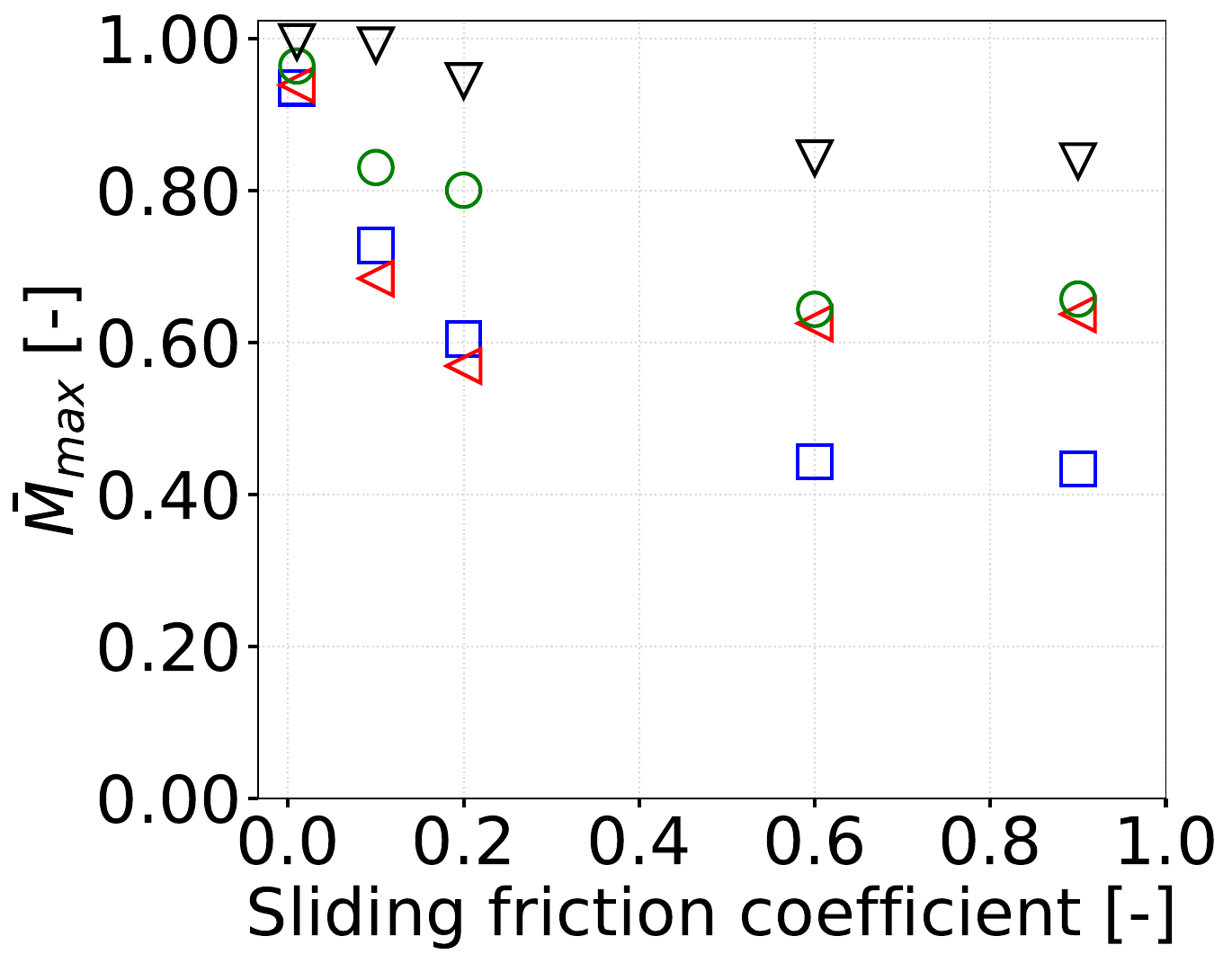}}
    \hfill
    \subcaptionbox{Particle C - Low - NNM.\label{fig::props_alumina_nnm}}{\includegraphics[width=.49\textwidth]{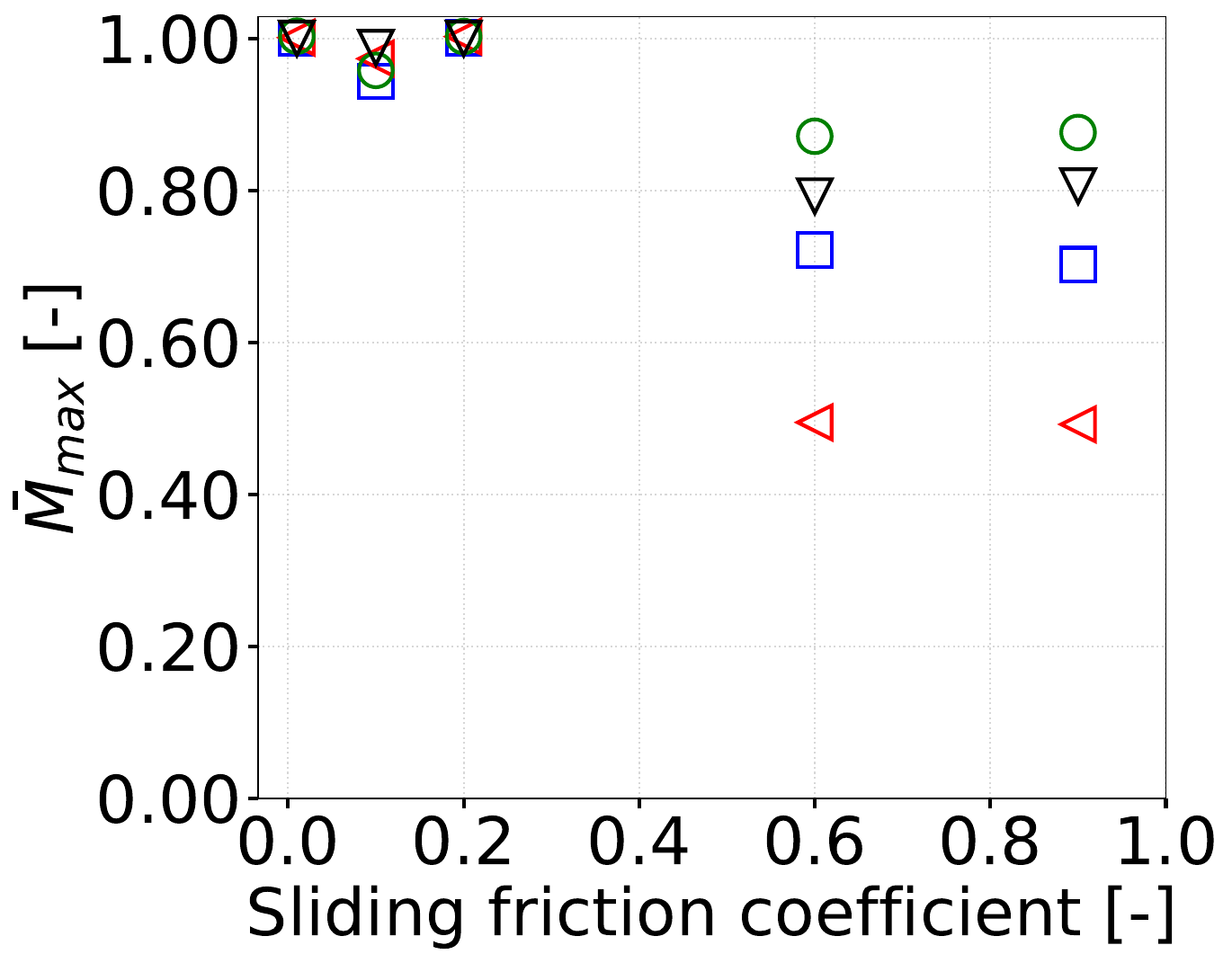}}
    \hfill

    \caption{Scatter plots of the analysis of the principal mixing component for particles A, B, and C, with various sliding friction coefficients.}
    \label{fig::props_sliding}
\end{figure}

As shown in Figure \ref{fig::props_sliding}, the results for particle A at the high and low inlet flow rate, and for particle B and C at the low inlet flow rates, there is a slight, yet, monotonous decrease in the mixing performance with the increase of the coefficient of sliding friction, prominently in $r$ for most results, and $\theta$ for the particle C. For the cases of particles B and C, the concentrated bed with slower particles facilitates longer tangential overlapping. For a very high friction coefficient, the excessive overlapping can cause a "sticking effect", virtually adding an inertial-like effect to the agglomerated particles. In the case of particle A, the very low Stokes number allows for longer overlaps such that this effect can be felt by looser beds.

This result corroborates with previous findings by \citet{Blais_2017} for a liquid-solid mixer. The authors report that the erosion speed is significantly decreased when lower sliding and rolling friction coefficients are applied, especially at higher impeller velocities. Still according to the authors, the pseudo-steady state suspension is not affected by the parameters. In the liquid-solid fluidized bed, the particles tend to have lower angular velocities given the multidimensional nature of their dynamics. As such, the rolling friction coefficient plays a less important part in the dynamics. On the contrary, the sliding friction coefficient can affect the bed in a similar fashion, as it does for the transitory erosion of the particles in the mixing tank.

\section{Conclusions}

In the present work, we leverage the CFD-DEM model implemented in Lethe \cite{Blais_2020, Golshan_2022, Geitani_2023a} and validated in a previous work of the research group \cite{Ferreira_2023} for the liquid-solid fluidized bed, using it to assess the LSFB mixing behavior. Simulations with a similar setup as the validation study were used to assess the particle dynamics inside the bed. More specifically, the method by \citet{Doucet_2008} and the Nearest Neighbors Method (NNM) were applied in the investigation of the bed's mixing performance in the pseudo-steady state. The mixing evolution was expressed in terms of the number of flows through (Eq. \eqref{eqn::n_flows}).

We demonstrate through principal component analysis that the particles mixing in the liquid-solid fluidized bed was isotropic. As mixing progresses, the eigenvectors are constantly interchanging in a random behavior, which resembles what is observed for particles following the Brownian motion pattern. The nearest neighbors method highlights the same trend, given the proximity between the indices for each of the components throughout simulations (Figure \ref{fig::mix_nflows_high_low}).

Lastly, changing the particles' interaction properties including Young's modulus, coefficient of restitution, coefficient of sliding friction, and coefficient of rolling friction did not have a significant effect on the mixing dynamics compared with the changes in the flow regimes. Nevertheless, the sliding friction coefficient was more influential than the other properties, especially in concentrated beds and beds with slow-moving particles. Our hypothesis is that the more concentrated the bed is, the higher the clustering provoked by the property, which results in a more prominent mixing resistance. Still, this effect vanishes for higher inlet flow rates.


\section{Acknowledgments}

Thanks for financial support are due to S\~ao Paulo Research Foundation (FAPESP, grant \#2019/19173-9, and grant \#2020/14567-6); and the Brazilian National Council for Scientific and Technological Development (CNPq, process number 408618/2018-3). The authors also thank the Federal University of S\~ao Carlos (UFSCar) and Polytechnique Montr\'eal for the infrastructure and all the technical support. This study was financed in part by the Coordena\c{c}\~ao de Aperfei\c{c}oamento de Pessoal de N\'ivel Superior - Brasil (CAPES) - Finance Code 001.

We would like to acknowledge the financial support from the Natural Sciences and Engineering Research Council of Canada (NSERC) through the RGPIN-2020-04510 Discovery Grant. The authors would also like to acknowledge the technical support and computing time provided by Compute Canada and Calcul Qu\'ebec.


\bibliography{bibliography}

\pagebreak

\appendix

\section{Principal component analysis of Brownian motion}
\label{brownian}
\begin{figure}[!htpb]
	\centering
     \includegraphics[width=0.6\linewidth]{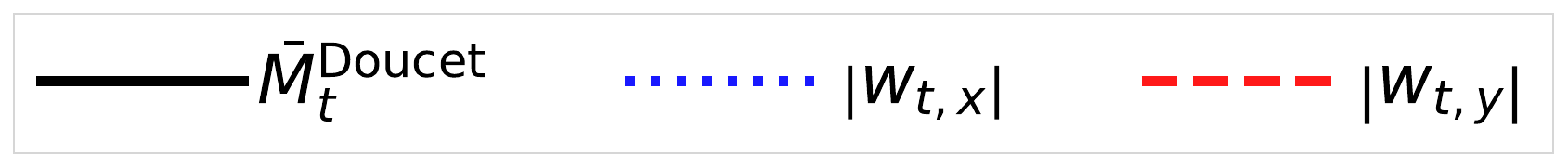}
    \begin{subfigure}{.50\textwidth}
		\centering
		\includegraphics[width=1\linewidth]{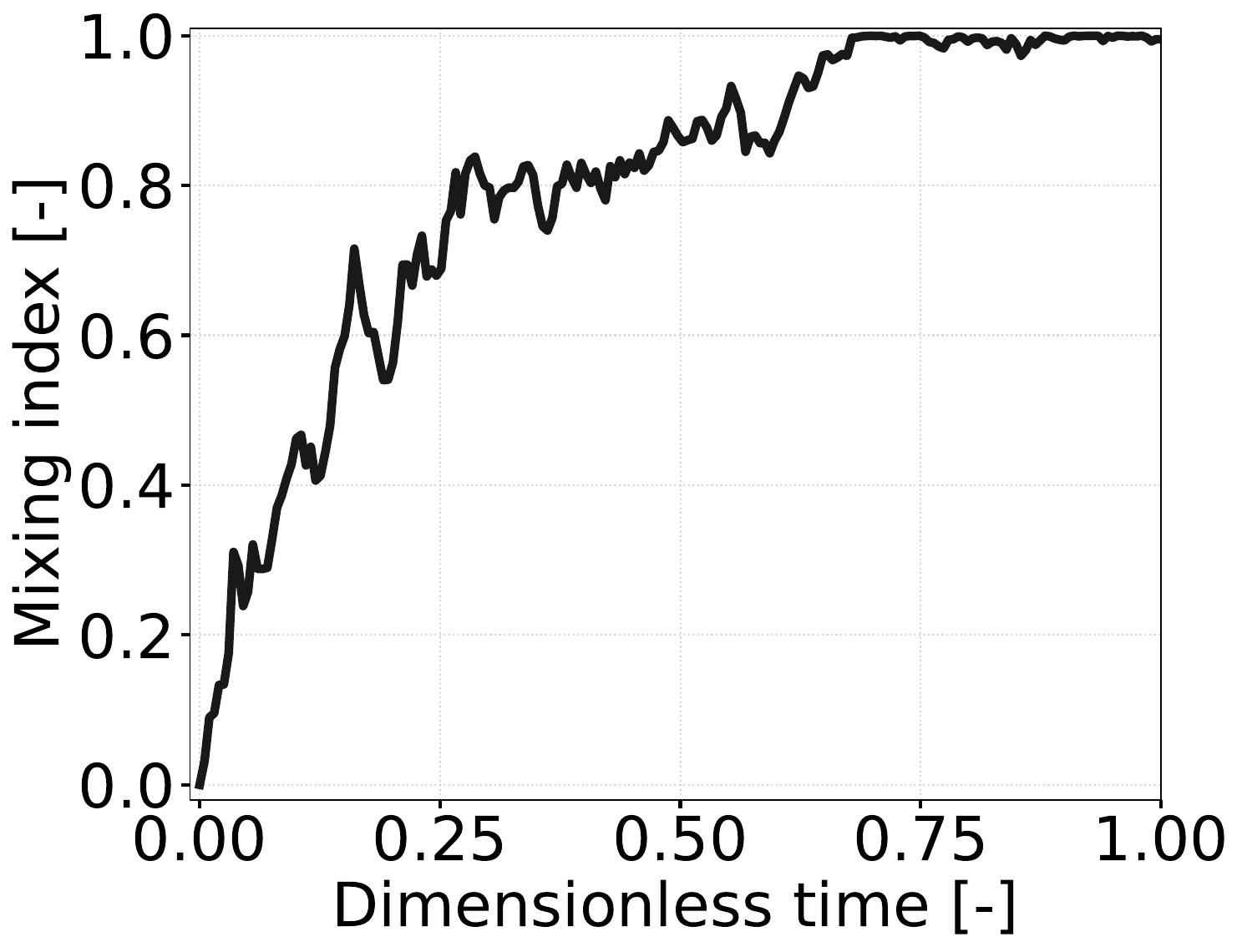}
        \caption{Doucet's mixing index with time.}
		\label{fig::brownian_mix}
	\end{subfigure}%
	\begin{subfigure}{.50\textwidth}
		\centering
		\includegraphics[width=1\linewidth]{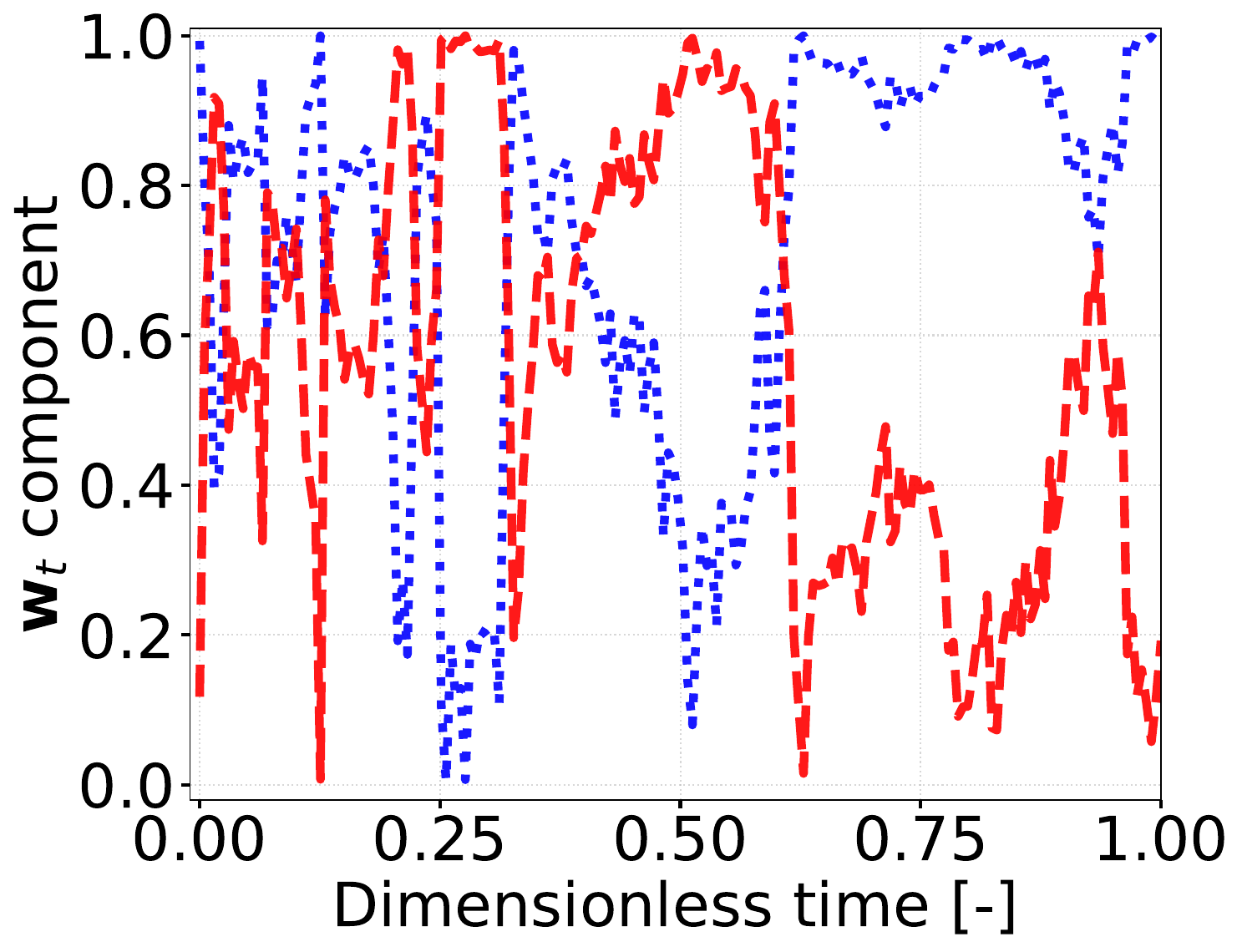}
        \caption{Corresponding eigenvectors.}
		\label{fig::brownian_eigen}
	\end{subfigure}
	\caption{Principal component analysis applied to Brownian motion.}
	\label{fig::brownian}
\end{figure}

Let us take the Brownian motion of 25 enclosed particles in bi-dimensional (x, y) domain. Performing the analysis of the principal components of mixing, we can observe an increase in mixing with time, as shown in Figure \ref{fig::brownian_mix}. By definition, given enough time for particles to diffuse, the Brownian motion will lead to a well-mixed system in all directions, with no principal component. This is evident in the interchangeable behavior of Doucet's eigenvectors with time shown in Figure \ref{fig::brownian_eigen}. In other words, the closer we are to an isotropic mixing, the less stable the eigenvectors will be.

\pagebreak

\section{Supplementary material}
\begin{flushleft}

\begin{landscape}

\vspace*{\fill}

\begin{table}[!htpb]
\centering
\resizebox{\linewidth}{!}{\begin{tabular}{cccccccccccccccccccc}
\toprule
\multicolumn{2}{c}{\bfseries \multirow{2}{*}{$N_{flows}^{\%}$}} & \bfseries \multirow{2}{*}{Standard} & \multicolumn{3}{c}{\bfseries Coefficient of restitution} & \multicolumn{4}{c}{\bfseries Coefficient of rolling friction} & \multicolumn{4}{c}{\bfseries Coefficient of sliding friction} & \multicolumn{2}{c}{\bfseries Young's modulus} \\
\bfseries  & \bfseries  & \bfseries  & \bfseries 0.1 & \bfseries 0.3 & \bfseries 0.6 & \bfseries 0.01 & \bfseries 0.1 & \bfseries 0.6 & \bfseries 0.9 & \bfseries 0.01 & \bfseries 0.2 & \bfseries 0.6 & \bfseries 0.9 & \bfseries 1e5 & 1e9 \\
\midrule
\multirow{6}{*}{\rotatebox[origin=c]{90}{NNM - $r$}} & 95\% & --- & --- & --- & --- & --- & --- & --- & --- & --- & --- & --- & --- & --- & --- \\
 & 90\% & --- & --- & --- & --- & --- & --- & --- & --- & --- & --- & --- & --- & --- & --- \\
 & 80\% & 0.234 & 0.261 & 0.250 & 0.256 & --- & 0.229 & 0.256 & 0.261 & 0.234 & --- & --- & --- & --- & --- \\
 & 70\% & 0.185 & 0.207 & 0.207 & 0.201 & 0.229 & 0.185 & 0.207 & 0.223 & 0.196 & --- & --- & --- & --- & --- \\
 & $\bar{M}_{max}$ & 0.848 & 0.822 & 0.831 & 0.835 & 0.767 & 0.868 & 0.833 & 0.818 & 0.862 & 0.608 & 0.476 & 0.574 & 0.548 & --- \\
 & $\bar{M}_{max}/N_{flows}^{max}$ & 3.117 & 3.019 & 3.054 & 3.066 & 2.820 & 3.188 & 3.059 & 3.006 & 3.167 & 2.233 & 1.750 & 2.110 & 2.013 & --- \\
\multirow{6}{*}{\rotatebox[origin=c]{90}{NNM - $\theta$}} & 95\% & --- & --- & --- & --- & --- & --- & --- & --- & --- & --- & --- & --- & --- & --- \\
 & 90\% & --- & --- & --- & --- & --- & --- & --- & --- & --- & --- & --- & --- & --- & --- \\
 & 80\% & 0.272 & --- & --- & --- & --- & 0.250 & 0.261 & --- & --- & --- & --- & --- & --- & --- \\
 & 70\% & 0.218 & 0.223 & 0.234 & 0.223 & 0.256 & 0.207 & 0.212 & 0.245 & 0.256 & 0.272 & --- & --- & --- & --- \\
 & $\bar{M}_{max}$ & 0.804 & 0.793 & 0.777 & 0.791 & 0.735 & 0.838 & 0.831 & 0.755 & 0.758 & 0.707 & 0.437 & 0.557 & 0.520 & --- \\
 & $\bar{M}_{max}/N_{flows}^{max}$ & 2.955 & 2.915 & 2.855 & 2.906 & 2.700 & 3.078 & 3.054 & 2.773 & 2.785 & 2.599 & 1.606 & 2.046 & 1.909 & --- \\
\multirow{6}{*}{\rotatebox[origin=c]{90}{NNM - $z$}} & 95\% & --- & --- & --- & --- & --- & --- & --- & --- & --- & --- & --- & --- & --- & --- \\
 & 90\% & --- & --- & --- & --- & --- & --- & --- & --- & --- & --- & --- & --- & --- & --- \\
 & 80\% & --- & --- & --- & --- & --- & --- & --- & --- & 0.256 & --- & --- & --- & --- & --- \\
 & 70\% & 0.250 & 0.245 & 0.256 & 0.272 & 0.261 & 0.245 & 0.261 & --- & 0.218 & --- & --- & --- & --- & --- \\
 & $\bar{M}_{max}$ & 0.740 & 0.770 & 0.748 & 0.704 & 0.732 & 0.765 & 0.728 & 0.682 & 0.831 & 0.657 & 0.575 & 0.619 & 0.674 & --- \\
 & $\bar{M}_{max}/N_{flows}^{max}$ & 2.718 & 2.829 & 2.749 & 2.586 & 2.688 & 2.810 & 2.677 & 2.504 & 3.054 & 2.415 & 2.113 & 2.273 & 2.477 & --- \\
\multirow{6}{*}{\rotatebox[origin=c]{90}{Doucet}} & 95\% & 0.239 & --- & 0.191 & 0.212 & --- & 0.212 & 0.191 & 0.223 & 0.234 & --- & --- & --- & 0.239 & --- \\
 & 90\% & 0.125 & 0.125 & 0.109 & 0.131 & 0.212 & 0.114 & 0.174 & 0.191 & 0.180 & 0.158 & 0.239 & --- & 0.163 & --- \\
 & 80\% & 0.087 & 0.082 & 0.082 & 0.082 & 0.147 & 0.093 & 0.136 & 0.136 & 0.109 & 0.136 & 0.191 & 0.163 & 0.142 & --- \\
 & 70\% & 0.076 & 0.071 & 0.071 & 0.065 & 0.054 & 0.076 & 0.098 & 0.076 & 0.076 & 0.120 & 0.152 & 0.136 & 0.131 & --- \\
 & $\bar{M}_{max}$ & 0.960 & 0.934 & 0.983 & 0.984 & 0.946 & 0.977 & 0.981 & 0.966 & 0.978 & 0.948 & 0.922 & 0.891 & 0.968 & --- \\
 & $\bar{M}_{max}/N_{flows}^{max}$ & 3.529 & 3.431 & 3.612 & 3.617 & 3.696 & 3.591 & 3.606 & 3.551 & 3.592 & 3.482 & 3.762 & 3.275 & 3.630 & --- \\
\bottomrule
\end{tabular}
}
\caption{Particle A - High inlet velocity.}
\label{tab::part_a_high}
\end{table}

\vspace*{\fill}

\end{landscape}

\begin{landscape}

\vspace*{\fill}
\begin{table}[!htpb]
\centering

\resizebox{\linewidth}{!}{\begin{tabular}{cccccccccccccccccccc}
\toprule
\multicolumn{2}{c}{\bfseries \multirow{2}{*}{$N_{flows}^{\%}$}} & \bfseries \multirow{2}{*}{Standard} & \multicolumn{3}{c}{\bfseries Coefficient of restitution} & \multicolumn{4}{c}{\bfseries Coefficient of rolling friction} & \multicolumn{4}{c}{\bfseries Coefficient of sliding friction} & \multicolumn{2}{c}{\bfseries Young's modulus} \\
\bfseries  & \bfseries  & \bfseries  & \bfseries 0.1 & \bfseries 0.3 & \bfseries 0.6 & \bfseries 0.01 & \bfseries 0.1 & \bfseries 0.6 & \bfseries 0.9 & \bfseries 0.01 & \bfseries 0.2 & \bfseries 0.6 & \bfseries 0.9 & \bfseries 1e5 & 1e9 \\
\midrule
\multirow{6}{*}{\rotatebox[origin=c]{90}{NNM - $r$}} & 95\% & --- & --- & --- & --- & --- & --- & --- & --- & --- & --- & --- & --- & --- & --- \\
 & 90\% & --- & --- & --- & --- & --- & --- & --- & --- & --- & --- & --- & --- & --- & --- \\
 & 80\% & --- & --- & --- & --- & --- & --- & --- & --- & --- & --- & --- & --- & --- & --- \\
 & 70\% & --- & --- & --- & --- & --- & --- & --- & --- & --- & --- & --- & --- & --- & --- \\
 & $\bar{M}_{max}$ & 0.185 & 0.173 & 0.184 & 0.187 & 0.199 & 0.187 & 0.180 & 0.179 & 0.241 & 0.163 & 0.123 & 0.128 & 0.145 & --- \\
 & $\bar{M}_{max}/N_{flows}^{max}$ & 1.680 & 1.571 & 1.670 & 1.698 & 1.801 & 1.697 & 1.635 & 1.688 & 2.181 & 1.473 & 1.118 & 1.163 & 1.318 & --- \\
\multirow{6}{*}{\rotatebox[origin=c]{90}{NNM - $\theta$}} & 95\% & --- & --- & --- & --- & --- & --- & --- & --- & --- & --- & --- & --- & --- & --- \\
 & 90\% & --- & --- & --- & --- & --- & --- & --- & --- & --- & --- & --- & --- & --- & --- \\
 & 80\% & --- & --- & --- & --- & --- & --- & --- & --- & --- & --- & --- & --- & --- & --- \\
 & 70\% & --- & --- & --- & --- & --- & --- & --- & --- & --- & --- & --- & --- & --- & --- \\
 & $\bar{M}_{max}$ & 0.162 & 0.096 & 0.116 & 0.125 & 0.182 & 0.145 & 0.138 & 0.151 & 0.245 & 0.085 & 0.092 & 0.096 & 0.050 & --- \\
 & $\bar{M}_{max}/N_{flows}^{max}$ & 1.472 & 0.866 & 1.053 & 1.135 & 1.648 & 1.310 & 1.254 & 1.421 & 2.224 & 0.772 & 0.836 & 0.867 & 0.451 & --- \\
\multirow{6}{*}{\rotatebox[origin=c]{90}{NNM - $z$}} & 95\% & --- & --- & --- & --- & --- & --- & --- & --- & --- & --- & --- & --- & --- & --- \\
 & 90\% & --- & --- & --- & --- & --- & --- & --- & --- & --- & --- & --- & --- & --- & --- \\
 & 80\% & --- & --- & --- & --- & --- & --- & --- & --- & --- & --- & --- & --- & --- & --- \\
 & 70\% & --- & --- & --- & --- & --- & --- & --- & --- & --- & --- & --- & --- & --- & --- \\
 & $\bar{M}_{max}$ & 0.179 & 0.167 & 0.167 & 0.170 & 0.208 & 0.178 & 0.174 & 0.170 & 0.220 & 0.142 & 0.094 & 0.099 & 0.171 & --- \\
 & $\bar{M}_{max}/N_{flows}^{max}$ & 1.619 & 1.516 & 1.512 & 1.537 & 1.889 & 1.615 & 1.580 & 1.607 & 1.992 & 1.286 & 0.851 & 0.900 & 1.549 & --- \\
\multirow{6}{*}{\rotatebox[origin=c]{90}{Doucet}} & 95\% & --- & --- & --- & --- & --- & --- & --- & --- & --- & --- & --- & --- & --- & --- \\
 & 90\% & --- & --- & --- & --- & --- & --- & --- & --- & --- & --- & --- & --- & --- & --- \\
 & 80\% & --- & --- & --- & --- & --- & --- & --- & --- & 0.090 & --- & --- & --- & --- & --- \\
 & 70\% & --- & --- & --- & --- & --- & --- & --- & --- & 0.068 & --- & --- & --- & --- & --- \\
 & $\bar{M}_{max}$ & 0.512 & 0.578 & 0.531 & 0.560 & 0.638 & 0.533 & 0.559 & 0.556 & 0.827 & 0.578 & 0.576 & 0.511 & 0.460 & --- \\
 & $\bar{M}_{max}/N_{flows}^{max}$ & 4.643 & 9.351 & 4.815 & 5.071 & 5.777 & 4.830 & 5.066 & 5.251 & 7.807 & 5.241 & 5.222 & 4.634 & 4.165 & --- \\
\bottomrule
\end{tabular}
}

\caption{Particle A - Low inlet velocity.}
\label{tab::part_a_low}
\end{table}
\vspace*{\fill}

\end{landscape}


\begin{landscape}

\vspace*{\fill}
\begin{table}[!htpb]
\centering

\resizebox{\linewidth}{!}{\begin{tabular}{cccccccccccccccccccc}
\toprule
\multicolumn{2}{c}{\bfseries \multirow{2}{*}{$N_{flows}^{\%}$}} & \bfseries \multirow{2}{*}{Standard} & \multicolumn{3}{c}{\bfseries Coefficient of restitution} & \multicolumn{4}{c}{\bfseries Coefficient of rolling friction} & \multicolumn{4}{c}{\bfseries Coefficient of sliding friction} & \multicolumn{2}{c}{\bfseries Young's modulus} \\
\bfseries  & \bfseries  & \bfseries  & \bfseries 0.1 & \bfseries 0.3 & \bfseries 0.6 & \bfseries 0.01 & \bfseries 0.1 & \bfseries 0.6 & \bfseries 0.9 & \bfseries 0.01 & \bfseries 0.2 & \bfseries 0.6 & \bfseries 0.9 & \bfseries 1e5 & 1e9 \\
\midrule
\multirow{6}{*}{\rotatebox[origin=c]{90}{NNM - $r$}} & 95\% & 0.726 & 0.682 & 0.682 & 0.594 & 0.704 & 0.660 & 0.748 & 0.660 & 0.726 & 0.660 & 0.616 & 0.836 & 0.660 & --- \\
 & 90\% & 0.550 & 0.550 & 0.550 & 0.484 & 0.550 & 0.528 & 0.616 & 0.550 & 0.594 & 0.528 & 0.484 & 0.660 & 0.572 & --- \\
 & 80\% & 0.418 & 0.418 & 0.418 & 0.374 & 0.418 & 0.418 & 0.484 & 0.418 & 0.440 & 0.396 & 0.374 & 0.484 & 0.440 & --- \\
 & 70\% & 0.330 & 0.352 & 0.330 & 0.308 & 0.330 & 0.330 & 0.374 & 0.330 & 0.352 & 0.330 & 0.308 & 0.396 & 0.374 & --- \\
 & $\bar{M}_{max}$ & 0.993 & 1.000 & 0.997 & 1.000 & 1.000 & 1.000 & 0.997 & 1.000 & 1.000 & 1.000 & 1.000 & 0.994 & 0.996 & --- \\
 & $\bar{M}_{max}/N_{flows}^{max}$ & 0.602 & 0.623 & 0.731 & 0.631 & 0.688 & 0.689 & 0.604 & 0.689 & 0.606 & 0.606 & 0.640 & 0.619 & 0.686 & --- \\
\multirow{6}{*}{\rotatebox[origin=c]{90}{NNM - $\theta$}} & 95\% & 0.968 & 0.770 & 0.858 & 0.748 & 0.704 & 0.704 & 0.946 & 0.902 & 0.748 & 0.814 & 0.814 & 0.902 & 0.748 & --- \\
 & 90\% & 0.770 & 0.638 & 0.682 & 0.638 & 0.594 & 0.550 & 0.704 & 0.682 & 0.594 & 0.660 & 0.660 & 0.748 & 0.616 & --- \\
 & 80\% & 0.594 & 0.484 & 0.506 & 0.484 & 0.462 & 0.440 & 0.528 & 0.506 & 0.462 & 0.506 & 0.506 & 0.572 & 0.506 & --- \\
 & 70\% & 0.484 & 0.418 & 0.396 & 0.374 & 0.374 & 0.352 & 0.418 & 0.418 & 0.374 & 0.418 & 0.418 & 0.462 & 0.418 & --- \\
 & $\bar{M}_{max}$ & 0.998 & 0.999 & 0.997 & 0.999 & 1.000 & 0.999 & 0.995 & 1.000 & 1.000 & 1.000 & 0.999 & 0.998 & 1.000 & --- \\
 & $\bar{M}_{max}/N_{flows}^{max}$ & 0.622 & 0.605 & 0.604 & 0.606 & 0.710 & 0.614 & 0.603 & 0.649 & 0.614 & 0.622 & 0.640 & 0.639 & 0.699 & --- \\
\multirow{6}{*}{\rotatebox[origin=c]{90}{NNM - $z$}} & 95\% & 1.298 & 1.386 & 1.056 & 0.792 & 0.880 & 1.056 & 1.474 & 1.166 & 1.034 & 0.968 & 0.968 & 1.364 & 1.276 & --- \\
 & 90\% & 0.990 & 1.122 & 0.924 & 0.682 & 0.748 & 0.814 & 1.210 & 0.924 & 0.858 & 0.814 & 0.792 & 1.122 & 1.078 & --- \\
 & 80\% & 0.726 & 0.858 & 0.748 & 0.550 & 0.594 & 0.638 & 0.924 & 0.660 & 0.704 & 0.616 & 0.594 & 0.880 & 0.836 & --- \\
 & 70\% & 0.616 & 0.726 & 0.616 & 0.462 & 0.484 & 0.528 & 0.770 & 0.528 & 0.616 & 0.506 & 0.506 & 0.748 & 0.682 & --- \\
 & $\bar{M}_{max}$ & 0.983 & 0.983 & 0.997 & 1.000 & 1.000 & 0.990 & 0.964 & 0.994 & 0.999 & 1.000 & 0.997 & 0.990 & 0.990 & --- \\
 & $\bar{M}_{max}/N_{flows}^{max}$ & 0.596 & 0.596 & 0.657 & 0.770 & 0.649 & 0.616 & 0.592 & 0.611 & 0.614 & 0.622 & 0.604 & 0.600 & 0.600 & --- \\
\multirow{6}{*}{\rotatebox[origin=c]{90}{Doucet}} & 95\% & 1.122 & 1.056 & 0.814 & 0.308 & 0.396 & 0.770 & 1.144 & 1.056 & 0.462 & 0.528 & 0.528 & 1.034 & 0.814 & --- \\
 & 90\% & 0.616 & 0.770 & 0.638 & 0.264 & 0.330 & 0.528 & 0.924 & 0.770 & 0.440 & 0.440 & 0.374 & 0.858 & 0.726 & --- \\
 & 80\% & 0.462 & 0.528 & 0.396 & 0.198 & 0.286 & 0.352 & 0.550 & 0.418 & 0.396 & 0.286 & 0.286 & 0.528 & 0.550 & --- \\
 & 70\% & 0.374 & 0.440 & 0.286 & 0.176 & 0.242 & 0.286 & 0.374 & 0.220 & 0.352 & 0.242 & 0.220 & 0.396 & 0.396 & --- \\
 & $\bar{M}_{max}$ & 0.991 & 0.993 & 0.999 & 1.000 & 1.000 & 0.998 & 0.986 & 0.998 & 1.000 & 1.000 & 0.999 & 0.999 & 0.997 & --- \\
 & $\bar{M}_{max}/N_{flows}^{max}$ & 0.601 & 0.602 & 0.606 & 0.640 & 0.710 & 0.605 & 0.623 & 0.605 & 0.606 & 0.623 & 0.605 & 0.622 & 0.604 & --- \\
\bottomrule
\end{tabular}
}

\caption{Particle B - High inlet velocity.}
\end{table}
\vspace*{\fill}

\end{landscape}

\begin{landscape}

\vspace*{\fill}
\begin{table}[!htpb]
\centering
\resizebox{\linewidth}{!}{\begin{tabular}{cccccccccccccccccccc}
\toprule
\multicolumn{2}{c}{\bfseries \multirow{2}{*}{$N_{flows}^{\%}$}} & \bfseries \multirow{2}{*}{Standard} & \multicolumn{3}{c}{\bfseries Coefficient of restitution} & \multicolumn{4}{c}{\bfseries Coefficient of rolling friction} & \multicolumn{4}{c}{\bfseries Coefficient of sliding friction} & \multicolumn{2}{c}{\bfseries Young's modulus} \\
\bfseries  & \bfseries  & \bfseries  & \bfseries 0.1 & \bfseries 0.3 & \bfseries 0.6 & \bfseries 0.01 & \bfseries 0.1 & \bfseries 0.6 & \bfseries 0.9 & \bfseries 0.01 & \bfseries 0.2 & \bfseries 0.6 & \bfseries 0.9 & \bfseries 1e5 & 1e9 \\
\midrule
\multirow{6}{*}{\rotatebox[origin=c]{90}{NNM - $r$}} & 95\% & --- & --- & --- & --- & --- & 0.611 & --- & --- & --- & --- & --- & --- & --- & --- \\
 & 90\% & --- & --- & --- & --- & 0.492 & 0.586 & --- & --- & 0.560 & --- & --- & --- & --- & --- \\
 & 80\% & --- & --- & --- & --- & 0.357 & 0.577 & --- & --- & 0.433 & --- & --- & --- & 0.603 & --- \\
 & 70\% & 0.603 & --- & --- & 0.586 & 0.306 & 0.509 & --- & 0.603 & 0.357 & --- & --- & --- & 0.475 & --- \\
 & $\bar{M}_{max}$ & 0.727 & 0.665 & 0.688 & 0.741 & 0.941 & 0.976 & 0.675 & 0.726 & 0.935 & 0.605 & 0.443 & 0.434 & 0.830 & --- \\
 & $\bar{M}_{max}/N_{flows}^{max}$ & 1.143 & 1.045 & 1.081 & 1.164 & 1.479 & 1.534 & 1.060 & 1.140 & 1.469 & 0.950 & 0.696 & 0.681 & 1.303 & --- \\
\multirow{6}{*}{\rotatebox[origin=c]{90}{NNM - $\theta$}} & 95\% & --- & --- & --- & --- & --- & --- & --- & --- & --- & --- & --- & --- & --- & --- \\
 & 90\% & --- & --- & --- & --- & 0.509 & 0.620 & --- & --- & 0.552 & --- & --- & --- & --- & --- \\
 & 80\% & --- & --- & --- & --- & 0.382 & 0.577 & --- & --- & 0.441 & --- & --- & --- & --- & --- \\
 & 70\% & --- & --- & --- & 0.577 & 0.340 & 0.560 & --- & 0.577 & 0.373 & --- & --- & --- & 0.560 & --- \\
 & $\bar{M}_{max}$ & 0.684 & 0.644 & 0.602 & 0.736 & 0.944 & 0.936 & 0.633 & 0.745 & 0.939 & 0.569 & 0.625 & 0.637 & 0.762 & --- \\
 & $\bar{M}_{max}/N_{flows}^{max}$ & 1.075 & 1.011 & 0.946 & 1.156 & 1.483 & 1.471 & 0.995 & 1.170 & 1.475 & 0.894 & 0.982 & 1.001 & 1.196 & --- \\
\multirow{6}{*}{\rotatebox[origin=c]{90}{NNM - $z$}} & 95\% & --- & --- & --- & --- & --- & 0.611 & --- & --- & 0.603 & --- & --- & --- & --- & --- \\
 & 90\% & --- & --- & --- & --- & --- & 0.586 & --- & --- & 0.509 & --- & --- & --- & --- & --- \\
 & 80\% & 0.594 & 0.594 & 0.594 & 0.594 & 0.518 & 0.569 & 0.577 & 0.628 & 0.416 & 0.637 & --- & --- & 0.594 & --- \\
 & 70\% & 0.484 & 0.475 & 0.467 & 0.475 & 0.416 & 0.484 & 0.467 & 0.492 & 0.357 & 0.560 & --- & --- & 0.467 & --- \\
 & $\bar{M}_{max}$ & 0.830 & 0.826 & 0.833 & 0.821 & 0.864 & 0.972 & 0.843 & 0.808 & 0.964 & 0.800 & 0.644 & 0.657 & 0.831 & --- \\
 & $\bar{M}_{max}/N_{flows}^{max}$ & 1.304 & 1.298 & 1.309 & 1.289 & 1.357 & 1.527 & 1.324 & 1.269 & 1.514 & 1.257 & 1.011 & 1.032 & 1.306 & --- \\
\multirow{6}{*}{\rotatebox[origin=c]{90}{Doucet}} & 95\% & 0.450 & 0.467 & 0.475 & 0.475 & 0.212 & 0.577 & 0.441 & 0.450 & 0.289 & --- & --- & --- & 0.501 & --- \\
 & 90\% & 0.365 & 0.424 & 0.424 & 0.357 & 0.195 & 0.314 & 0.399 & 0.340 & 0.246 & 0.552 & --- & --- & 0.280 & --- \\
 & 80\% & 0.280 & 0.289 & 0.306 & 0.272 & 0.153 & 0.255 & 0.297 & 0.263 & 0.161 & 0.433 & 0.603 & 0.586 & 0.212 & --- \\
 & 70\% & 0.136 & 0.229 & 0.238 & 0.212 & 0.102 & 0.119 & 0.229 & 0.127 & 0.127 & 0.331 & 0.450 & 0.475 & 0.170 & --- \\
 & $\bar{M}_{max}$ & 0.991 & 0.979 & 0.981 & 0.981 & 0.997 & 0.989 & 0.976 & 0.989 & 0.996 & 0.944 & 0.843 & 0.839 & 0.968 & --- \\
 & $\bar{M}_{max}/N_{flows}^{max}$ & 1.743 & 1.537 & 1.778 & 1.628 & 1.836 & 1.554 & 1.576 & 1.792 & 1.564 & 1.483 & 1.324 & 1.317 & 1.521 & --- \\
\bottomrule
\end{tabular}
}

\caption{Particle B - Low inlet velocity.}
\end{table}
\vspace*{\fill}

\end{landscape}


\begin{landscape}

\vspace*{\fill}
\begin{table}[!htpb]
\centering
\resizebox{\linewidth}{!}{\begin{tabular}{cccccccccccccccccccc}
\toprule
\multicolumn{2}{c}{\bfseries \multirow{2}{*}{$N_{flows}^{\%}$}} & \bfseries \multirow{2}{*}{Standard} & \multicolumn{3}{c}{\bfseries Coefficient of restitution} & \multicolumn{4}{c}{\bfseries Coefficient of rolling friction} & \multicolumn{4}{c}{\bfseries Coefficient of sliding friction} & \multicolumn{2}{c}{\bfseries Young's modulus} \\
\bfseries  & \bfseries  & \bfseries  & \bfseries 0.1 & \bfseries 0.3 & \bfseries 0.6 & \bfseries 0.01 & \bfseries 0.1 & \bfseries 0.6 & \bfseries 0.9 & \bfseries 0.01 & \bfseries 0.2 & \bfseries 0.6 & \bfseries 0.9 & \bfseries 1e5 & 1e9 \\
\midrule
\multirow{6}{*}{\rotatebox[origin=c]{90}{NNM - $r$}} & 95\% & 0.628 & 0.534 & 0.565 & 0.565 & 0.534 & 0.597 & 0.628 & 0.534 & 0.565 & 0.534 & 0.785 & 0.597 & 0.597 & 0.534 \\
 & 90\% & 0.503 & 0.440 & 0.471 & 0.503 & 0.471 & 0.503 & 0.503 & 0.408 & 0.471 & 0.440 & 0.628 & 0.471 & 0.503 & 0.408 \\
 & 80\% & 0.345 & 0.377 & 0.377 & 0.408 & 0.377 & 0.377 & 0.408 & 0.314 & 0.345 & 0.345 & 0.471 & 0.377 & 0.377 & 0.345 \\
 & 70\% & 0.283 & 0.314 & 0.314 & 0.314 & 0.314 & 0.314 & 0.345 & 0.251 & 0.283 & 0.283 & 0.377 & 0.314 & 0.314 & 0.283 \\
 & $\bar{M}_{max}$ & 1.000 & 1.000 & 1.000 & 1.000 & 1.000 & 1.000 & 1.000 & 1.000 & 1.000 & 1.000 & 1.000 & 1.000 & 1.000 & 1.000 \\
 & $\bar{M}_{max}/N_{flows}^{max}$ & 0.468 & 0.601 & 0.816 & 0.482 & 0.430 & 0.777 & 0.442 & 0.612 & 0.637 & 0.692 & 0.522 & 0.624 & 0.650 & 0.692 \\
\multirow{6}{*}{\rotatebox[origin=c]{90}{NNM - $\theta$}} & 95\% & 0.597 & 0.503 & 0.503 & 0.565 & 0.565 & 0.628 & 0.660 & 0.597 & 0.503 & 0.534 & 0.911 & 0.628 & 0.534 & 0.628 \\
 & 90\% & 0.471 & 0.440 & 0.440 & 0.503 & 0.471 & 0.534 & 0.565 & 0.503 & 0.408 & 0.440 & 0.754 & 0.503 & 0.440 & 0.534 \\
 & 80\% & 0.345 & 0.345 & 0.345 & 0.377 & 0.377 & 0.440 & 0.471 & 0.377 & 0.314 & 0.314 & 0.597 & 0.377 & 0.345 & 0.408 \\
 & 70\% & 0.283 & 0.283 & 0.283 & 0.314 & 0.345 & 0.377 & 0.408 & 0.283 & 0.251 & 0.283 & 0.503 & 0.314 & 0.283 & 0.345 \\
 & $\bar{M}_{max}$ & 1.000 & 1.000 & 1.000 & 1.000 & 1.000 & 1.000 & 1.000 & 1.000 & 1.000 & 1.000 & 1.000 & 1.000 & 1.000 & 1.000 \\
 & $\bar{M}_{max}/N_{flows}^{max}$ & 0.498 & 0.514 & 0.430 & 0.724 & 0.579 & 0.724 & 0.468 & 0.475 & 0.816 & 0.579 & 0.461 & 0.498 & 0.590 & 0.861 \\
\multirow{6}{*}{\rotatebox[origin=c]{90}{NNM - $z$}} & 95\% & 0.628 & 0.565 & 0.565 & 0.597 & 0.628 & 0.660 & 0.628 & 0.722 & 0.597 & 0.628 & 0.848 & 0.660 & 0.565 & 0.628 \\
 & 90\% & 0.503 & 0.471 & 0.471 & 0.534 & 0.534 & 0.534 & 0.534 & 0.597 & 0.503 & 0.503 & 0.691 & 0.565 & 0.471 & 0.534 \\
 & 80\% & 0.377 & 0.408 & 0.408 & 0.440 & 0.440 & 0.440 & 0.440 & 0.471 & 0.408 & 0.440 & 0.534 & 0.440 & 0.377 & 0.440 \\
 & 70\% & 0.314 & 0.345 & 0.345 & 0.377 & 0.377 & 0.345 & 0.377 & 0.377 & 0.345 & 0.377 & 0.440 & 0.377 & 0.314 & 0.377 \\
 & $\bar{M}_{max}$ & 1.000 & 1.000 & 1.000 & 1.000 & 1.000 & 1.000 & 1.000 & 1.000 & 1.000 & 1.000 & 1.000 & 1.000 & 1.000 & 1.000 \\
 & $\bar{M}_{max}/N_{flows}^{max}$ & 0.425 & 0.663 & 0.677 & 0.758 & 0.482 & 0.601 & 0.724 & 0.677 & 0.796 & 0.637 & 0.590 & 0.540 & 0.430 & 0.777 \\
\multirow{6}{*}{\rotatebox[origin=c]{90}{Doucet}} & 95\% & 0.283 & 0.251 & 0.314 & 0.377 & 0.377 & 0.471 & 0.440 & 0.408 & 0.251 & 0.283 & 0.534 & 0.377 & 0.283 & 0.345 \\
 & 90\% & 0.220 & 0.220 & 0.283 & 0.220 & 0.283 & 0.314 & 0.377 & 0.251 & 0.220 & 0.251 & 0.440 & 0.314 & 0.251 & 0.283 \\
 & 80\% & 0.188 & 0.188 & 0.220 & 0.157 & 0.220 & 0.188 & 0.188 & 0.188 & 0.188 & 0.220 & 0.345 & 0.220 & 0.188 & 0.220 \\
 & 70\% & 0.157 & 0.157 & 0.188 & 0.126 & 0.220 & 0.157 & 0.157 & 0.157 & 0.157 & 0.188 & 0.188 & 0.188 & 0.188 & 0.188 \\
 & $\bar{M}_{max}$ & 1.000 & 1.000 & 1.000 & 1.000 & 1.000 & 1.000 & 1.000 & 1.000 & 1.000 & 1.000 & 1.000 & 1.000 & 1.000 & 1.000 \\
 & $\bar{M}_{max}/N_{flows}^{max}$ & 0.692 & 0.740 & 0.624 & 0.531 & 0.677 & 0.590 & 0.637 & 0.590 & 0.442 & 0.475 & 0.514 & 0.663 & 0.448 & 0.637 \\
\bottomrule
\end{tabular}
}
\caption{Particle C - High inlet velocity.}
\end{table}
\vspace*{\fill}

\end{landscape}

\begin{landscape}

\vspace*{\fill}
\begin{table}[!htpb]
\centering
\resizebox{\linewidth}{!}{\begin{tabular}{cccccccccccccccccccc}
\toprule
\multicolumn{2}{c}{\bfseries \multirow{2}{*}{$N_{flows}^{\%}$}} & \bfseries \multirow{2}{*}{Standard} & \multicolumn{3}{c}{\bfseries Coefficient of restitution} & \multicolumn{4}{c}{\bfseries Coefficient of rolling friction} & \multicolumn{4}{c}{\bfseries Coefficient of sliding friction} & \multicolumn{2}{c}{\bfseries Young's modulus} \\
\bfseries  & \bfseries  & \bfseries  & \bfseries 0.1 & \bfseries 0.3 & \bfseries 0.6 & \bfseries 0.01 & \bfseries 0.1 & \bfseries 0.6 & \bfseries 0.9 & \bfseries 0.01 & \bfseries 0.2 & \bfseries 0.6 & \bfseries 0.9 & \bfseries 1e5 & 1e9 \\
\midrule
\multirow{6}{*}{\rotatebox[origin=c]{90}{NNM - $r$}} & 95\% & --- & 1.292 & --- & --- & 1.071 & 1.343 & --- & 1.632 & 0.646 & 0.969 & --- & --- & 1.173 & 1.530 \\
 & 90\% & 1.292 & 1.258 & 1.360 & 1.275 & 0.833 & 1.020 & 1.360 & 1.241 & 0.527 & 0.935 & --- & --- & 0.850 & 1.224 \\
 & 80\% & 0.918 & 0.969 & 0.986 & 0.935 & 0.595 & 0.748 & 0.969 & 0.901 & 0.391 & 0.918 & --- & --- & 0.629 & 0.918 \\
 & 70\% & 0.714 & 0.748 & 0.782 & 0.765 & 0.493 & 0.629 & 0.782 & 0.714 & 0.306 & 0.867 & 1.513 & 1.700 & 0.544 & 0.731 \\
 & $\bar{M}_{max}$ & 0.943 & 0.998 & 0.934 & 0.945 & 0.982 & 0.976 & 0.943 & 0.956 & 1.000 & 1.000 & 0.722 & 0.702 & 0.984 & 1.000 \\
 & $\bar{M}_{max}/N_{flows}^{max}$ & 0.566 & 0.631 & 0.549 & 0.556 & 0.590 & 0.574 & 0.555 & 0.562 & 0.520 & 0.668 & 0.451 & 0.413 & 0.597 & 0.429 \\
\multirow{6}{*}{\rotatebox[origin=c]{90}{NNM - $\theta$}} & 95\% & 1.258 & 1.190 & 1.683 & 1.326 & 1.173 & 1.598 & 1.122 & 1.343 & 0.731 & 0.935 & --- & --- & 1.122 & 1.258 \\
 & 90\% & 0.935 & 0.935 & 1.156 & 0.969 & 0.901 & 1.292 & 0.901 & 1.054 & 0.595 & 0.901 & --- & --- & 0.884 & 0.986 \\
 & 80\% & 0.697 & 0.697 & 0.799 & 0.714 & 0.697 & 0.799 & 0.697 & 0.765 & 0.442 & 0.714 & --- & --- & 0.629 & 0.731 \\
 & 70\% & 0.544 & 0.561 & 0.612 & 0.578 & 0.544 & 0.595 & 0.561 & 0.595 & 0.340 & 0.578 & --- & --- & 0.527 & 0.578 \\
 & $\bar{M}_{max}$ & 0.973 & 1.000 & 0.950 & 0.968 & 0.978 & 0.961 & 0.975 & 0.971 & 1.000 & 1.000 & 0.495 & 0.492 & 0.989 & 1.000 \\
 & $\bar{M}_{max}/N_{flows}^{max}$ & 0.572 & 0.653 & 0.564 & 0.575 & 0.587 & 0.565 & 0.573 & 0.571 & 0.442 & 0.653 & 0.310 & 0.290 & 0.582 & 0.463 \\
\multirow{6}{*}{\rotatebox[origin=c]{90}{NNM - $z$}} & 95\% & 1.428 & 1.292 & 1.615 & 1.700 & 1.394 & 1.496 & 1.632 & 1.700 & 0.782 & 0.952 & --- & --- & 1.241 & 1.632 \\
 & 90\% & 1.054 & 1.156 & 1.088 & 1.122 & 1.105 & 1.071 & 1.054 & 1.224 & 0.663 & 0.935 & --- & --- & 0.969 & 1.241 \\
 & 80\% & 0.765 & 0.799 & 0.782 & 0.799 & 0.748 & 0.748 & 0.731 & 0.816 & 0.544 & 0.833 & 1.054 & 1.037 & 0.680 & 0.816 \\
 & 70\% & 0.595 & 0.612 & 0.629 & 0.612 & 0.544 & 0.595 & 0.578 & 0.612 & 0.425 & 0.629 & 0.765 & 0.765 & 0.544 & 0.612 \\
 & $\bar{M}_{max}$ & 0.958 & 1.000 & 0.956 & 0.951 & 0.974 & 0.967 & 0.952 & 0.951 & 1.000 & 1.000 & 0.871 & 0.876 & 0.984 & 1.000 \\
 & $\bar{M}_{max}/N_{flows}^{max}$ & 0.569 & 0.588 & 0.568 & 0.559 & 0.579 & 0.569 & 0.560 & 0.559 & 0.507 & 0.700 & 0.545 & 0.515 & 0.590 & 0.470 \\
\multirow{6}{*}{\rotatebox[origin=c]{90}{Doucet}} & 95\% & 0.459 & 0.629 & 0.629 & 0.629 & 0.323 & 0.459 & 0.510 & 0.476 & 0.357 & 0.833 & --- & --- & 0.442 & 0.476 \\
 & 90\% & 0.442 & 0.442 & 0.425 & 0.442 & 0.289 & 0.340 & 0.459 & 0.340 & 0.170 & 0.595 & --- & --- & 0.323 & 0.442 \\
 & 80\% & 0.272 & 0.272 & 0.323 & 0.323 & 0.170 & 0.255 & 0.340 & 0.255 & 0.136 & 0.476 & --- & 1.632 & 0.238 & 0.255 \\
 & 70\% & 0.238 & 0.238 & 0.255 & 0.272 & 0.136 & 0.221 & 0.255 & 0.153 & 0.102 & 0.374 & 0.850 & 1.088 & 0.221 & 0.170 \\
 & $\bar{M}_{max}$ & 0.988 & 1.000 & 0.987 & 0.986 & 0.999 & 0.997 & 0.987 & 0.994 & 1.000 & 1.000 & 0.793 & 0.806 & 0.999 & 1.000 \\
 & $\bar{M}_{max}/N_{flows}^{max}$ & 1.452 & 0.588 & 0.611 & 1.137 & 0.587 & 0.604 & 0.611 & 0.596 & 0.452 & 0.606 & 0.496 & 0.483 & 0.816 & 0.414 \\
\bottomrule
\end{tabular}
}
\caption{Particle C - Low inlet velocity.}
\end{table}
\vspace*{\fill}

\end{landscape}

\end{flushleft}

\end{document}